\documentclass[a4paper,fleqn,amssymb,usenatbib,useAMS,usedcolumns]{mnras}
\usepackage{times} 
\usepackage{latexsym,amssymb,amsmath,amsfonts}
\usepackage{rotfloat} 
\usepackage{rotating} 
\usepackage{float}
\usepackage{graphicx}
\usepackage{natbib}
\usepackage{supertabular}
\usepackage{longtable}
\usepackage{url}
\usepackage{dcolumn}
\usepackage{placeins}
\usepackage{multirow} 
\usepackage{textcomp}
\usepackage{multicol}
\usepackage{bm}
\usepackage{pdflscape}
\usepackage[utf8]{inputenc}
\usepackage[T1]{fontenc}
\usepackage{ae,aecompl}
\usepackage[caption=false]{subfig}
\begin{document}
\newcommand{\hi}{\mbox{H\,{\sc i}}}
\newcommand{\mgii}{\mbox{Mg\,{\sc ii}}}
\newcommand{\mgi}{\mbox{Mg\,{\sc i}}}
\newcommand{\feii}{\mbox{Fe\,{\sc ii}}}
\newcommand{\mnii}{\mbox{Mn\,{\sc ii}}}
\newcommand{\crii}{\mbox{Cr\,{\sc ii}}}
\newcommand{\tii}{\mbox{Ti\,{\sc ii}}}
\newcommand{\sii}{\mbox{Si\,{\sc ii}}}
\newcommand{\znii}{\mbox{Zn\,{\sc ii}}}
\newcommand{\caii}{\mbox{Ca\,{\sc ii}}}
\newcommand{\nai}{\mbox{Na\,{\sc i}}}
\def\lya{\ensuremath{{\rm Ly}\alpha}}
\def\lymana{\ensuremath{{\rm Ly}\alpha}}
\def\chis{$\chi^2$}
\def\kms{km\,s$^{-1}$}
\def\cms{cm$^{-2}$}
\def\cc{cm$^{-3}$}
\def\zabs{$z_{\rm abs}$}
\def\zem{$z_{\rm em}$}
\def\nhi{$N$($\hi$)}
\def\nh{$n_{\rm H}$}
\def\ne{$n_{e}$}
\def\ll{$\lambda\lambda$}
\def\l{$\lambda$}
\def\ha{H\,$\alpha$}
\def\hb{H\,$\beta$}
\def\oi{[O\,{\sc i}]}
\def\oii{[O\,{\sc ii}]}
\def\oiii{[O\,{\sc iii}]}
\def\nii{[N\,{\sc ii}]}
\def\sii{[S\,{\sc ii}]}
\def\wmg{$\rm W_{\rm r}^{\mgii}$}
\def\v90{$\Delta v_{90}$}
\def\ergs{$\rm erg\,s^{-1}$}
\def\ergscm{$\rm erg\,s^{-1}\,cm^{-2}$}
\def\ergscmarc{$\rm erg\,s^{-1}\,cm^{-2}\,arcsec^{-2}$}
\def\mstar{$\rm M_*$}
\def\msun{$\rm M_\odot$}
\def\msunyr{$\rm M_\odot yr^{-1}$}
%
%=============================================================================================
%
\title[Metal-enriched haloes around $z\sim1$ galaxies]{
{MUSE Analysis of Gas around Galaxies (MAGG) - II: Metal-enriched halo gas around $z\sim1$ galaxies}
\author[R. Dutta et al.]{Rajeshwari Dutta$^{1,2}$\thanks{E-mail: rajeshwari.dutta@durham.ac.uk}, Michele Fumagalli$^{3,1,2}$, Matteo Fossati$^{3,1,2}$, Emma K. Lofthouse$^{3,1,2}$,
\newauthor{J. Xavier Prochaska$^{4,5}$, Fabrizio Arrigoni Battaia$^6$, Richard M. Bielby$^{2,7}$, Sebastiano} 
\newauthor{Cantalupo$^8$, Ryan J. Cooke$^2$, Michael T. Murphy$^9$, John M. O'Meara$^{10}$} \\ 
$^1$Institute for Computational Cosmology, Durham University, South Road, Durham, DH1 3LE, UK \\
$^2$Centre for Extragalactic Astronomy, Durham University, South Road, Durham, DH1 3LE, UK \\
$^3$Dipartimento di Fisica G. Occhialini, Universit\`a degli Studi di Milano Bicocca, Piazza della Scienza 3, 20126 Milano, Italy \\
$^4$Department of Astronomy \& Astrophysics, UCO/Lick Observatory, University of California, 1156 High Street, Santa Cruz, CA 95064, USA \\
$^5$Kavli Institute for the Physics and Mathematics of the Universe (Kavli IPMU), 5-1-5 Kashiwanoha, Kashiwa, 277-8583, Japan \\
$^6$Max-Planck-Institut f\"ur Astrophysik, Karl-Schwarzschild-Str 1, D-85748
Garching bei M\"unchen, Germany \\
$^7$Institute for Data Science, Durham University, South Road, DH1 3LE, UK \\
$^8$Department of Physics, ETH Zurich, Wolgang-Pauli-Strasse 27, 8093, Zurich \\
$^9$Centre for Astrophysics and Supercomputing, Swinburne University of Technology, Hawthorn, Victoria 3122, Australia \\
$^{10}$W. M. Keck Observatory, 65-1120 Mamalahoa Hwy. Kamuela, HI 96743 \\
} 
}
\date{Accepted. Received; in original form }
\pubyear{}
\maketitle
\label{firstpage}
\pagerange{\pageref{firstpage}--\pageref{lastpage}}
%
%============================== ABSTRACT =================================================================================
%
\begin {abstract}  
\par\noindent
We present a study of the metal-enriched cool halo gas traced by \mgii\ absorption around 228 galaxies at $z\sim0.8-1.5$ within 28 quasar fields from the MUSE Analysis of Gas around Galaxies (MAGG) survey. We observe no significant evolution in the \mgii\ equivalent width versus impact parameter relation and in the \mgii\ covering fraction compared to surveys at $z\lesssim 0.5$. The stellar mass, along with distance from galaxy centre,  appears to be the dominant factor influencing the \mgii\ absorption around galaxies. With a sample that is 90\% complete down to a star formation rate of $\approx0.1$\,\msunyr\ and up to impact parameters  $\approx250-350$\,kpc from quasars, we find that the majority ($67^{+12}_{-15}$\% or 14/21) of the \mgii\ absorption systems are associated with more than one galaxy. The complex distribution of metals in these richer environments adds substantial scatter to previously-reported correlations. Multiple galaxy associations show on average five times stronger absorption and three times higher covering fraction within twice the virial radius than isolated galaxies. The dependence of \mgii\ absorption on galaxy properties disfavours the scenario in which a widespread intra-group medium dominates the observed absorption. 
This leaves instead gravitational interactions among group members or hydrodynamic interactions of the galaxy haloes with the intra-group medium as favoured mechanisms to explain the observed enhancement in the \mgii\ absorption strength and cross section in rich environments.
\end {abstract}  
%
%=========================== KEY WORDS ===================================================================================== 
%
%
%=========================== INTRODUCTION ================================================================================== 
%
\section{Introduction} 
\label{sec_intro}  
In the current theoretical paradigm of $\Lambda$ cold dark matter, which is supported by extensive observations, structures form hierarchically with galaxies forming first in collapsed dark matter haloes \citep{gunn1972,white1978}. The subsequent evolution of galaxies through cosmic time is closely governed by gas flow processes occurring in and around them. For instance, galaxies grow by accreting fuel for star formation either via cooling of hot halo gas or from cold gas streams of the cosmic web depending on their halo mass \citep{keres2005,dekel2006}. Outflows from star forming regions and nuclear activity in galaxies in turn replenish their surroundings with metals and inhibit further star formation in some cases \citep{oppenheimer2008,lilly2013,shull2014}. In this framework, the observed properties of galaxies like their colour and morphology are regulated by the competition of the above accretion and feedback processes. The immediate environment of galaxies where the processing of baryons to and from the intergalactic medium takes place is termed as the circumgalactic medium (CGM). Over the last few decades, it has become increasingly evident that a successful theory of galaxy formation and evolution requires us to understand the gas flow processes occurring within the CGM and their connection with the galaxies \citep{tumlinson2017}.

Beyond the CGM, the environment plays a fundamental role in shaping the way galaxies evolve. In the hierarchical structure formation model, galaxies assemble into more massive structures like groups and clusters, and only a small fraction of galaxies are observed to be in isolated environments. It is well established that morphology and physical properties of galaxies vary with the local galaxy density \citep{dressler1980,whitmore1993,boselli2006,wetzel2012}. These differences have been explained to be arising due to interactions between galaxies themselves or between galaxies and the hot and dense intra-cluster medium \citep{gunn1972,merritt1983,steinhauser2016,vandevoort2017}. As galaxies undergo complex encounters, the gas in their CGM, being more diffuse and extended, should be even more susceptible to the effects of gravitational interactions like tidal stripping, or to hydrodynamic interactions. Indeed at low redshifts ($z<0.1$), the effects of galaxy interactions within groups and clusters are manifested on the cool neutral hydrogen and molecular gas in the form of tidal streams, plumes, fountains, high-velocity clouds and warped discs \citep{fraternali2002,sancisi2008,mihos2012,jachym2014,ramatsoku2019,moretti2020}, as well as on the ionized gas \citep{fumgalli2014,poggianti2017,vulcani2018,fossati2019a}. Galaxy mergers are also more likely in denser environments, and merger-driven outflows could affect the gas content and distribution in the CGM \citep{hani2018}.

At high redshifts, it becomes observationally challenging to directly detect the diffuse CGM in emission \citep{wisotzki2018,umehata2019}. On the other hand, absorption lines arising from this diffuse gas, detected in the spectra of bright background sources like quasars, have long been powerful tools to study the kinematics, ionization, chemical content and physical properties of the CGM. Large samples of background quasars at small projected separation from foreground galaxies have been used to statistically map different phases of the CGM and the baryon content around different galaxy types, especially at $z<1$ \citep[e.g.][]{prochaska2011,stocke2013,tumlinson2013,werk2014,chen2018}. The \mgii\,\ll2796, 2803 absorption doublet traces the cool ($T\sim10^4$\,K) photo-ionized gas, and being accessible from the ground over a large redshift range ($0.2\lesssim z \lesssim2$), has been used extensively to study the galaxy-CGM connection \citep[e.g.][]{bergeron1986,steidel1992,churchill2000,chen2010,lehner2013,nielsen2013a,rubin2018,lan2020,huang2020}.

The above studies have found that the amount and cross section of metals in the CGM as traced by \mgii\ absorption decline with increasing distance from galaxies. The \mgii\ absorbing haloes around $z<0.5$ galaxies extends out to $\sim100$\,kpc and the metal distribution is patchy within the halo radius with a covering fraction of $\sim0.5-0.7$ for \mgii\ absorbers with rest-frame equivalent width $\ge0.3$\,\AA\ \citep{kacprzak2008,chen2010}. Further, the properties of the cool CGM have been found to be different around passive and star forming galaxies. The radial extent and covering fraction of \mgii\ absorbing gas have been observed to depend on galaxy luminosity and colour, with blue, more luminous galaxies having more extended haloes and higher covering fractions \citep{chen2010,bordoloi2011,nielsen2013b,lan2014}. 

In addition, the \mgii\ absorbing gas has been observed to exhibit a non-symmetric geometric distribution around the galaxies, with the strongest absorption occurring close to the galaxy minor axis and likely to originate in outflows \citep{bouche2012,kacprzak2012,schroetter2016,lan2018}. However, the majority of the studies involving \mgii\ absorption have focused on galaxy-absorber pairs, without a full characterization  of the environment. Presence of galaxy interactions in denser environments is most likely to complicate the simple picture of the CGM in which metal-poor gas accretes onto galaxies along the galaxy major axis and metal-enriched gas is expelled along the minor axis in the form of winds and outflows.

In one of the initial works on \mgii\ absorption in denser environments, \citet{lopez2008} found that strong \mgii\ absorbers are significantly overabundant in cluster environments, especially closer to cluster centres and near more massive clusters. This excess of strong absorbers could be related to the absorption arising in clusters having larger velocity spread due to galaxy interactions as compared to absorption arising around field galaxies. The study also found relatively fewer weak \mgii\ absorbers, suggesting cluster galaxies have truncated haloes due to stripping of the cold gas by the hot intra-cluster medium \citep{padilla2009}. 

When it comes to less massive structures like groups, statistical studies have found that \mgii\ absorption detected around galaxies in a group environment is less correlated with distance from the galaxy centres and more radially extended compared to isolated galaxies \citep{chen2010,bordoloi2011,nielsen2018}. Further, \citet{nielsen2018} have found that groups show stronger \mgii\ absorption and higher covering fraction compared to isolated galaxies. They suggest that the \mgii\ absorption is associated with the intra-group medium rather than any individual galaxies in the group. Besides the above, there have been studies of single \mgii\ systems that are associated with pairs and groups of galaxies \citep{whiting2006,kacprzak2010,nestor2011,gauthier2013,bielby2017,peroux2017,peroux2019,klitsch2018,rahmani2018}. Different origins for the observed absorption in these systems have been proposed, that include tidal interactions, intra-group medium, outflows driven by starburst that is triggered by interactions, and cold-flow accretion onto a warped disc of a member galaxy.

In addition to studies of \mgii\ in group environment, there have been extensive studies on \mgii\ absorption associated with luminous red galaxies (LRGs). The bias estimated for LRGs \citep{padmanabhan2007} indicate that they reside in massive haloes (halo mass, M$_{\rm h}\gtrsim 10^{13}$\,\msun), and also in relatively denser environments \citep{tal2012}. \mgii\ absorbers are found to cluster strongly with LRGs \citep{bouche2006,lundgren2009,gauthier2009,zhu2014}, and the mean covering fraction of \mgii\ gas within $\approx100$\,kpc of LRGs is found to be $\approx15$\% \citep{huang2016}. In the absence of starburst-driven outflows in these quiescent haloes, environmental effects like tidal and ram-pressure stripping of cool gas from satellite galaxies could contribute to the wide-spread presence of metals around LRGs.

Most of the large statistical studies of \mgii\ absorption around galaxies until recently have been based on identifying galaxies close to quasar sightlines in wide-field images and following up with long-slit or multi-object spectroscopy of the galaxies. Wide-field optical integral field unit (IFU) spectrographs like the Multi Unit Spectroscopic Explorer \citep[MUSE;][]{bacon2010} at the Very Large Telescope (VLT) have revolutionised the study of CGM by enabling efficient and sensitive spectroscopic surveys complete down to a given flux limit around quasar sightlines \citep[e.g.][]{fumagalli2016,fumagalli2017,schroetter2016,mackenzie2019,fossati2019b,muzahid2020,zabl2019,lofthouse2020}. Such type of surveys allow a more complete characterization of the small-scale galaxy environment, a complete (flux limited) analysis of the CGM-galaxy connection up to $\lesssim250$\,kpc from the quasar sightlines, and enable studies of CGM to lower mass galaxies (\mstar$<10^{10}$\,\msun) at $z\ge1$.

While the MUSE guaranteed time observations (GTO) survey of \mgii\ absorbers, MusE GAs FLOw and Wind (MEGAFLOW), is studying outflows and gas accretion traced by strong \mgii\ absorption around individual galaxies at $0.5<z<1.5$ \citep{schroetter2016,zabl2019}, the recent MUSE survey of \citet{hamanowicz2020} have found that most of the \mgii\ absorption line systems in their sample are associated with two or more galaxies. Further, \citet{fossati2019b} have found that associations and groups (M$_{\rm h}\approx10^{11-13.5}$\,\msun) detected in the MUSE Ultra Deep Field are linked to stronger \mgii\ absorption compared to isolated galaxies. While they do not find evidence for the presence of widespread intra-group gas, their results suggest that gravitational interactions within groups could be responsible for stripping gas from member galaxy haloes and thus boosting the cross section of \mgii\ absorbing gas.

In this work, we study the gaseous haloes as traced by \mgii\ absorption around $z\sim1$ galaxies also in relation to their environment within the MUSE Analysis of Gas around Galaxies (MAGG) survey. This survey comprises MUSE observations of 28 fields centred on $z=3.2-4.5$ quasars that show strong \ion{H}{I} absorption lines at $z>3$ in their high-resolution optical spectra, which are therefore ideal to address the \mgii-galaxy connection in an unbiased way at lower redshift. The MUSE data are primarily from our VLT large programme (ID 197.A$-$0384, PI: M. Fumagalli), which are supplemented by data from the MUSE GTO \citep[PI: J. Schaye;][]{muzahid2020}. The survey strategy and methodology, including sample selection and data processing, have been presented in \citet{lofthouse2020}. The main thrust of this project is the study of the CGM of $z=3-4$ galaxies detected via \lymana\ emission and strong \hi\ absorption (\nhi\ $\gtrsim10^{17}$\,\cms; Lofthouse et al. in prep.). The rich MUSE datasets are further being utilised to study the environment around the $z>3$ quasars (Fossati et al. in prep.), and the CGM around lower redshift galaxies as in this work.

This paper is organized as follows. First we describe the sample of \mgii\ absorption line systems and galaxies used in this work in Sections~\ref{sec_analysis_mgii} and \ref{sec_analysis_galaxies}, respectively. Then in Section~\ref{sec_results}, we study the connection between \mgii\ absorption and galaxies, including the detection rate of associated galaxies (Section~\ref{sec_results_detection}), radial (Section~\ref{sec_results_radial}) and azimuthal distribution (Section~\ref{sec_results_azimuthal}) of \mgii\ absorption around galaxies, dependence of \mgii\ absorption on galaxy properties (Section~\ref{sec_results_galprop}), the covering fraction of \mgii\ absorbing gas (Section~\ref{sec_results_covfrac}), and redshift evolution in \mgii-galaxy trends (Section~\ref{sec_results_redshift}). Next, we investigate in detail the effect of environment on the observed \mgii\ absorption in Section~\ref{sec_environment}. Finally, we discuss and summarize the results of this work in Section~\ref{sec_summary}. Throughout this paper we use a Planck 15 cosmology with $H_{\rm 0}$ = 67.7\,\kms\,Mpc$^{-1}$ and $\Omega_{\rm M}$ = 0.307 \citep{planck2016}.
%
%==================================================== ANALYSIS =========================================================

%
\begin{figure}
    \centering
    \includegraphics[width=0.5\textwidth]{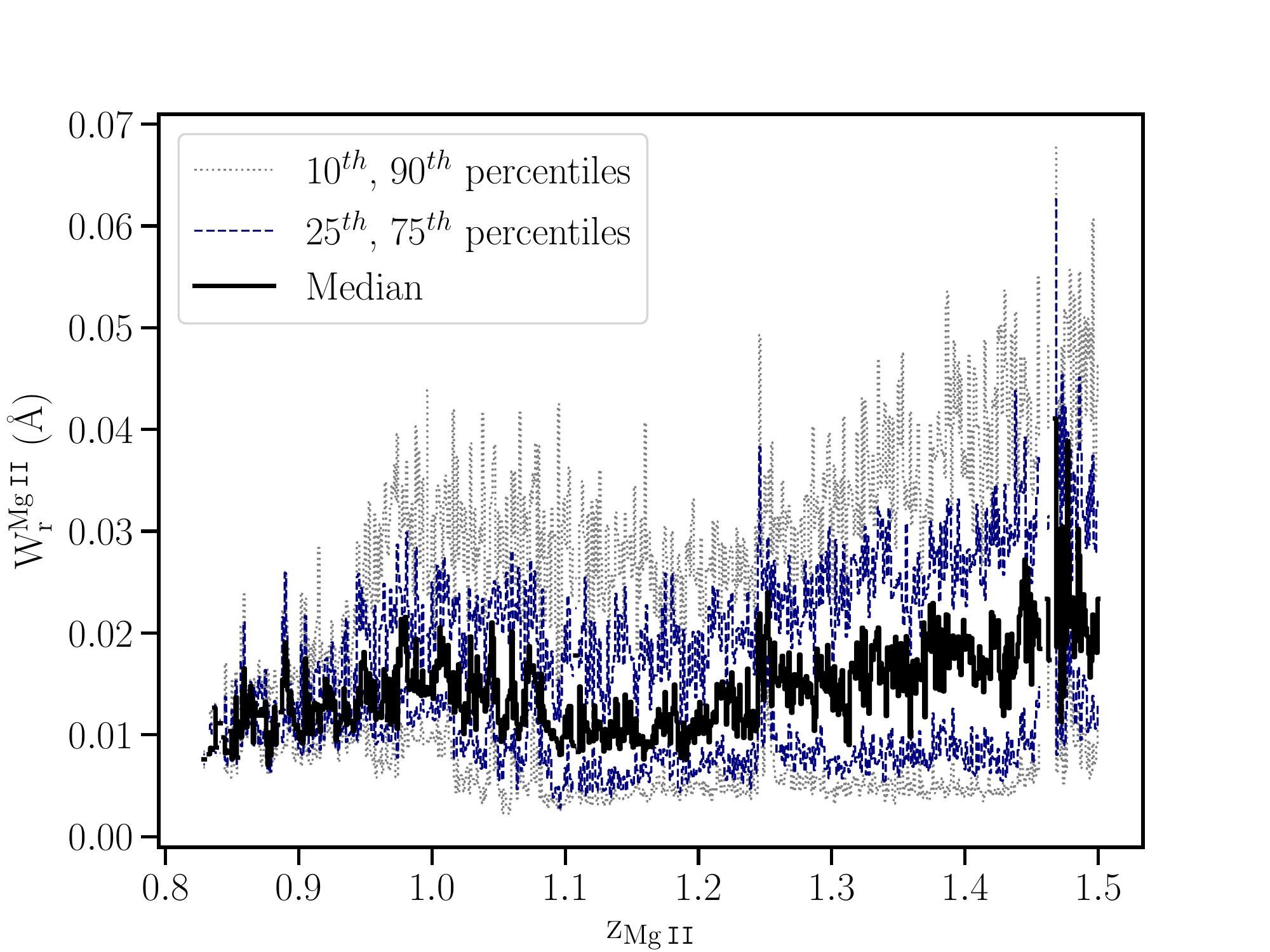}
    \caption{The $10^{\rm th}, 25^{\rm th}, 50^{\rm th}, 75^{\rm th}$ and $90^{\rm th}$ percentiles of the $3\sigma$ sensitivity of rest-frame equivalent width of \mgii\ absorption for the sample of 28 quasar spectra, as a function of redshift, from redward of the \lya\ forest up to $z=1.5$.}
    \label{fig:ew_limits}
\end{figure}  
\section{Data Analysis}
\label{sec_analysis}
\subsection{\mgii\ absorption line systems}
\label{sec_analysis_mgii}
All the quasars in our sample have high resolution spectra obtained with the Ultraviolet and Visual Echelle Spectrograph \citep[UVES;][]{dekker2000} at the VLT, the High Resolution Echelle Spectrometer \citep[HIRES;][]{vogt1994} at Keck, or the Magellan Inamori Kyocera Echelle \citep[MIKE;][]{bernstein2003} at the Magellan telescopes. These are often supplemented with medium resolution spectra from X-SHOOTER \citep{vernet2011} at the VLT and Echellette Spectrograph and Imager \citep[ESI;][]{sheinis2002} at Keck. Details of the reduction of these archival spectra and their properties are summarised in section 3.1 and table 2 of \citet{lofthouse2020}, respectively. 

We searched for the doublet lines of \mgii\,\ll2796, 2803 by visually inspecting the quasar spectra, restricting to wavelengths redward of the quasar \lymana\ forest for clarity of identification (i.e. $z\gtrsim0.8$). We identified in total 114 \mgii\ absorption line systems over $0.8<z<3.8$. For the purpose of this paper, we focus on the 27 systems present in the redshift range $0.8<z<1.5$, where we are able to search for the corresponding \oii\ emission line in the MUSE spectra. Note that the quasar fields in the MAGG sample were chosen on the presence of Lyman Limits Systems (LLS) at $z>3$. Therefore, the \mgii\ sample presented here is largely thought to be unbiased and likely representative of the underlying absorber population. This is in contrast to the other recent MUSE studies based on pre-selection of known \mgii\ systems \citep[e.g.][]{schroetter2016,hamanowicz2020}. 

To assess our sensitivity to detect \mgii\ absorption, we estimate the $3\sigma$ upper limits on the rest-frame equivalent width of the \mgii\ \l2796 line (\wmg) over 100\,\kms, after masking out strong absorption lines in the quasar spectra, from redward of the \lya\ forest up to $z=1.5$. In Fig.~\ref{fig:ew_limits}, we plot the $10^{\rm th}, 25^{\rm th}, 50^{\rm th}, 75^{\rm th}$ and $90^{\rm th}$ percentiles of the estimated \wmg\ sensitivity of the quasar sample as a function of redshift. The median \wmg\ sensitivity is $\approx 0.01$\,\AA, and ranges from $\approx 0.001$\,\AA\ to $\approx 0.1$\,\AA\, with 90\% of the quasar spectra having a sensitivity $\lesssim0.03$\,\AA. 

\begin{figure}
    \centering
    \includegraphics[width=0.5\textwidth]{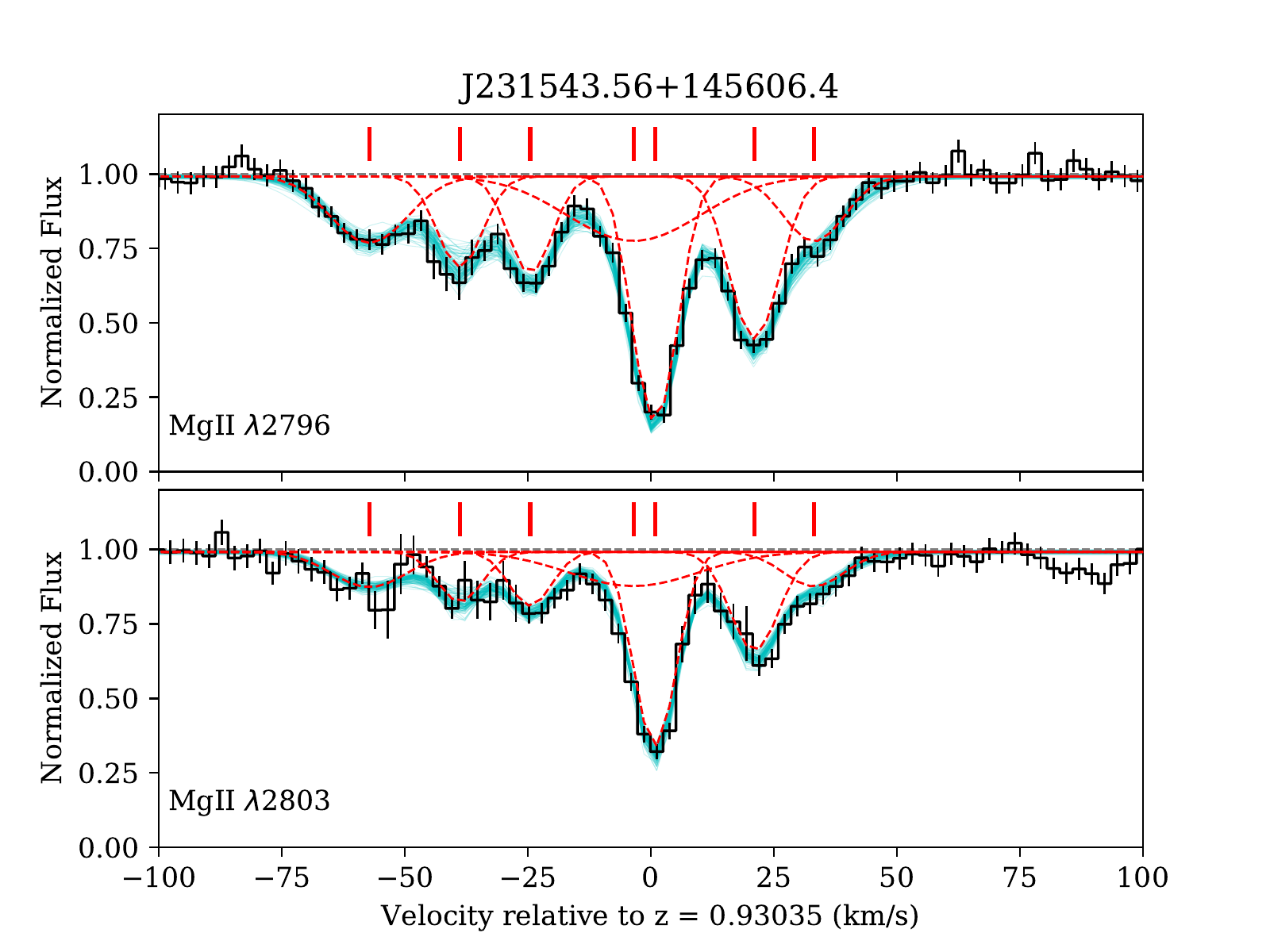}
    \caption{Example of \mgii\ doublet absorption line systems in our sample. The normalized spectrum is shown as black histogram and the black ticks are the $1\sigma$ error bars. The Voigt profile models, sampled from the posterior distribution estimated by MC-ALF, are shown as solid cyan lines. Individual \mgii\ absorption components are shown as dashed red lines and their positions are marked by vertical red ticks.}
    \label{fig:mgii_absorption}
\end{figure}

The properties of the \mgii\ systems are summarized in Appendix~A  (available online). In our working definition, all absorption within 500\,\kms\ of the line centroid is considered to be part of one system. To put this into perspective, note that the virial velocity of $z\sim1$ galaxies with stellar mass in the range $10^{9-11}$\,\msun\ will be $\sim100-300$\,\kms. In two cases, we find two adjacent absorption clumps with their peaks separated by $\sim200-300$\,\kms. Also in these cases, we consider them as part of one system. In case of more than one spectra being available, we use the one with the highest spectral dispersion, and in case of comparable dispersion, the one with the highest signal-to-noise ratio (S/N) in the region of the \mgii\ absorption.

We define the redshift of the \mgii\ systems as that where the cumulative optical depth profile of the absorption reaches 50\% of the total value. For saturated pixels, we take their flux value to be three times the error. The \wmg\ lies in the range of 0.016\,\AA\ to 3.23\,\AA, with a median value of 0.12\,\AA. The velocity width containing 90\% of the total optical depth  of the absorption profile (\v90) varies from $\sim10$\,\kms\ to $\sim350$\,\kms. The total redshift path covered over $z=0.8-1.5$ is $\Delta z$ = 12, which gives the number of \mgii\ absorbers per unit redshift as $dN/dz = 2.2\pm0.4$. The $dN/dz$ for the population of weak (\wmg\ $\le0.3$\,\AA) and strong (\wmg\ $\ge1$\,\AA) \mgii\ absorbers is $1.6\pm0.4$ and $0.3\pm0.2$, respectively. These values are consistent with those that have been estimated from statistical samples of \mgii\ absorbers in the literature over a similar redshift range \citep{nestor2005,prochter2006,narayanan2007,seyffert2013}.

We fit the \mgii\ doublet lines with Voigt profiles using the code MC-ALF (Fossati et al. in prep), as described in \citet{fossati2019b}. In brief, the code uses Bayesian statistics to estimate the minimum number of Voigt profile components required to model the absorption. Each component is defined by redshift, Doppler parameter ($b$) and column density ($N$). We define priors on $b$ and $N$ to be uniform in the range $1\le b$ (\kms) $\le 35$ and $11.5 \le$ log\,$N$(\cms) $\le 17.0$, respectively. The prior on redshift is defined visually by the wavelength range over which the absorption occurs. The continuum around each absorption line is first normalised \citep[see][for details]{lofthouse2020}, and we then include a multiplicative constant as a free parameter (allowed to vary between 0.98 and 1.02) to account for continuum fitting uncertainty. We also include `filler' Voigt components, wherever required based on visual inspection, to account for blended absorption lines that are physically unrelated to the \mgii\ absorption system of interest. Among the optimal fits (those with Akaike Information Criterion within 5 of the lowest value), we choose the one with the minimum number of components. The number of components required to model the absorption vary from the simplest systems with single component to the most complex system comprising 18 components. If the absorption is saturated ($N\gtrsim10^{14}$\,\cms), we take the column density of the \mgii\,\l2803 line estimated from the apparent optical depth method \citep{savage1991} as a lower limit to the total column density. An example of a \mgii\ system is shown in Fig.~\ref{fig:mgii_absorption}. All the \mgii\ systems along with their best Voigt profile fits are shown in the online Appendix~A.

In addition to analysing the above \mgii\ absorption line systems, we estimate $3\sigma$ upper limits on \wmg\ for all continuum-detected galaxies in our catalogue that are not associated with an \mgii\ absorber (Section~\ref{sec_analysis_continuum}), and whose redshifts fall redward of the quasar \lya\ forest and below $z\sim1.5$. We measure the \wmg\ upper limits within a velocity window of 100\,\kms\ (the median \v90\ of the \mgii\ lines in our sample) centred at the redshifts of the galaxies.
\begin{figure}
    \centering
    \includegraphics[width=0.5\textwidth]{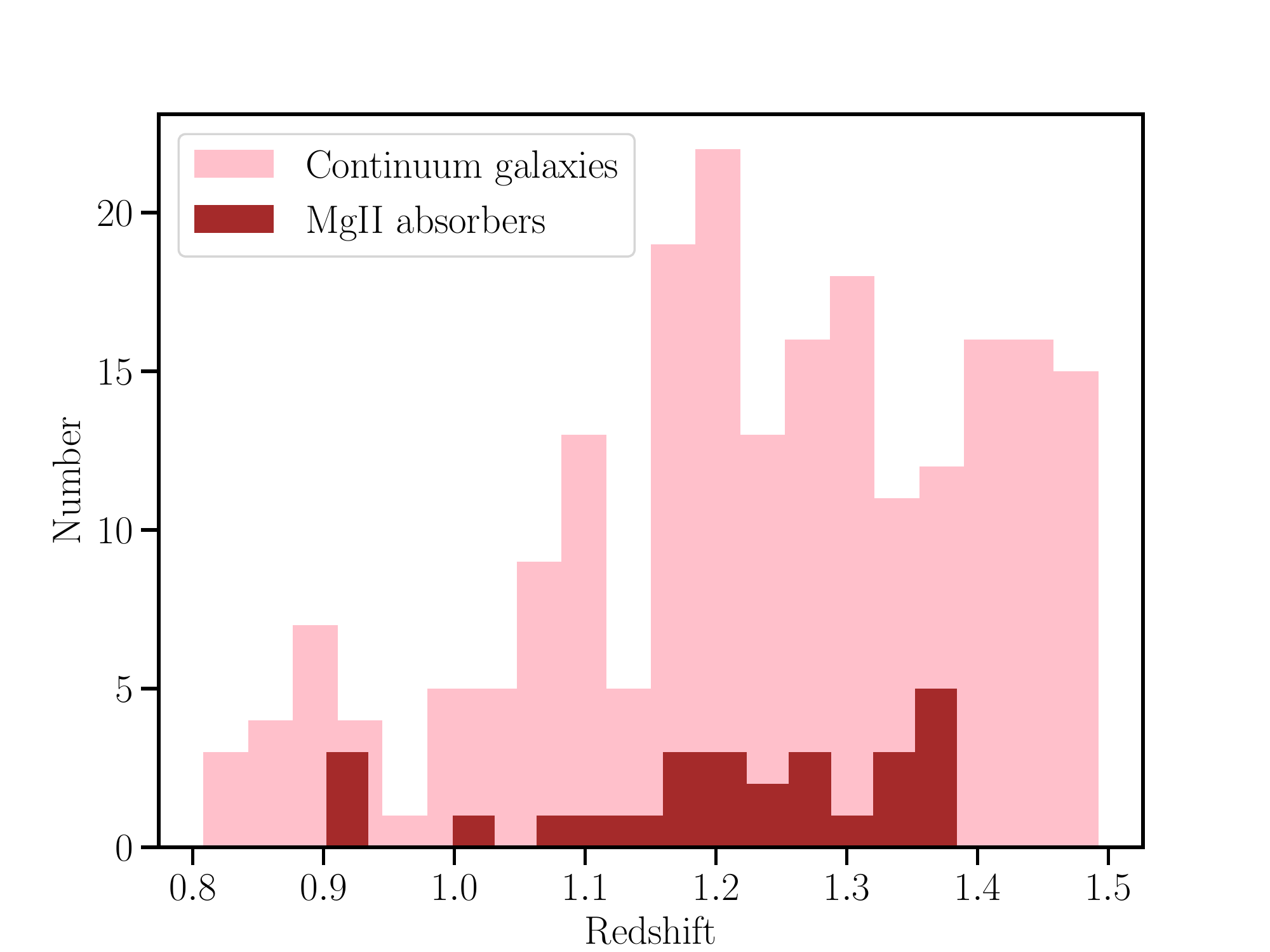}
    \caption{Histogram (pink) of the redshifts of the continuum-detected galaxies in MAGG fields at $0.8<z<1.5$, for which we can measure the associated \mgii\ absorption properties in the quasar spectra. Also shown in red is the histogram of the redshifts of the \mgii\ absorbers (with \wmg\ $\gtrsim0.02$\,\AA) identified in the quasar spectra over this redshift range. }
    \label{fig:z_histogram}
\end{figure}  
\begin{figure*}
    \centering
     \begin{minipage}{10cm}
      \centering
      \includegraphics[height=0.35\textheight]{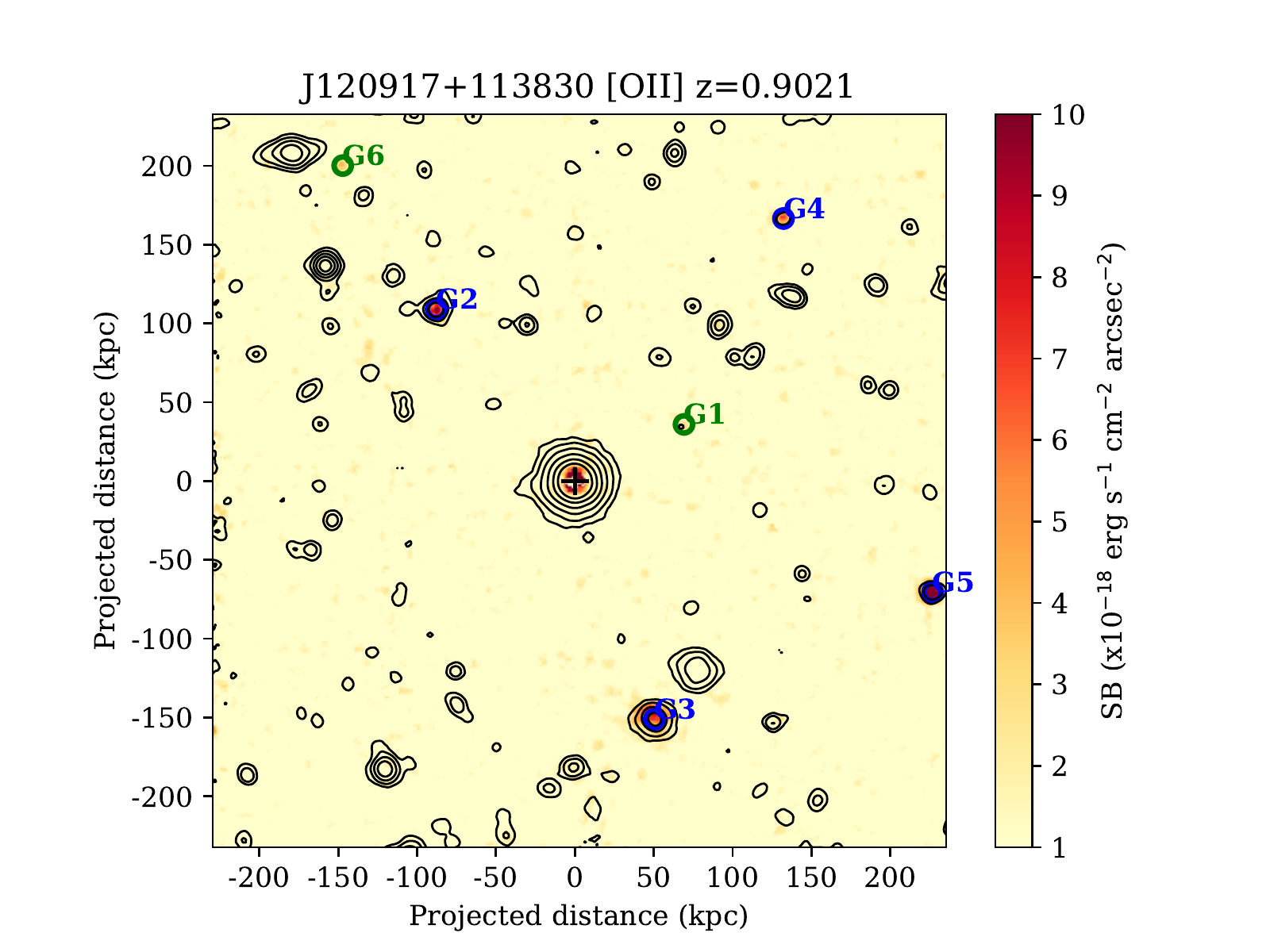}
    \end{minipage}
    \hspace{1mm}
    \begin{minipage}{3cm}
      \centering
      \includegraphics[height=0.12\textheight]{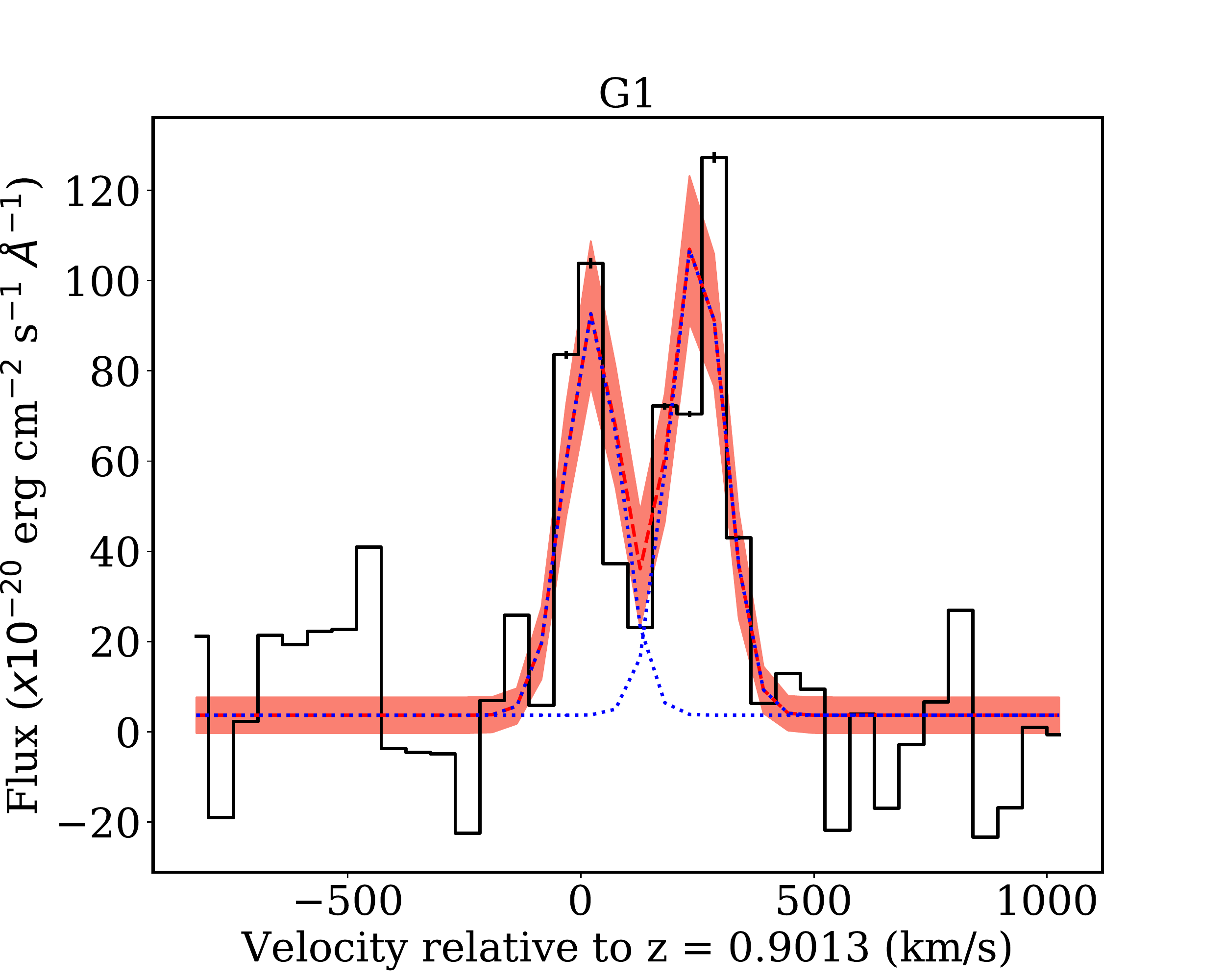} \\
      \includegraphics[height=0.12\textheight]{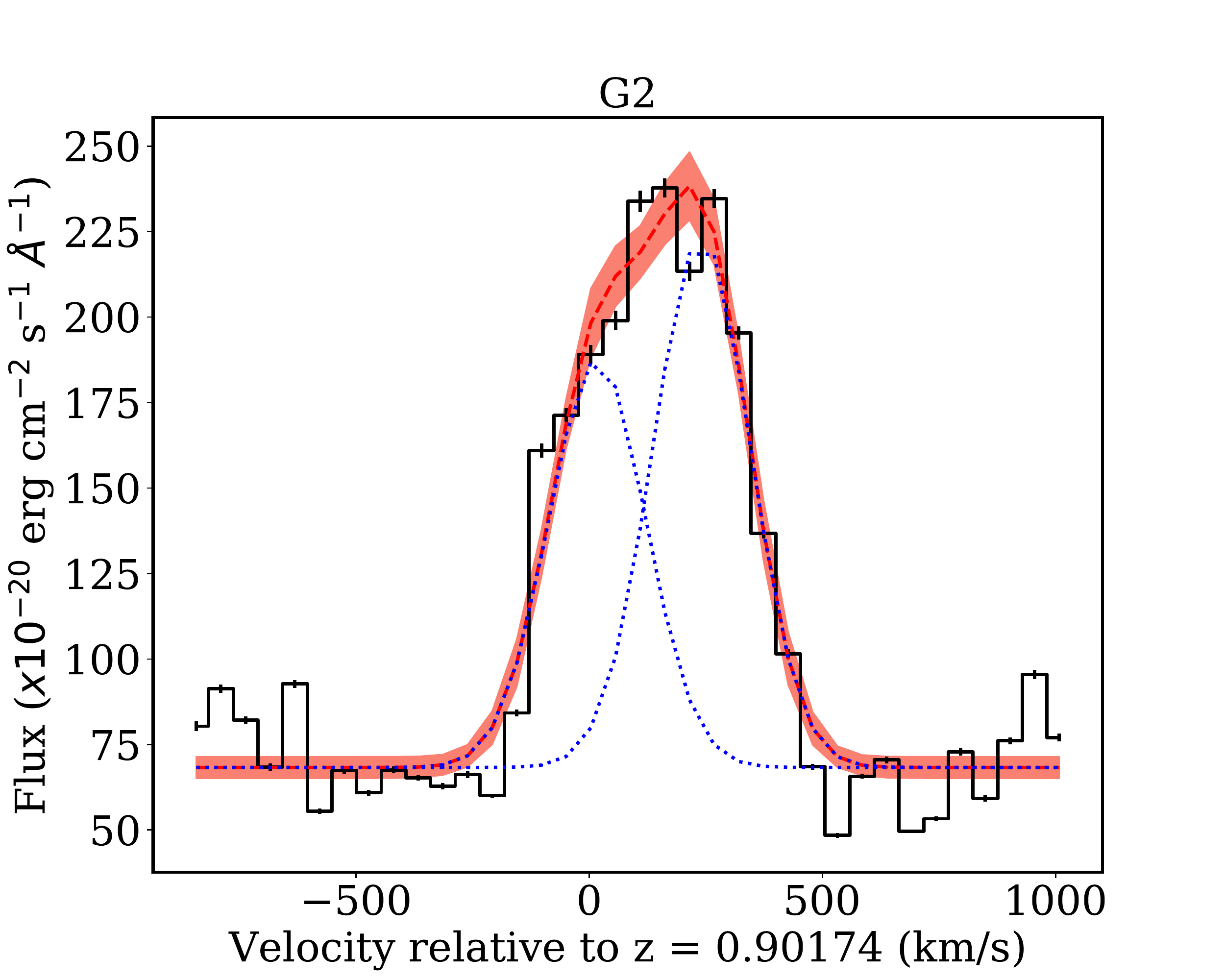} \\
      \includegraphics[height=0.12\textheight]{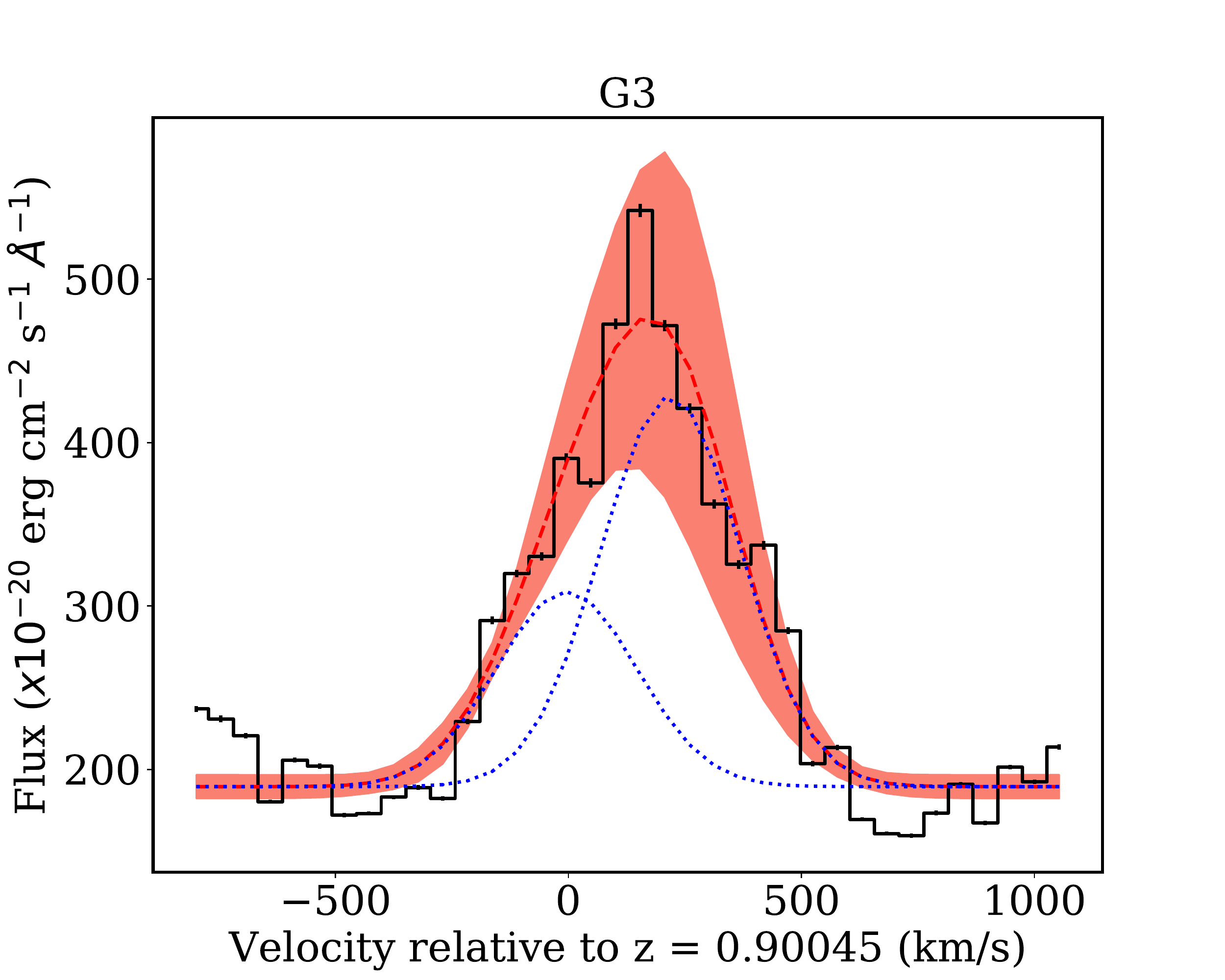}
    \end{minipage}
    \hspace{1mm}
    \begin{minipage}{3cm}
      \centering
      \includegraphics[height=0.12\textheight]{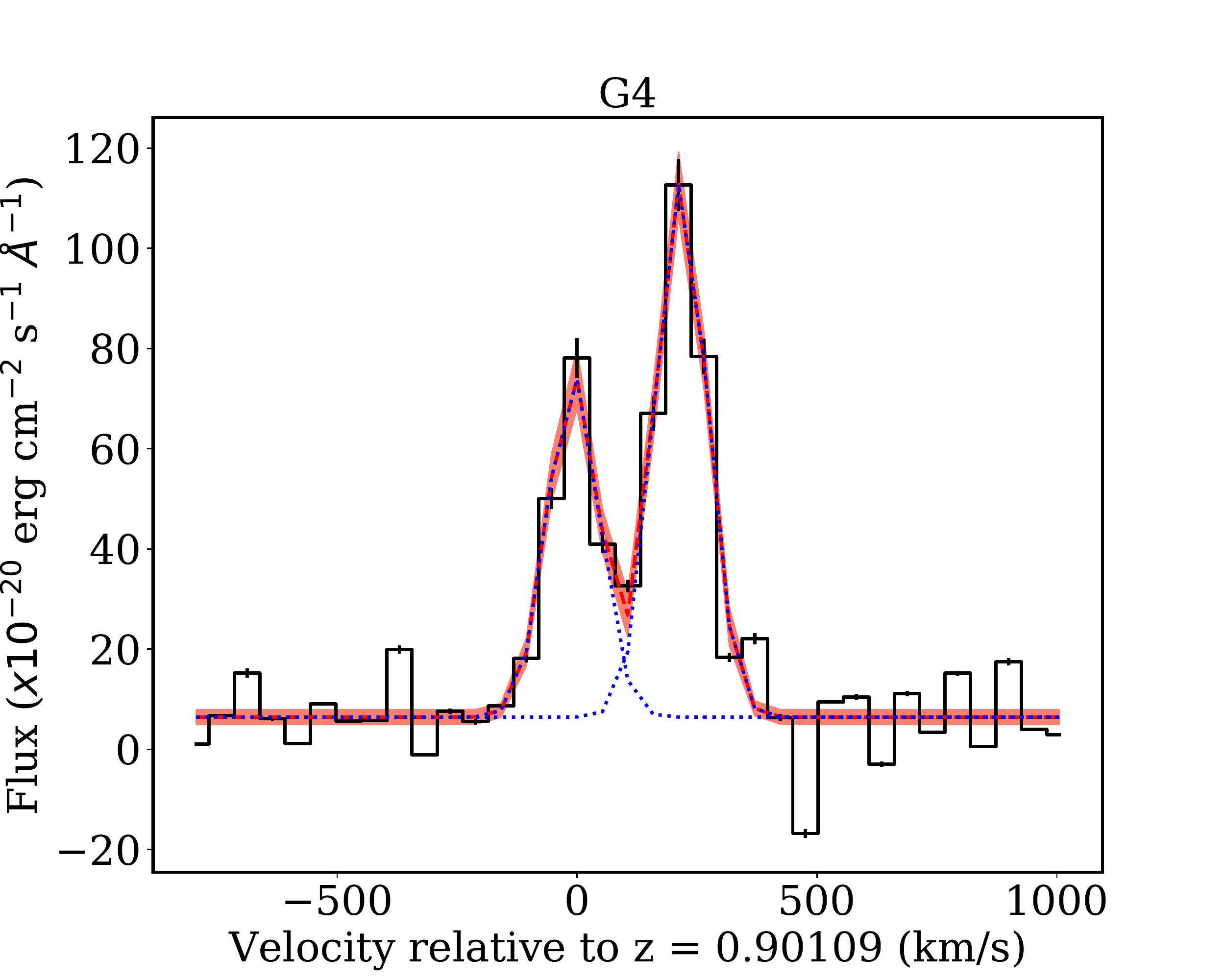} \\
      \includegraphics[height=0.12\textheight]{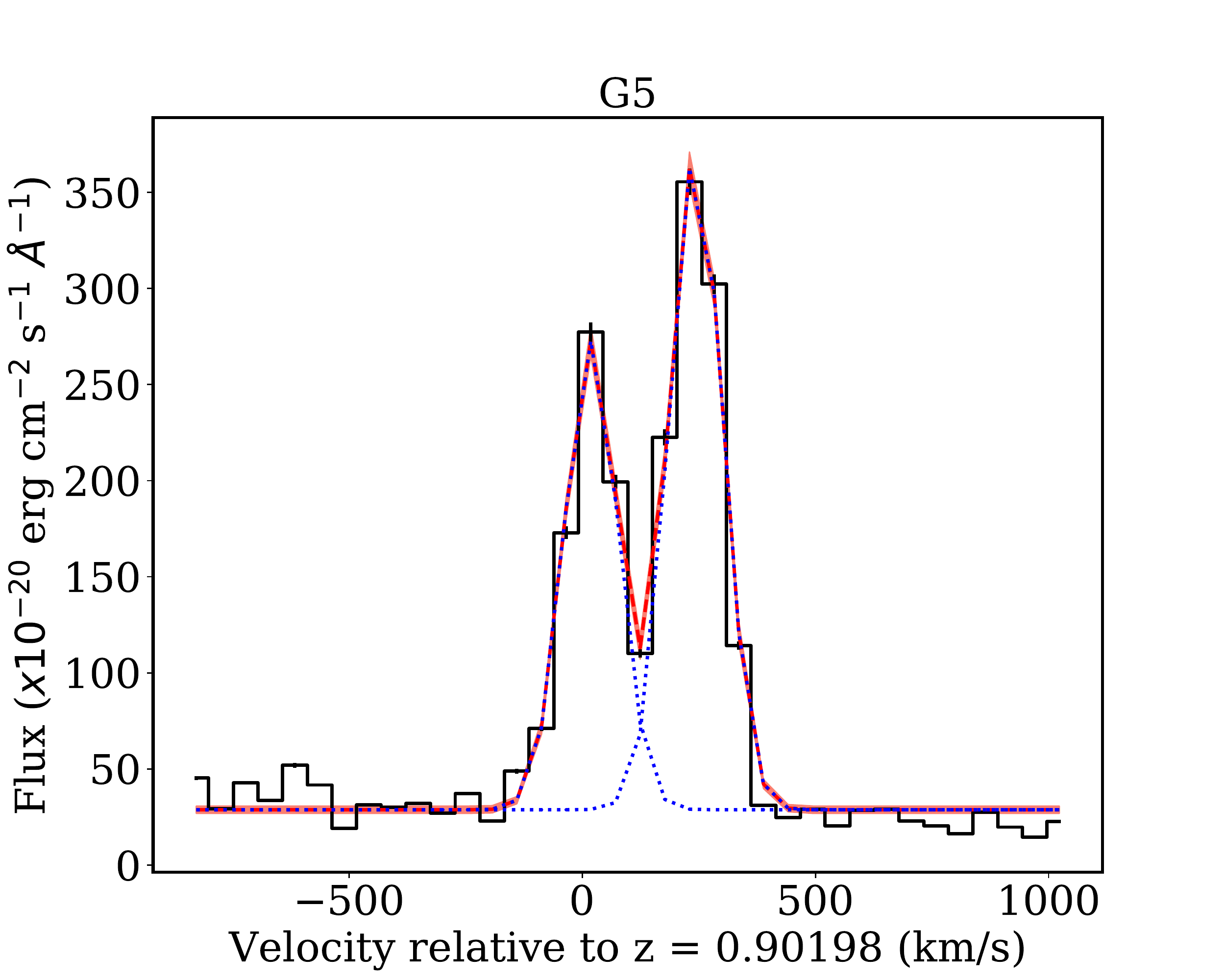} \\
      \includegraphics[height=0.12\textheight]{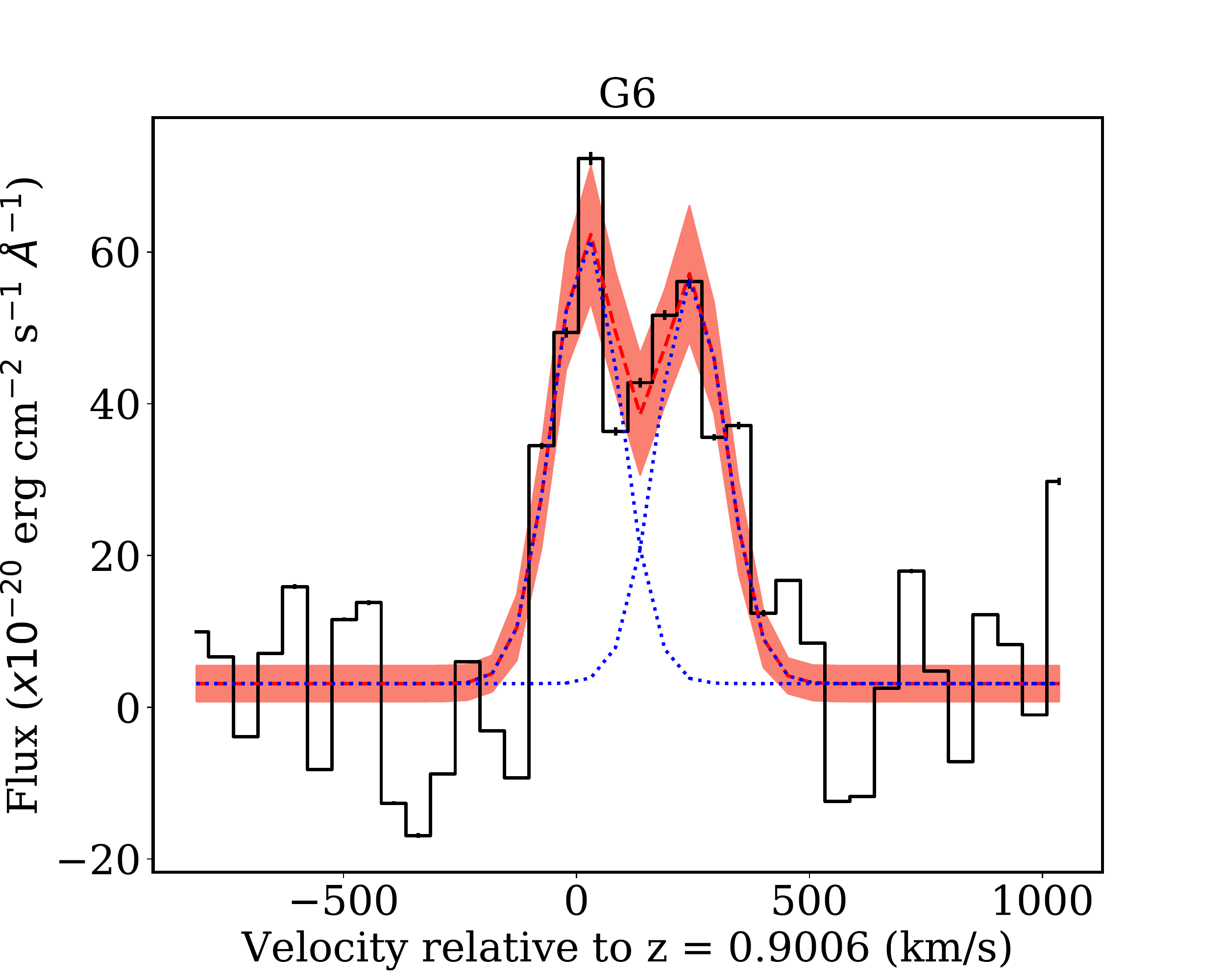}
    \end{minipage}
    \caption{Example of galaxy detection in continuum and \oii\ emission. {\it Left:} \oii\ pseudo-narrow band image (over $\sim1000$\,\kms) centred at the redshift of an \mgii\ absorber ($z=0.9021$) in the MAGG field J120917$+$113830. The galaxies detected in continuum are marked with blue circles, and those detected in line emission only are marked with green circles. The quasar position is marked by a plus sign. The black contours show all the continuum sources detected in this field at levels of 22, 23, 24, 25, 26 and 27 mag arcsec$^{-2}$. The image has been smoothed by a top-hat kernel of width $0.4~$arcsec for display purpose.
    {\it Right:} 1D spectra of the galaxies marked in the narrow-band image to the left, showing the \oii\ emission. The double Gaussian fit to the \oii\ line is shown by solid red lines. Fits to each component are shown by dotted blue lines. The shaded region shows the $1\sigma$ uncertainty in the fit.}
    \label{fig:cont_galaxies_map}
\end{figure*}

\subsection{Galaxies in the MUSE data}
\label{sec_analysis_galaxies}
In each of the MAGG fields we identify galaxies detected in continuum in the MUSE white-light images as well as in line emission in the MUSE 3D cubes. For the purpose of this paper, we focus on the continuum-detected galaxies within $0.8<z<1.5$. We cross-match these with the \mgii\ absorbers described in Section~\ref{sec_analysis_mgii}. In addition, we specifically search for \oii\ line-emitting galaxies around the redshift of the \mgii\ absorbers at $0.8<z<1.5$, that are too faint to be detected in the continuum. A blind search for line-emitting galaxies over the full MUSE wavelength range will be presented in a future work. We refer to \citet[][section 5]{lofthouse2020} for a detailed description of the methodology adopted to detect continuum and line emitting galaxies in the MAGG fields. Below we describe the analysis of these galaxies.
\begin{figure*}
    \centering
    \includegraphics[width=1.0\textwidth]{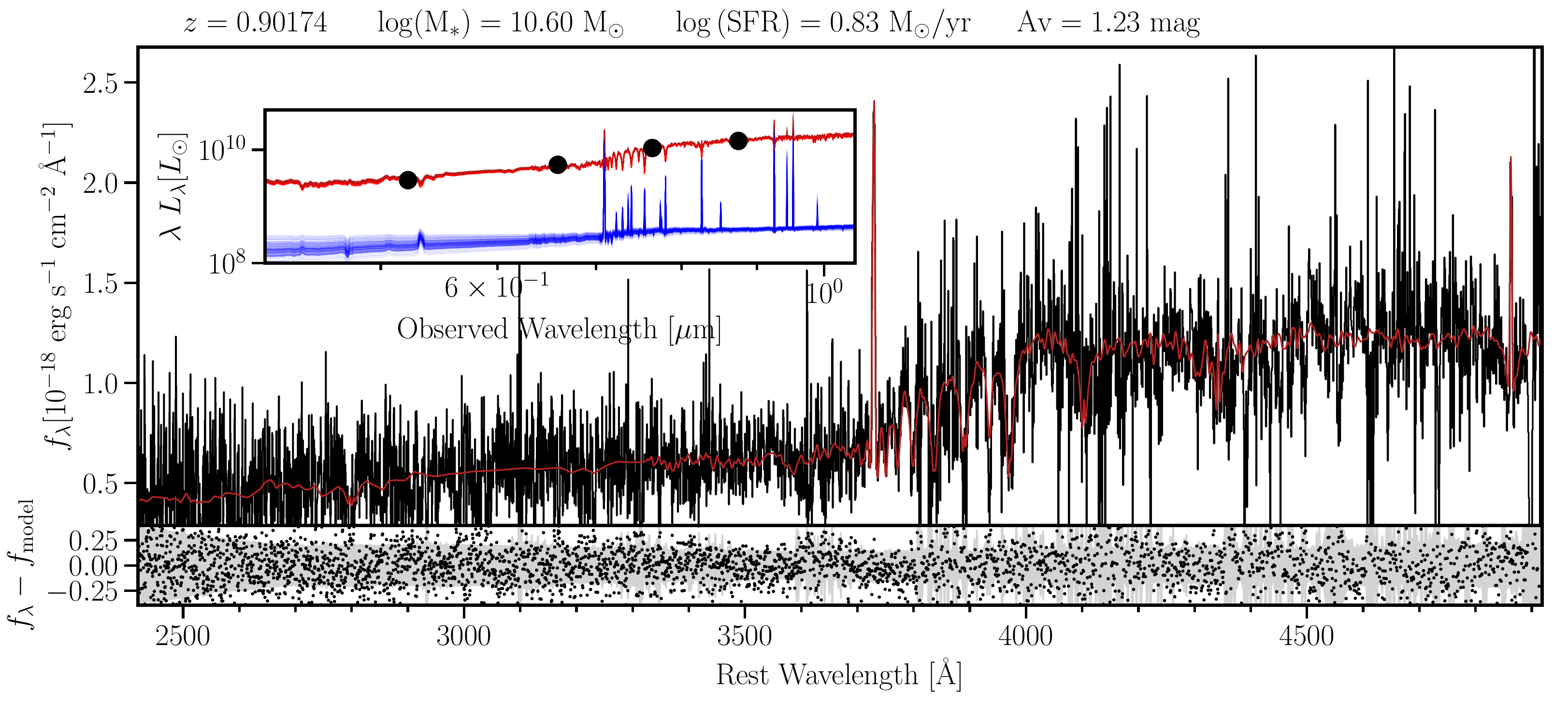}
    \caption{Example of SPS fitting results using MC-SPF on one of the galaxies in our continuum sample. The MUSE spectrum (black line) of the galaxy is plotted with the best fit model (red line). Parameters derived from the fit are listed on the top of the plot. The residuals of the fit are shown in the bottom panel, with the shaded region showing the noise in the data. The inset shows the MUSE photometry in four bands overlaid on the spectral energy distributions (red lines) obtained from MC-SPF. The contribution of young stars ($<10$ Myr) to the spectral template is shown in blue.}
    \label{fig:sps_fit}
\end{figure*}

\subsubsection{Continuum-detected galaxies}
\label{sec_analysis_continuum}

The MAGG catalogue of continuum-detected galaxies is 100\% complete down to an $r$-band magnitude of $\sim25.5$~mag and 90\% complete down to $\sim26.3$~mag \citep[see fig. 6 of][]{lofthouse2020}. This completeness is based on an empirical determination of the number counts of continuum-detected sources as a function of $r$-band magnitude. There are 411 galaxies in this catalogue at $0.8<z<1.5$ out of a total of $\approx$1000 galaxies with reliable spectroscopic redshifts measured using {\sc marz} \citep{hinton2016}. Note that these galaxies have redshifts with confidence flag 3 (one high S/N emission or absorption line and other low S/N emission and/or absorption features) and flag 4 (multiple high S/N emission and/or absorption lines), according to the categorization scheme listed in \citet{bielby2019} and \citet{lofthouse2020}. Hence, we are confident of the \oii\ doublet line identification in these galaxies based on other emission and/or absorption features in the spectra, even when the \oii\ line is only partially resolved. 

Among the above, we identify 214 galaxies for which we can obtain a clean measurement of the associated \mgii\ absorption in the quasar spectra, i.e. outside of the quasar \lya\ forest and free from contamination by other strong absorption lines at different redshifts. The distribution of redshifts of these galaxies along with that of the \mgii\ absorbers in our sample is shown in Fig.~\ref{fig:z_histogram}. Fig.~\ref{fig:cont_galaxies_map} shows an example of the galaxies detected in one of the MAGG fields. We extract the 1D spectra of the galaxies using the 2D segmentation maps created by SE{\sc xtractor} \citep{bertin1996}. We derive the \oii\ luminosity by integrating the 1D spectra within $\sim500$\,\kms\ around the \oii\ line, after subtracting a constant mean continuum level estimated around the \oii\ line.

Next, we estimate the physical properties of the continuum-detected galaxies by fitting the MUSE spectra and photometry with stellar population synthesis (SPS) models. We use the Monte Carlo Spectro-Photometric Fitter (MC-SPF) developed by \citet{fossati2018}, following the procedure described in \citet{fossati2019b}. Briefly, the code uses the \citet{bruzual2003} models at solar metallicity and the Chabrier initial mass function \citep[IMF;][]{chabrier2003}. Nebular emission lines are added to the stellar templates using a grid of emission line luminosities from the models of \citet{byler2018}. The parameters of the grid are the ionization parameter of the gas and the age of the ionizing spectrum. The line luminosities are then scaled by the number of Lyman continuum photons in the stellar models and converted into flux at the galaxy's redshift. The shape of the emission lines is assumed to be Gaussian with the width treated as a free parameter.
MC-SPF then jointly fits the MUSE spectra and photometry estimated in four bands from the MUSE data to derive the stellar mass (\mstar), star formation rate (SFR) and dust extinction ($A_{\rm v}$) of the galaxies (see Fig.~\ref{fig:sps_fit}). We caution that in the absence of near-infrared (NIR) photometry for the galaxies, there remains some degeneracy between the model parameters, and we plan to refine these measurements once NIR data for all the fields become available.

\begin{figure}
    \centering
    \includegraphics[width=0.5\textwidth]{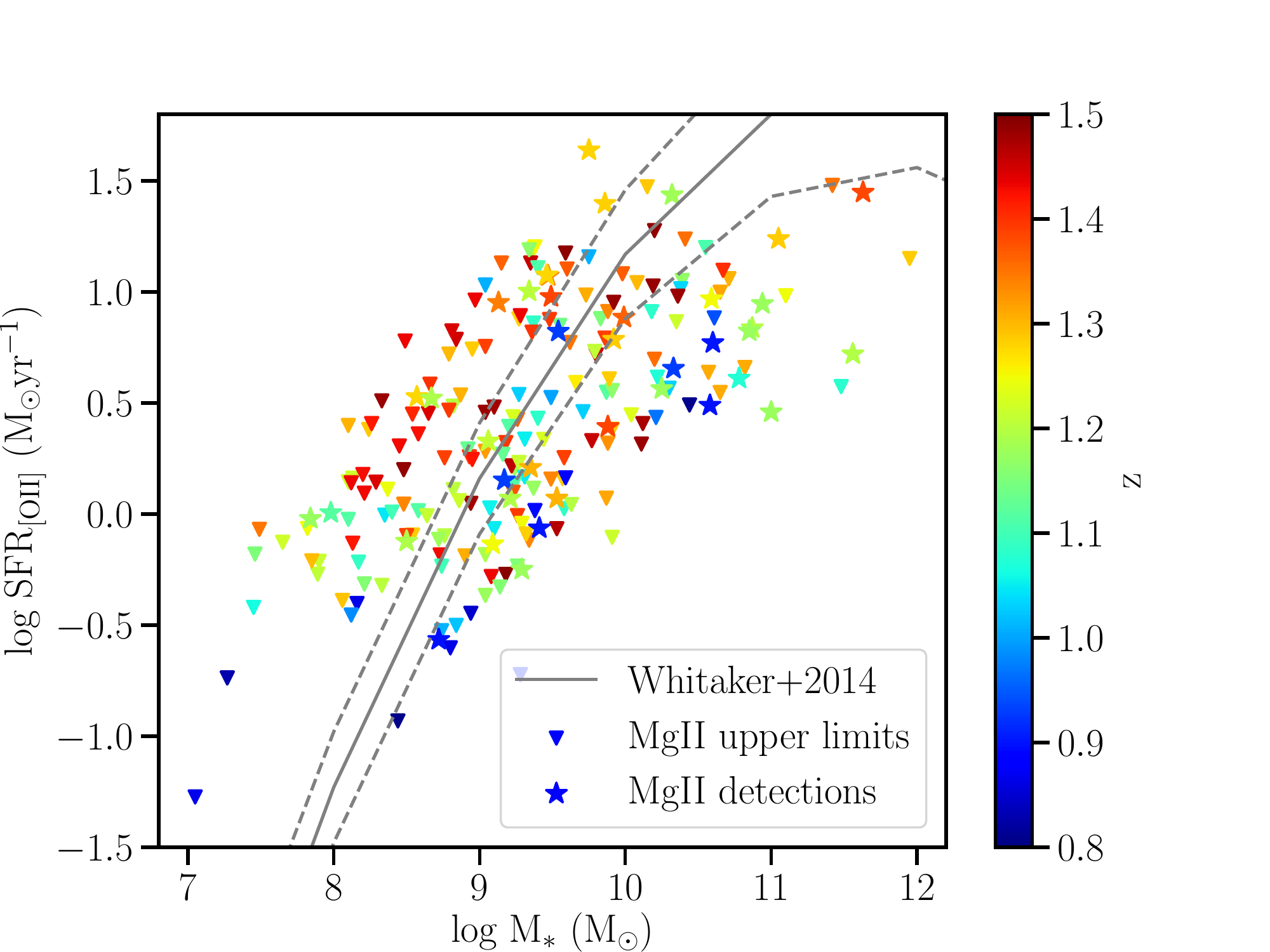}
    \caption{SFR derived from \oii\ versus the stellar mass of the continuum-detected galaxies in our sample. Galaxies with associated \mgii\ absorption are shown as stars, while those without are shown as downward triangles. The symbols are colour-coded by the galaxy redshifts. The solid line is the fit to the main sequence obtained by \citet{whitaker2014} in the redshift range $1<z<1.5$, while the dashed lines represent the $1\sigma$ uncertainty in the fit.}
    \label{fig:mainseq}
\end{figure}

For the purpose of this paper we primarily use the stellar masses estimated by MC-SPF, which appear to be well constrained by the MUSE photometry. The galaxies in our sample have stellar masses between $10^7$ and $10^{12}$\,\msun, with a median of $2\times10^{9}$\,\msun. The typical error in the estimates of the logarithm of stellar mass is $\approx\pm0.1$ dex. With the stellar mass in hand, we also derive the halo masses of the galaxies using the stellar-halo mass relation from \citet{moster2010}. The halo masses range from $3\times10^{10}$ to $4\times10^{14}$\,\msun, with a median value of $3\times10^{11}$\,\msun. We additionally use the dust extinction estimated by MC-SPF to correct the \oii\ luminosity for dust using the extinction law of \citet{calzetti2000}. This includes an attenuation factor of 2.27 to account for the effect of extra extinction in young ($<10$ Myr) stellar populations, $A_{\rm young} = 2.27\times A_{\rm old}$. We then estimate the star formation rate based on \oii\ using the relation of \citet{kennicutt1998}. We use a factor of 1.7 to convert from \citet{salpeter1955} IMF used in the Kennicutt relation to Chabrier IMF used by MC-SPF \citep{zahid2012}. The SFR as a function of the galaxy stellar mass is shown in Fig.~\ref{fig:mainseq}. The galaxies in our sample overall appear to follow the SFR versus \mstar\ relation shown by star forming galaxies at this redshift range \citep{whitaker2014}. The properties of the galaxies in our sample are summarized in the online Appendix~B.

\begin{figure*}
    \centering
    \begin{minipage}{10cm}
      \centering
      \includegraphics[height=0.35\textheight]{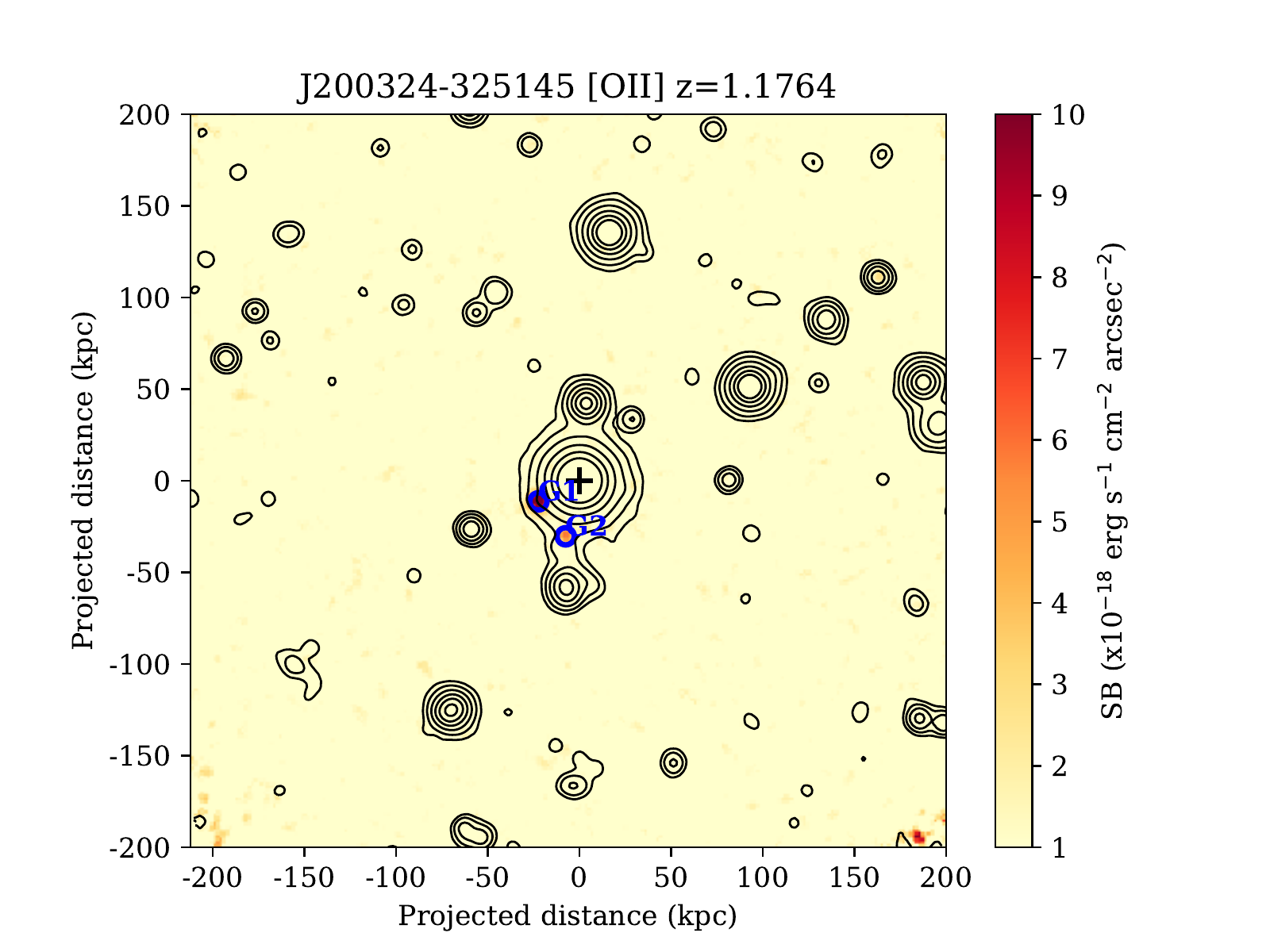}
    \end{minipage}
    \hspace{1mm}
    \begin{minipage}{6cm}
      \centering
      \includegraphics[height=0.15\textheight]{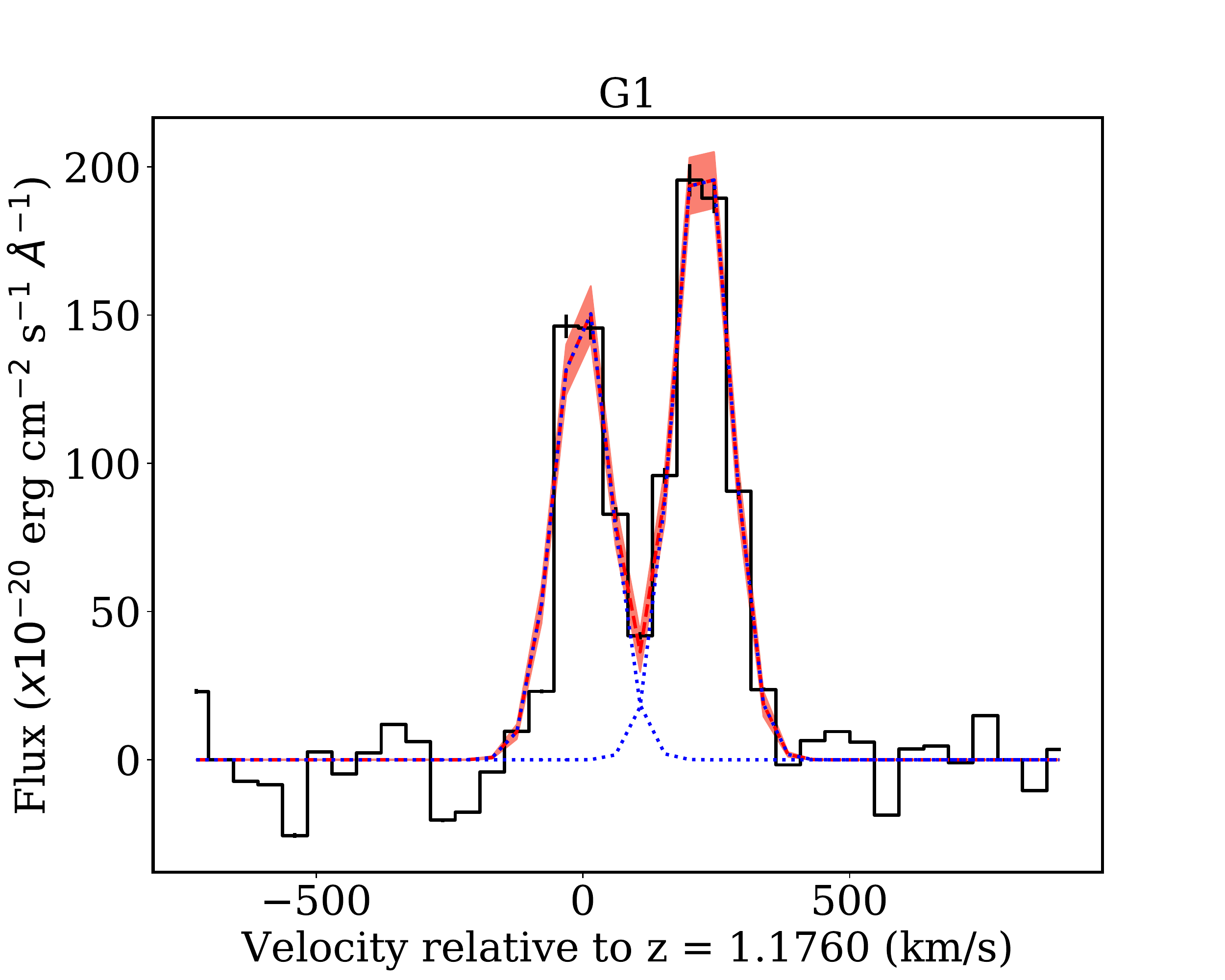} \\
      \includegraphics[height=0.15\textheight]{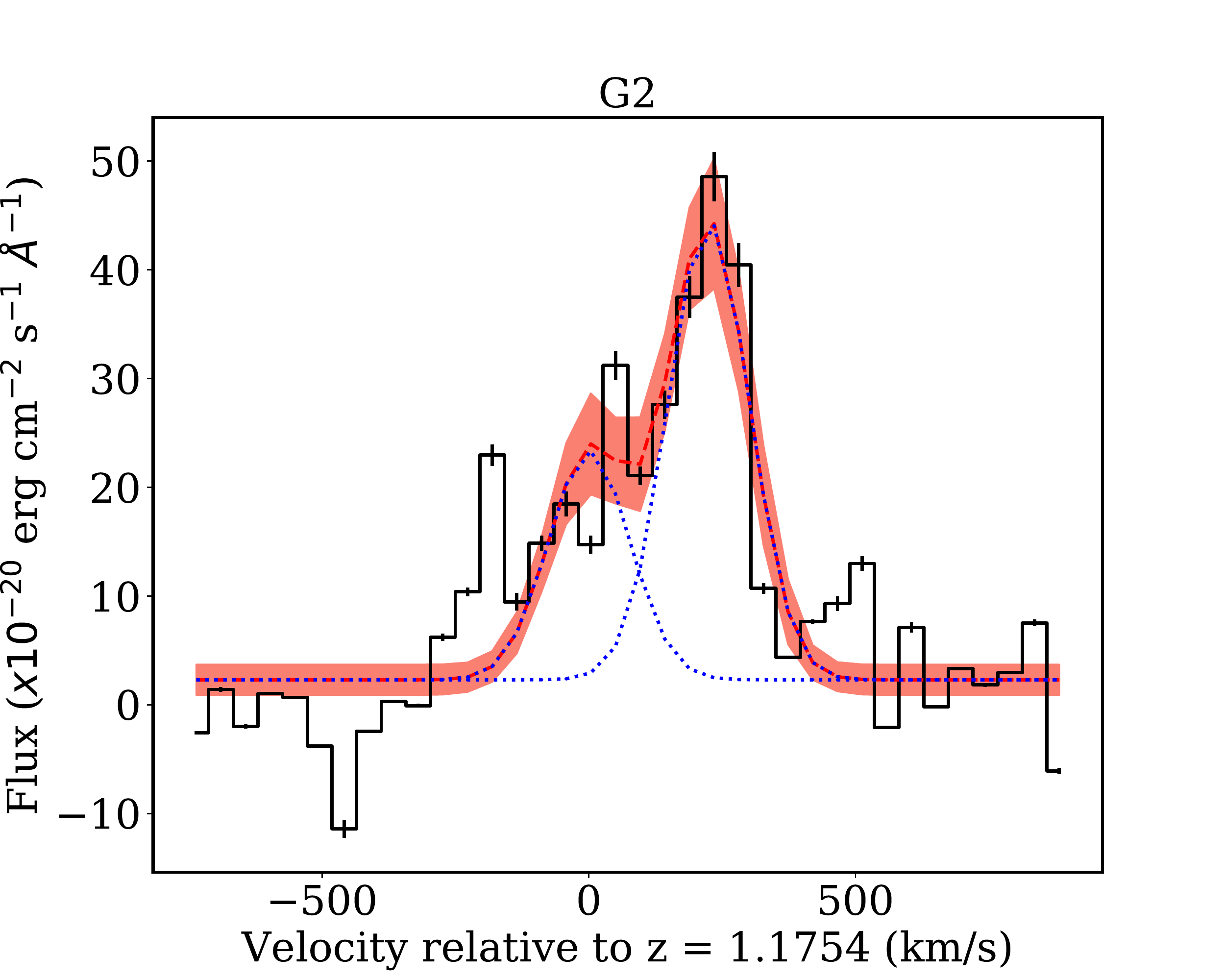}
    \end{minipage}
    \caption{Example of galaxy detection in \oii\ pseudo-narrow band image, in the field J200324$-$325145 around the \mgii\ absorber at $z=1.1764$. Notations and symbols are the same as in Fig.~\ref{fig:cont_galaxies_map}. The two galaxies marked `G1' and `G2' could be identified on the basis of the \oii\ pseudo-narrow band image, despite being in close proximity to the quasar PSF.}
    \label{fig:emit_galaxies_map}
\end{figure*}    

\subsubsection{Emission line-detected galaxies}
\label{sec_analysis_emitters}
In order to extend the association of \mgii\ systems in our sample to the population of galaxies that are faint in continuum but bright in line emission, we search for \oii\ line emitters in a velocity window of $\pm500$\,\kms\ (see Section~\ref{sec_results_detection}) around the \mgii\ redshifts using {\sc CubExtractor} \citep[{\sc CubEx};][Cantalupo in prep.]{cantalupo2019} and the procedure described in \citet{lofthouse2020} for \lya\ emitters.
Briefly, we first subtract the quasar point spread function (PSF) and remove continuum sources from the reduced MUSE 3D cubes using the {\sc CubePSFSub} and {\sc CubeBKGSub} methods in {\sc CubEx}, respectively \citep[see][for details]{cantalupo2019}.
We then run {\sc CubEx} on the continuum-subtracted cubes to identify and extract potential line emitters. {\sc CubEx} convolves the cubes with a two-pixel boxcar in the spatial direction and identifies groups of connected voxels (volumetric pixels) that each have S/N $>3$. The groups are extracted as line emitters if they satisfy the following conditions: (i) number of voxels $\ge27$, (ii) pixels spread over $\ge3.75$\,\AA\ in at least one spatial direction, and (iii) wavelength span of group $\le20$ channels to exclude residuals from continuum sources. 
We compare the two data cubes containing half of the exposure time to rule out residuals from cosmic rays.
For \oii\ line identification, we additionally require a doublet line to avoid ambiguity with other lines, like \lya\ at a higher redshift, in the classification. 

We find eleven \oii\ emitters associated with the \mgii\ absorbers that are not identified in the continuum catalogue. Among these, the integrated signal-to-noise ratio corrected for correlated noise \citep[ISN; see][]{lofthouse2020} of the entire source is $>7$ for seven of the line emitters and $5<ISN\le7$ for the remaining four emitters. The emitter classification has been visually confirmed by three of the authors (RD, MF, EL). We then extract the 1D spectra of these emitters along the full wavelength range using the 3D segmentation maps produced by {\sc CubEx}.

In addition to the above line emitter search, we construct pseudo-narrow band images around the \oii\ emission corresponding to the \mgii\ absorption systems. These narrow band images are obtained from the MUSE cubes from which the continuum emission and the quasar PSF have been subtracted, in a velocity window of $\pm500$\,\kms\ around the \mgii\ redshifts. We visually inspect these images for additional \oii\ emitters that could have been missed in our above search using {\sc CubEx}. We find three additional \oii\ emitters that are close to the quasar PSF (see Fig.~\ref{fig:emit_galaxies_map} for an example). The \oii\ emission lines in these cases were contaminated with residual quasar emission in the {\sc CubEx} extraction and hence could not be reliably identified with this method. We run instead {\sc SExtractor} on the narrow band images with the same parameters used for detecting the continuum galaxies (S/N threshold of 2, minimum area of 6 pixels for extraction, a minimum deblending parameter of 0.0001), and use the resulting 2D segmentation maps to extract the 1D spectra of these emitters. Thus in total we find 14 \oii\ emitting galaxies in the MUSE data associated to the \mgii\ absorbers. We additionally searched for spatially extended emission in the above narrow band images by smoothing them with a top-hat kernel of width 1 arcsec. We do not detect any extended emission down to a median $5\sigma$ surface brightness limit of $\approx3\times10^{-18}$\,\ergscmarc\ over an aperture of 1 arcsec$^2$.

\begin{figure}
    \centering
    \includegraphics[width=0.5\textwidth]{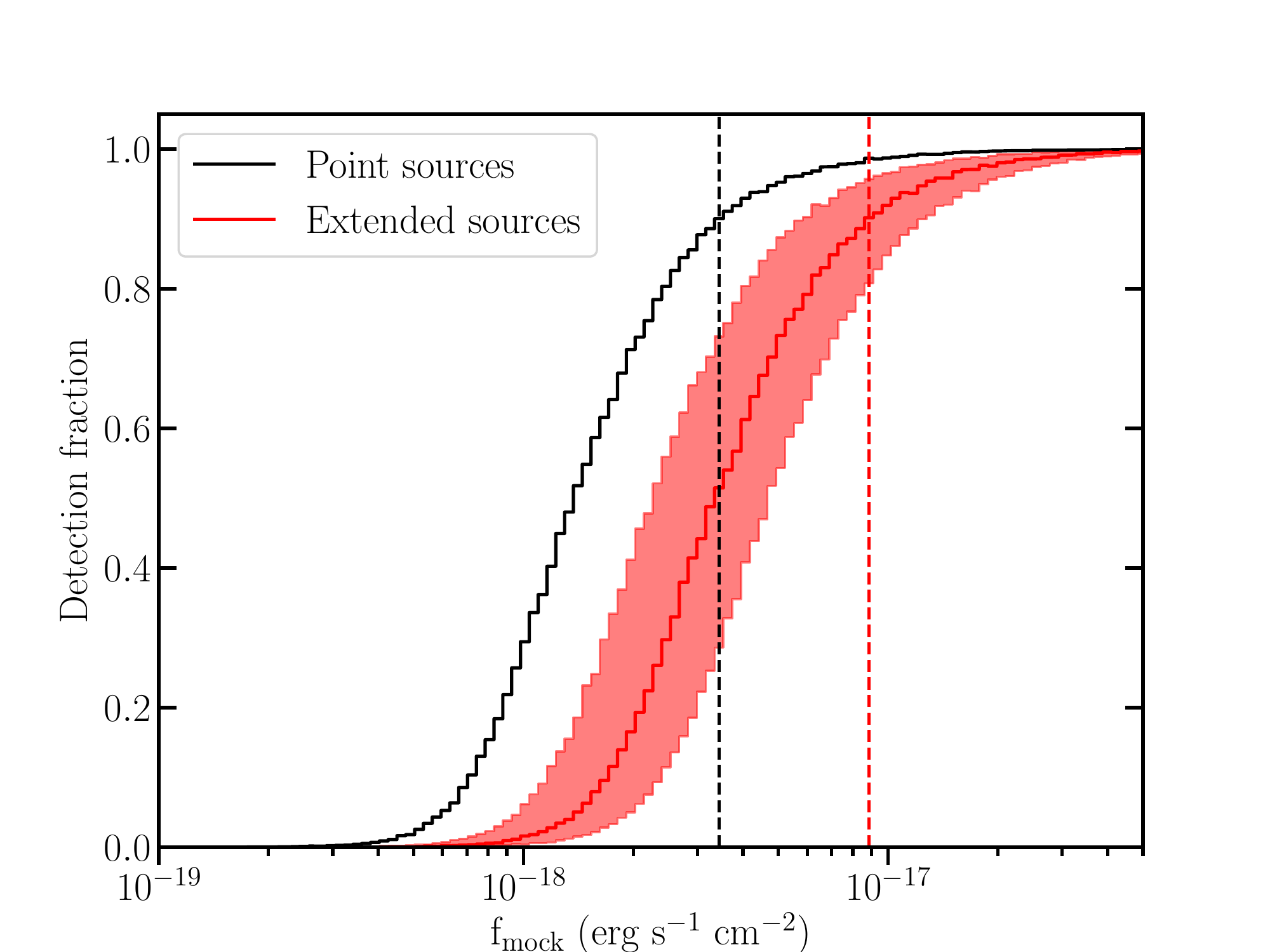}
    \caption{The median detection fraction of simulated emitters as a function of line flux of mock point (black) and extended (red) sources injected into the MUSE cubes with $\sim4$h exposure time. The solid red line shows the detection fraction for exponential disks with scale length of 3.9 kpc and the shaded region shows the same with scale lengths varying between 2.3 and 5.5 kpc. We are typically 90\% complete down to $3\times 10^{-18}$\,\ergscm\ for point sources (vertical black line) and $9\times10^{-18}$\,\ergscm\ for extended sources with scale length 3.9 kpc (vertical red line) at wavelengths between 6800 and 9300 \AA.}
    \label{fig:completeness}
\end{figure} 
To assess the completeness of our line emitter search, we inject mock sources into the MUSE cubes and repeat the detection experiment 1000 times. We inject 3000 sources at random spatial and spectral locations over the full MUSE wavelength range. We consider both point-like and extended sources. The point sources are characterized by a 1D Gaussian with full-width at half-maximum (FWHM) of 2.5 \AA\ spectrally and a 2D Gaussian with FWHM of 0.7 arcsec spatially. The extended sources are modeled by exponential discs with scale lengths of 2.3, 3.9 and 5.5 kpc (at $z=1$) and 1D Gaussian with same line spread function as for the point sources, and convolved with the seeing. We identify the mock sources that are detected at $ISN\ge5$ in a similar fashion as the real sources. The typical on-source exposure time for the MUSE fields is $\sim4$h, except for two fields that have been observed for $\sim10$h as part of the MUSE GTO. Fig.~\ref{fig:completeness} shows the median recovery fraction of the mock sources in the MUSE fields with $\sim4$h exposure in the wavelength range of interest, $\sim6800-9300$\,\AA. The median 90\% completeness limit is $\sim3\times 10^{-18}$\,\ergscm\ for the point sources and $\sim9\times10^{-18}$\,\ergscm\ for the extended sources with scale length of 3.9 kpc. Alternatively, for a fixed line flux of $5\times10^{-18}$\,\ergscm, the median recovery fraction is $\sim95$\% for point sources, and $\sim73$\% for the extended sources with scale length of 3.9 kpc. The average completeness limit in the MUSE fields with 10h exposure is lower by a factor of $\sim$1.2 due to the different conditions (e.g. seeing or brightness of these observations compared to the data from our Large Programme). For the purpose of this work, the completeness limit estimated for the $\sim4$h MUSE fields can be taken as the typical value.
%
%=================================================== RESULTS =============================================================== 
%
\section{Connecting \mgii\ absorbers and galaxies}
\label{sec_results}

With the powerful combination of a complete (flux-limited) redshift survey up to a radius of $\sim250$~kpc from quasar sightlines and of high-resolution spectroscopy that has not been pre-selected according to the presence of \mgii\ absorbers, we are able to conduct an unbiased census of the cool gas distribution around galaxies up to and beyond redshift $z\approx 1$, thus extending to moderate redshifts the study of correlations that have emerged from lower-redshift ($z\lesssim 0.5$) studies utilizing primarily multi-object spectrographs (see Section~\ref{sec_results_redshift}). Moreover, the possibility to exploit the full cosmological volume along the line of sight enables us to complete both galaxy-centric (Sections~\ref{sec_results_radial}, \ref{sec_results_galprop}, \ref{sec_results_covfrac}) and absorber-centric studies (Sections~\ref{sec_results_detection}, \ref{sec_results_azimuthal}), and in the following we alternate between both approaches. 
Finally, the possibility to trace the environment of the \mgii\ absorbers and of the galaxies identified along the line of sight adds a novel dimension to the study of cool gas in the CGM. While we proceed by first analysing previously-reported correlations between \mgii\ absorption and galaxies (Section~\ref{sec_results}) and then focus on the role of environment in shaping the cool gas within the CGM (Section~\ref{sec_environment}), it will become clear from the following analysis that it is not always straightforward (or not even at all possible) to completely disentangle the two effects.

\subsection{Association of galaxies to \mgii\ absorbers}
\label{sec_results_detection}
To study the incidence of \mgii\ absorption associated to galaxies, we take first an absorber-centric point of view and  cross-match the MAGG catalogue of continuum galaxies from Section~\ref{sec_analysis_continuum} with the \mgii\ absorption line sample from Section~\ref{sec_analysis_mgii}. For a match, we consider a velocity window of $\pm500$\,\kms\ centred on the \mgii\ redshift and the full MUSE field-of-view (FoV). For reference, the MUSE FoV of $1\times1~$arcmin$^2$ corresponds to $\sim500\times500$\,kpc$^2$ at $z=1$, which probes impact parameters ($R$, projected separation between quasar and galaxy centre) up to $\sim250-350$\,kpc. At $z=1$, for the median stellar mass of the continuum-detected galaxies in our sample of \mstar = $2\times10^9$\,\msun, the corresponding virial mass, radius and velocity are $\rm \sim3\times10^{11}$\,\msun, $\sim100$\,kpc and $\sim120$\,\kms. Therefore, we typically probe up to twice the virial radius and five times the virial velocity. For the more massive galaxies (\mstar = $10^{11}$\,\msun), we can probe up to the virial radius ($\sim280$\,kpc) and velocity ($\sim340$\,\kms).

\begin{figure}
    \centering
    \includegraphics[width=0.5\textwidth]{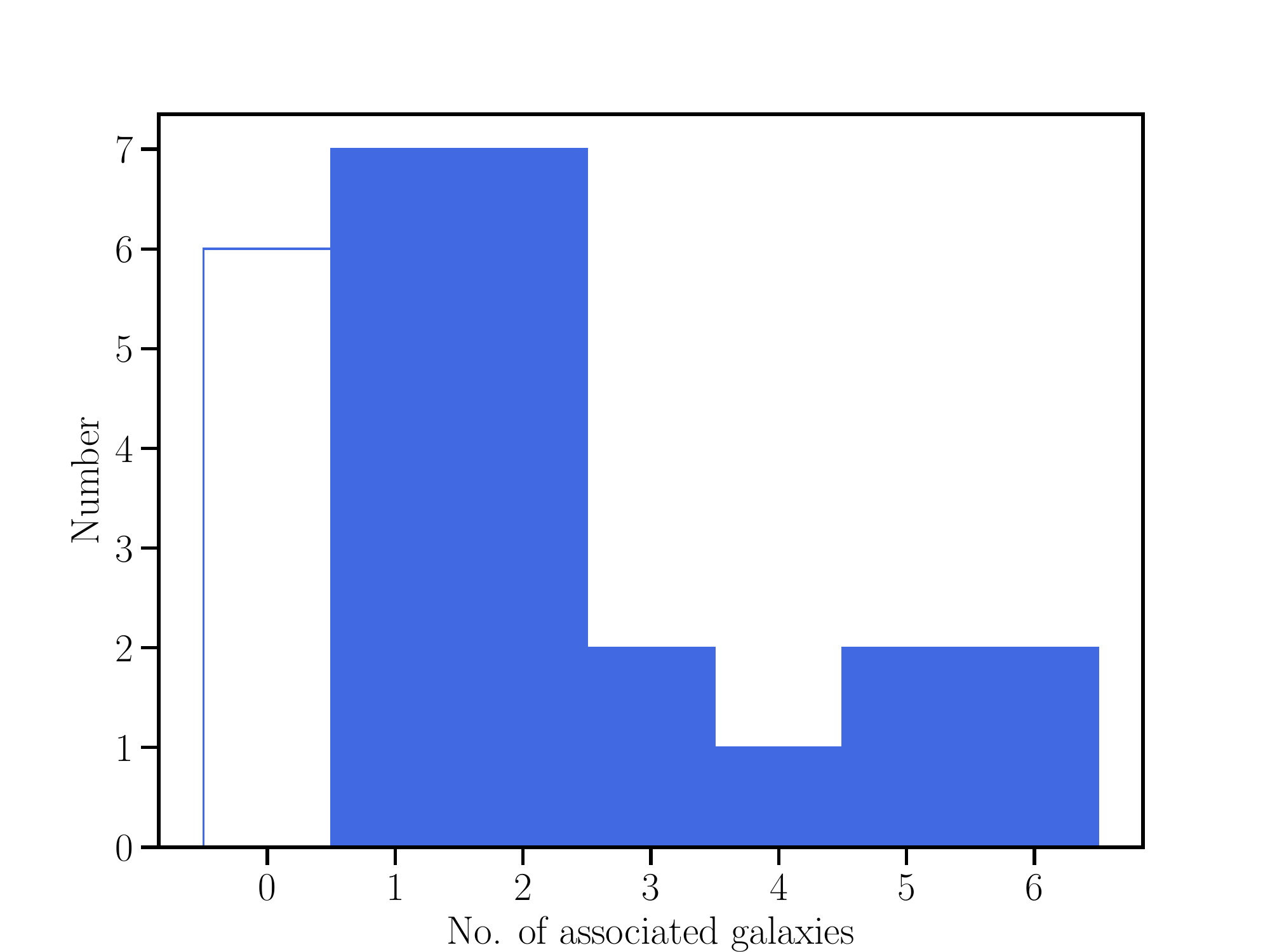}
    \caption{Histogram of the number of galaxies associated with the \mgii\ absorbers that are detected in the MUSE FoV within $\pm$500\,\kms\ of the absorber redshift. The solid bars are in cases of galaxy detections, while the open bar is when there is no galaxy identified. At least one galaxy is detected for 21 out of 27 absorbers, while 14 out of 21 absorbers have more than one associated galaxy.}
    \label{fig:ngal_histogram}
\end{figure}

We find 39 continuum-detected galaxies associated with the \mgii\ absorbers. As described in Section~\ref{sec_analysis_emitters}, we identified a further 14 \oii\ line emitting galaxies associated with the \mgii\ absorbers. Therefore, in total we find 53 galaxies associated with the 27 \mgii\ absorbers, as summarized in  Fig.~\ref{fig:ngal_histogram}. We detect at least one galaxy for 21 of the absorbers, giving a detection rate of $78^{+9}_{-13}$\%\footnote{The errors quoted here are $1\sigma$ Wilson score confidence intervals.}. For 14 of the absorbers, we detect more than one galaxy, i.e. a detection rate of $67^{+12}_{-15}$\% for multiple galaxies.
We do not find any significant dependence of the detection rate of galaxies on \wmg\ in our sample.

The average number of galaxies associated with \mgii\ absorption in the MUSE FoV within a velocity window of $\pm500$\,\kms\ is 2.5, showing that the presence of multiple galaxies around individual \mgii\ absorbers is a common occurrence. 
Based on the fit to the \oii\ luminosity function at $z=1.2$ from \citet{drake2013}, we expect to detect $\sim$0.3 galaxies in the volume searched around each absorber, indicating the presence of an over-density of galaxies associated with the \mgii\ absorbers.

\begin{figure}
    \centering
    \includegraphics[width=0.5\textwidth]{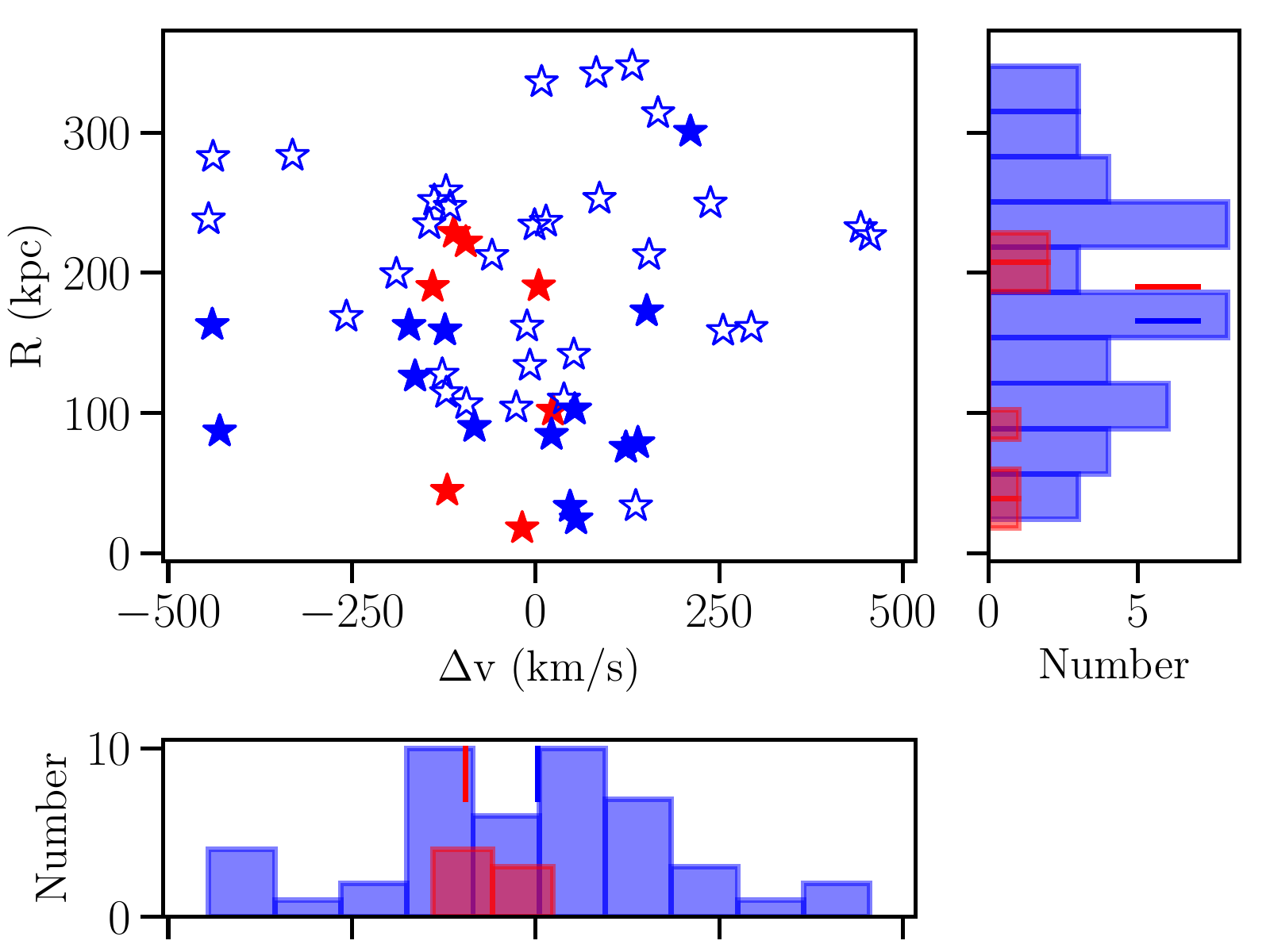}
    \caption{Distribution of impact parameters and velocity separations of the galaxies associated with the \mgii\ absorbers in our sample. Red stars denote single galaxies. Filled blue stars denote the galaxy with the smallest impact parameter in case of multiple galaxies associated with the same absorber, while the remaining galaxies are denoted by open blue stars. To the right and bottom, we show the histograms for the impact parameter and velocity separation, respectively. The median values of these parameters are marked by ticks. The histograms and ticks are red for the single galaxies and blue for the multiple galaxies.}
    \label{fig:imp_delv}
\end{figure}

The galaxies are detected at impact parameters ranging from 18 to 348 kpc (median $R = 169$~kpc), thus encompassing both galaxies that are sufficiently close to host the \mgii\ clouds in their inner CGM and galaxies at further distances that are likely to be associated with the \mgii\ absorbers via their large-scale environment. The median velocity separation between the galaxy and \mgii\ redshifts is $-8$\,\kms\ and the majority (75\%) of the galaxies lie within $\pm200$\,\kms\ of the absorber redshifts, indicative of a real association and not of a chance superposition along the line of sight. The distributions of the impact parameters and velocity separations are shown in Fig~\ref{fig:imp_delv}.

Given the significant radius probed by MUSE, the reason behind the non-detection of associated galaxies for six of the absorbers is likely attributed to the presence of galaxies that are fainter than our flux limit in the \oii\ line of $\lesssim3\times10^{-18}$\,\ergscm, which corresponds to a 90\% completeness down to an \oii\ luminosity of $\sim2\times10^{40}$\,\ergs\ for point sources at $z=1$, or an unobscured SFR of $\sim0.14$\,\msunyr. We note, however, that our survey does not exclude a priori passive galaxies that can still be identified via absorption lines. We do not however identify any passive galaxies close to the quasar sightlines ($R<50$\,kpc) associated to the \mgii\ absorbers.
Based on the 90\% $r$-band completeness limit of 26.3 mag of the continuum galaxy sample, and \citet{bruzual2003} SPS models with solar metallicity and Chabrier IMF, we estimate the limiting stellar mass to be $\sim2\times10^8$\,\mstar\ for a young star-forming galaxy and $\sim4\times10^9$\,\mstar\ for an old passive galaxy at $z\sim1$.

Despite our search for galaxies at close impact parameters via narrow-band imaging, we cannot completely exclude the presence of galaxies just below the quasar PSF ($R\lesssim10$\,kpc), where residual emission hampers the detectability even after the PSF subtraction. 
Based on the relation between \wmg\ and $R$ presented in Section~\ref{sec_results_radial}, we expect only $\approx 2$ cases in which galaxies lie at $R\lesssim10$\,kpc. 
We further note that for two of the absorbers, we identify galaxies outside our set velocity search window at velocity separations of $\sim800-900$\,\kms. However, in line with our definition, they are not included in the detections.

\begin{figure*}
    \centering
    \includegraphics[width=1.0\textwidth]{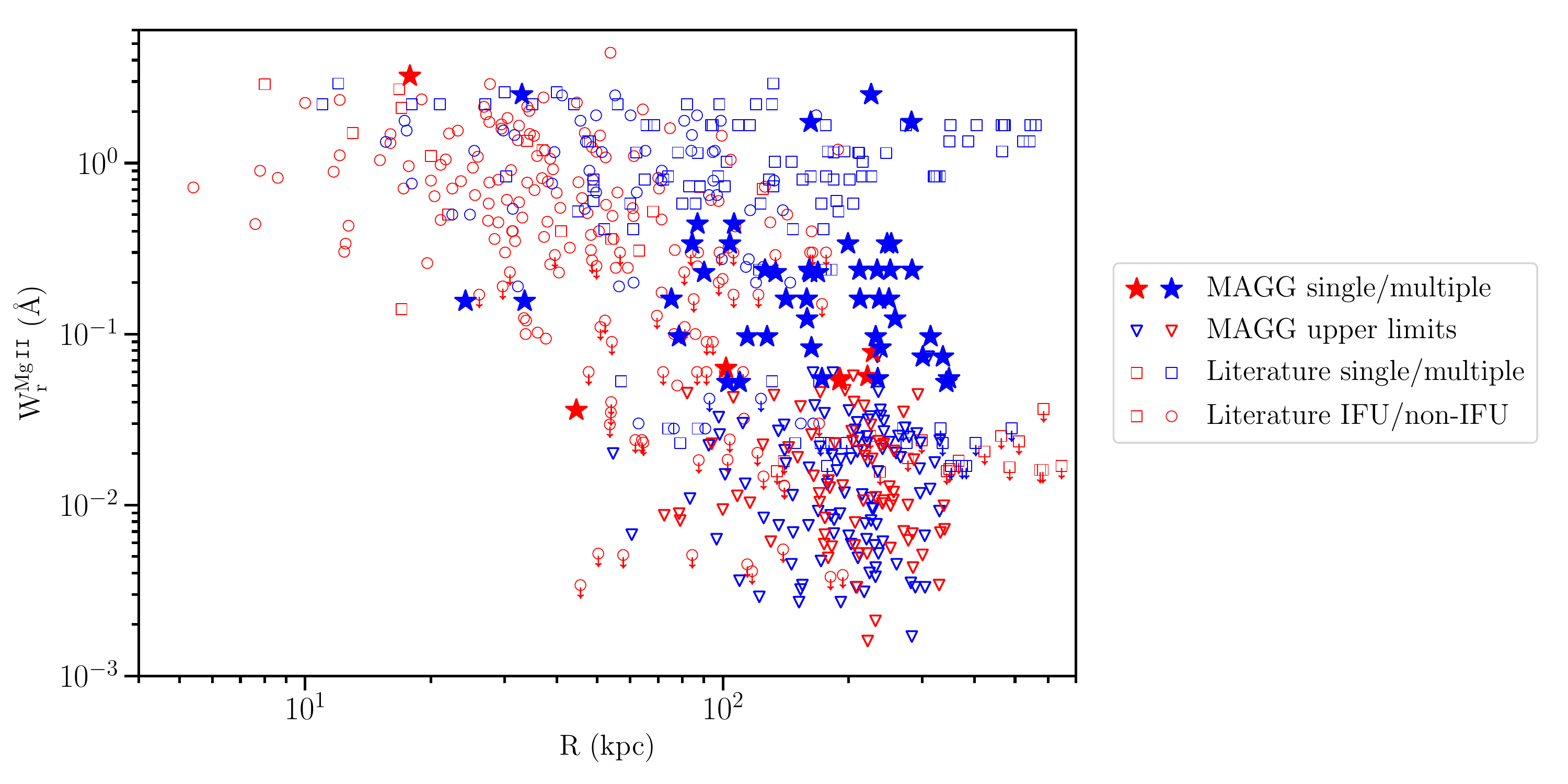}
    \caption{Rest-frame equivalent width of \mgii\ as a function of galaxy impact parameter in our sample and from the literature (see Section~\ref{sec_results_radial}). Error bars on the \wmg\ measurements in our sample are smaller than or similar to the size of the symbols. The different symbol notations differentiate between results of MAGG (stars for measurements and triangles for upper limits) and the literature (squares and circles, to distinguish between IFU and non-IFU studies). 
    With colours, we differentiate instead between galaxies that are reported as isolated (red) or in pairs and multiples (blue). 
    Note that our galaxy sample is not complete at $R\lesssim10$\,kpc where the quasar PSF could hinder reliable galaxy identification. While \wmg\ declines on average with $R$, there is considerable scatter at higher $R$ due to detection of multiple galaxies mostly using IFU studies.}
    \label{fig:ew_vs_imp}
\end{figure*}

While there have been extensive searches for galaxies associated to \mgii\ absorbers in the past \citep[e.g.][]{bergeron1991,steidel1994,gauthier2011,kacprzak2011}, our analysis is more directly comparable with other IFU surveys in terms of completeness and uniformity of the search in a wide FoV. Recently, \citet{hamanowicz2020} observed with MUSE five quasar fields with 14 strong \hi\ absorbers at $0.1<z<1$, among which 5 had known galaxy associations. They report a detection rate of 89\% (8/9) for serendipitous absorbers identified in the line of sight, and a detection rate of 57\% (8/14) for multiple galaxies associated to a single absorber. Differently from our work, their search employed a velocity window of $\pm$1000\,\kms and was 90\% complete for line flux greater than $5\times10^{-17}$\,\ergscm\ for point sources.  Considering these differences, the overall detection rate and the fraction of multiple galaxies reported by these authors are compatible with that of our sample. 

The MUSE GTO MEGALFOW survey \citep{schroetter2016,zabl2019} targets instead the fields around 22 quasars that are known to host three or more strong \mgii\ absorbers (\wmg\ $>0.5$\,\AA) at $0.5<z<1.5$. They searched for associated galaxies based on multiple pseudo-narrow-band images of width 400\,\kms, with a reported $5\sigma$ \oii\ detection limit of $\sim4\times10^{-18}$\,\ergscm\ for point sources at $z\sim0.9$. They detected one or more galaxy for 75\% of the absorbers at $R<100$\,kpc, i.e. with an higher detection rate for galaxies at close impact parameters compared to our study. This difference is clearly linked to their pre-selection based on \mgii\ that skews the statistics compared to a random sample (see also next section). Our sample is in fact more representative of a blind survey for galaxies associated with typical and not just the strongest \mgii\ absorbers, with a median \wmg\ of 0.1\,\AA. 

A third significant MUSE study focusing on \mgii\ is the analysis of \mgii-galaxy associations in the MUSE Ultra Deep Field (MUDF), a very deep survey of a $1.5\times1.2~$arcmin$^2$ region around two $z\sim3$ quasars with up to 150~h on-source observations \citep{lusso2019,fossati2019b}. Based on the first 44~h of data from this survey, \citet{fossati2019b} find that 5 out of 6 \mgii\ absorbers at $0.5<z<1.5$ detected in the spectrum of the bright quasar are associated with three or more galaxies.

Despite the variation in target selection and depth of these different observations, the emerging picture is that \mgii\ absorbers are nearly always associated with galaxies within 250\,kpc and more than 50\% of these galaxies are not in an isolated environment. Rather, the majority are part of pairs, multiple associations, and groups as detected down to the typical stellar mass ($10^9~\rm M_\odot$) probed in shallow/medium depth ($\sim2-10$h exposure) MUSE data. The incidence of \mgii\ absorbers in groups and multiple associations is in fact likely even higher once we probe the environment around absorbers more completely with deeper and wider coverage, as already hinted by the MUDF results \citep{fossati2019b}.

\subsection{Radial profile of \mgii\ gas around galaxies}
\label{sec_results_radial}

In this section, we move from the absorber-centric view taken above to a galaxy-centric view in which, starting with a known galaxy position in 3D we reconstruct the cool gas distribution in the CGM as traced by \mgii.
In particular, the relationship between the rest equivalent width of \mgii\ and galaxy impact parameter is well-studied in the literature, with numerous studies reporting a strong anti-correlation between \wmg\ and $R$ \citep[e.g.][]{lanzetta1990,chen2010,bordoloi2011,nielsen2013b,rubin2018}. 

Fig.~\ref{fig:ew_vs_imp} shows \wmg\ as a function of $R$ for the MAGG galaxies where we detect \mgii\ absorption (filled stars) or can derive a clean upper limit on \wmg\ (open downward triangles). To keep track of the different environment around these galaxies, we differentiate between single or multiple galaxy associations (red and blue symbols, respectively). Also shown in the same plot for comparison are measurements for \mgii-galaxy pairs compiled from the following literature: \citet{nielsen2013a,nielsen2018,schroetter2016,bielby2017,klitsch2018,zabl2019,fossati2019b,hamanowicz2020}. 
  
This literature compilation is unavoidably heterogeneous, with studies differing in observing sensitivity and definition of galaxy associations. To differentiate between the literature galaxy samples with different observing methods, we plot systems from single and multi-slit spectroscopic observations as circles, and those from IFU observations as squares. 
We further note that while the maximum impact parameter probed in a single MUSE pointing is $\sim350$kpc at $z\sim1$, the work by \citet{fossati2019b} is based on two overlapping MUSE pointings, which extends this type of analysis up to $\approx 600~$kpc from a galaxy. 

The perhaps more subtle and yet relevant difference among these literature studies is the starting point for the galaxy survey. While some authors start from a \mgii-centric approach in which galaxies have been searched around known \mgii\ absorbers to understand their origin \citep[e.g.][]{zabl2019}, others take a more galaxy-centric view and search for \mgii\ absorbers around the position of known galaxies \citep[e.g.][]{rubin2018}. The MAG{\sc ii}CAT sample \citep{nielsen2013a}, being in itself a compilation of \mgii-galaxy pairs from the literature, includes both \mgii-centric and galaxy-centric samples. 
Further, we note that the study of \citet{rubin2018}, unlike the other studies mentioned above, uses background galaxies rather than quasars to probe the gas around foreground galaxies.

The majority of the galaxy sample used in our work (94\%) is identified independently of the absorption properties, on the basis of continuum and \oii\ emission alone (Section~\ref{sec_analysis_continuum}). The remaining pure \oii\ line emitting galaxies (Sec~\ref{sec_analysis_emitters}) are instead identified on the basis of the known \mgii\ absorption systems. While we include these galaxies in our analysis, we note that our results do not change upon removing them from our sample.

\begin{figure*}
    \centering
    \includegraphics[width=0.48\textwidth]{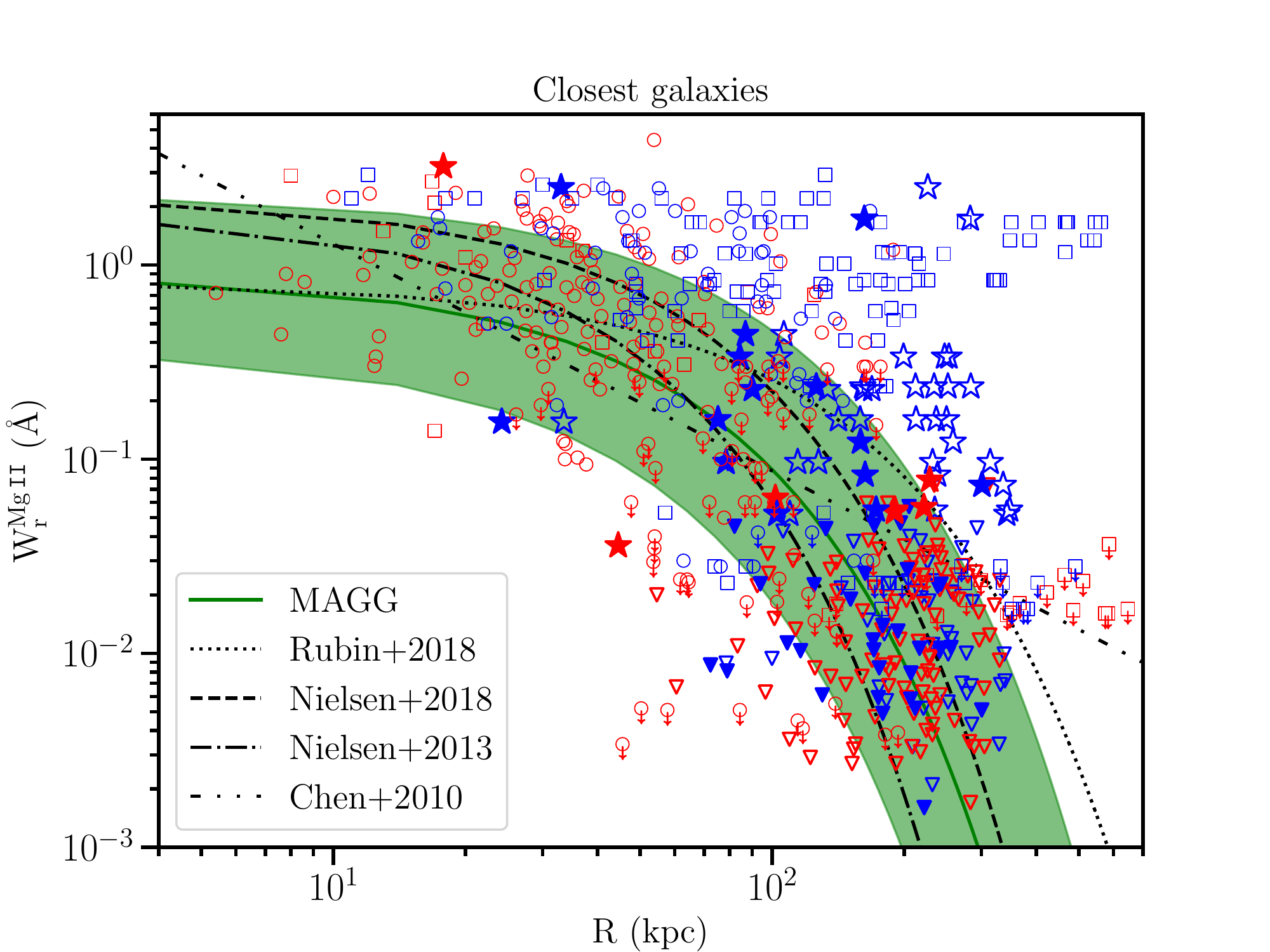}
    \includegraphics[width=0.48\textwidth]{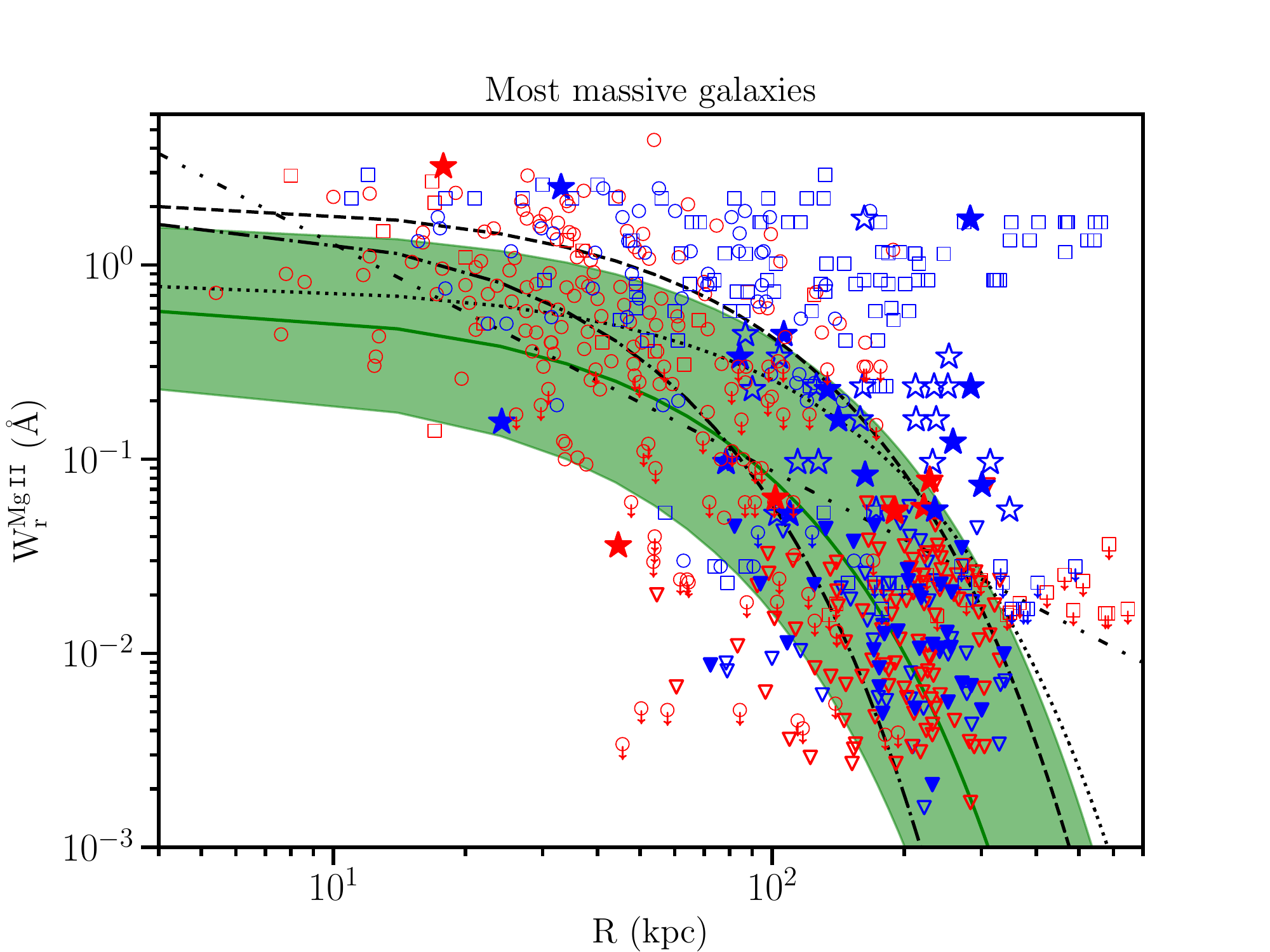}
    \caption{Fit to the rest equivalent width of \mgii\ versus galaxy impact parameter relation for the samples with the closest galaxies {(\it left)} and the most massive galaxies {(\it right)}. Measurements and upper limits from MAGG are plotted as stars and triangles, respectively. The symbols are color-coded same as in Fig.~\ref{fig:ew_vs_imp}. In the left panel, for multiple galaxies, the closest galaxies are plotted as solid symbols while the rest are plotted as open symbols. Similarly in the right panel, the most massive galaxies in case of multiple are plotted as solid symbols. The symbols for the literature data are same as in Fig.~\ref{fig:ew_vs_imp}. The green solid lines show the best-fit linear relation obtained between log\,\wmg\ and $R$. The 16$^{\rm th}$ and 84$^{\rm th}$ percentiles are indicated by the shaded regions. Various best-fit relationships from the literature \citep{chen2010,nielsen2013b,nielsen2018,rubin2018} are also plotted for reference as denoted in the left panel. In the left panel we show the fit to the closest galaxy in the groups from \citet{nielsen2018}, while in the right panel we show their fit to the most luminous galaxy in their groups. Our fits are consistent within the uncertainties with the log-linear fits obtained in the literature at $z\lesssim0.8$.}
    \label{fig:ew_vs_imp_fit}
\end{figure*}

Overall, from Fig.~\ref{fig:ew_vs_imp} we find that the MAGG galaxies occupy a similar parameter space in the \wmg\ versus $R$ plot as the literature systems, despite the fact that our observations probe an average higher redshift compared to most previous samples. This consistency hints at a lack of strong evolution in the \wmg\ around galaxies with time and thus with their SFRs, considering the redshift evolution of the galaxy main sequence.

While the (reportedly) single galaxies (in red) paint a picture of a strong anti-correlation between \wmg\ and $R$, multiple galaxy systems (in blue) appear to extend the parameter space to larger \wmg\ at larger $R$, broadening the scattering and weakening the overall correlation. 
Notably, these galaxies only scatter upward and/or rightward in this parameter space, with the single galaxies defining a lower envelope beyond which no galaxies are found despite the sensitivity to both the presence of galaxies and \mgii\ absorption. In other words, there are no systems that populate the parameter space defined by low \wmg\ and small $R$.
 
The above dichotomy between single and multiple galaxies appears clearly linked to the technique used to complete the galaxy survey. IFU surveys (stars, squares and triangles), which provide a more complete view of the galaxy population, appear in fact largely responsible for the detection of galaxies at larger impact parameters and of multiple galaxy associations. This trend raises the question on whether some of the reportedly isolated galaxies in the literature may in fact be part of pairs, multiples or groups. 
 
To investigate trends between \wmg\ and $R$ more statistically, we quantify the significance of the correlation including $3\sigma$ upper limits on \wmg\ for the continuum-detected galaxies in our sample, for which there is no detection of associated \mgii\ absorption.
The upper limits are taken into account by using standard survival analysis methods in {\sc r}\footnote{https://www.r-project.org/} \citep{R2020}.  Most of our absorption systems have more than one associated galaxy. To avoid double counting and to differentiate the role of environment from the effect of individual galaxies (see below for a more detailed discussion), 
we carry out the analysis on two samples: (i) single galaxies and galaxies with smallest impact parameter among multiple, and (ii) single galaxies and galaxies with highest stellar mass among multiple. We perform a correlation analysis between \wmg\ and $R$ using the {\sc cenken} function in {\sc r}, which computes the non-parametric Kendall's tau correlation coefficient for censored data. 

For the sample with closest galaxies, the Kendall rank correlation coefficient between \wmg\ and $R$ is $\tau_{\rm k} = -0.13$ and the probability of the correlation arising by chance is $p_{\rm k} = 0.01$. Dividing the sample into two sub-samples at the median $R=169$\,kpc, we find the probability that they are arising from the same parent distribution is $P=9\times10^{-5}$. This is estimated using the {\sc cendiff} function in {\sc r}, which tests if there is a difference between two empirical cumulative distribution functions using the Peto \& Peto modification of the Gehan-Wilcoxon test. For the sample with the most massive galaxies, the anti-correlation between \wmg\ and $R$ is weaker, with $\tau_{\rm k} = -0.08$ and $p_{\rm k} = 0.15$. The probability that the sub-samples at higher and lower impact parameters are arising from the same parent distribution is $P=0.003$. The above indicates that the strength of \mgii\ absorption declines primarily with increasing distance from the galaxy centre.

Next, to characterize the relationship between \wmg\ and $R$, we fit a log-linear model to the data by assuming a linear dependence between log\,\wmg\ and $R$ of the form
\begin{equation}
 \rm log~W_{r}^{Mg\,\textsc{ii}} (\mbox{\normalfont\AA}) = a + b\times{\it R}~{\rm (kpc)}.
\label{eqn:loglinear}
\end{equation}
We define the likelihood function as the product of the likelihood functions for the $n$ measurements and $m$ upper limits, following \citet{chen2010} and \citet{rubin2018}, as:
\begin{multline*}
  \mathcal{L}(W) = \left ( \prod_{i=1}^{n} \frac{1}{\sqrt{2\pi \sigma_i^2}} \exp \left \{- \frac{1}{2} \left [ \frac{W_i - W(R_i)}{\sigma_i} \right ]^2 \right \} \right ) \\
  \times \left ( \prod_{i=1}^{m} \int_{-\infty}^{W_i} \frac{dW'}{\sqrt{2\pi \sigma_i^2}} \exp \left \{- \frac{1}{2} \left [ \frac{W' - W(R_i)}{\sigma_i} \right ]^2 \right \} \right ), ~~~~~~~~~~(2) 
\label{eqn:likelihood}
\end{multline*}
where $W_i$ is the log~\wmg\ for each measurement, $W(R_i)$ is the value of log~\wmg\ expected from the model (Eqn.~\ref{eqn:loglinear}) at each impact parameter $R_i$, and the total error is given by the sum in quadrature of each measurement error ($\sigma_{mi}$) and an intrinsic scatter ($\sigma_{sc}$) to account for intrinsic variation in galaxy properties, $\sigma_i^2 = \sigma_{mi}^2 + \sigma_{sc}^2$. 
To fit the model to the data, we define flat priors on the parameters such that $-10<a<10$, $-10<b<10$ and $0<\sigma_{sc}<10$. We sample the posterior probability density function using  {\sc pymultinest} \citep{buchner2014}, which is a python wrapper for {\sc multinest}  \citep{feroz2008}. The median value of this function and the 16$^{\rm th}$ and 84$^{\rm th}$ percentiles are plotted in Fig.~\ref{fig:ew_vs_imp_fit} for the sample of closest galaxies (left panel) and most massive galaxies (right panel). The fit to the sample of most massive galaxies is slightly flatter but consistent with that to the sample of closest galaxies. For the closest galaxies, the best-fit relation has coefficients $a = -0.05^{+0.42}_{-0.38}$, $b = -0.010 \pm 0.003$ and $\sigma_{sc} = 0.93^{+0.21}_{-0.17}$, while for the most massive galaxies the best-fit parameters are $a = -0.20^{+0.42}_{-0.39}$, $b = -0.009 \pm 0.003$ and $\sigma_{sc} = 1.18^{+0.34}_{-0.24}$.

We show for comparison in Fig.~\ref{fig:ew_vs_imp_fit}, the relationships between \wmg\ and $R$ given by different literature studies.
As noted, different studies are based on different sets of assumptions.
Both \citet{nielsen2013b} and \citet{rubin2018} have fitted a log-linear model to their data as in our analysis, whereas \citet{chen2010} have adopted a power-law model. The fits given by \citet{nielsen2013b} and \citet{chen2010} were performed on samples defined as isolated in the respective studies, while \citet{rubin2018} did not place any constraint on the galaxy environment in their sample. Further, \citet{nielsen2018}
differentiated on the basis of galaxy environment, finding a flatter fit to the closest and the most luminous galaxies in a sample of groups compared to their isolated sample in \citet{nielsen2013b}. 

Despite these differences, some common traits emerge.  Due to the presence of a large number of upper limits beyond $R\gtrsim 100~$kpc, we find that our data are not well-represented by a power-law model, like the one by \citet{chen2010}. Our best-fit relationships are instead consistent within the uncertainties with the log-linear fits previously obtained in the literature, albeit with small differences. The fit to the closest galaxies in our sample is slightly flatter than that obtained for isolated galaxies by \citet{nielsen2013b}, but it has similar slope as the fit to the closest galaxies in groups by \citet{nielsen2018}. Similarly, the fit to the most massive galaxies in our sample is flatter than that to the isolated galaxies and its slope is more consistent with that of the fit to the most luminous galaxies in groups by \citet{nielsen2018}.

Finally, we observe that the functional form in Eqn.~\ref{eqn:loglinear} clearly provides only an adequate description of the lower envelope of the distribution in the \wmg\ versus $R$ plane. 
As already noted above, there exists a substantial number of galaxies that up-scatter from, or scatter to the right of, this relation and occupy a region of higher \wmg\ at fixed impact parameter.
As these galaxies tend to almost exclusively lie in group-like environments, a single log-linear model appears inadequate to describe the emerging complexity of the \wmg\ versus $R$ plane.    

\begin{figure*}
    \centering
    \includegraphics[width=1.0\textwidth]{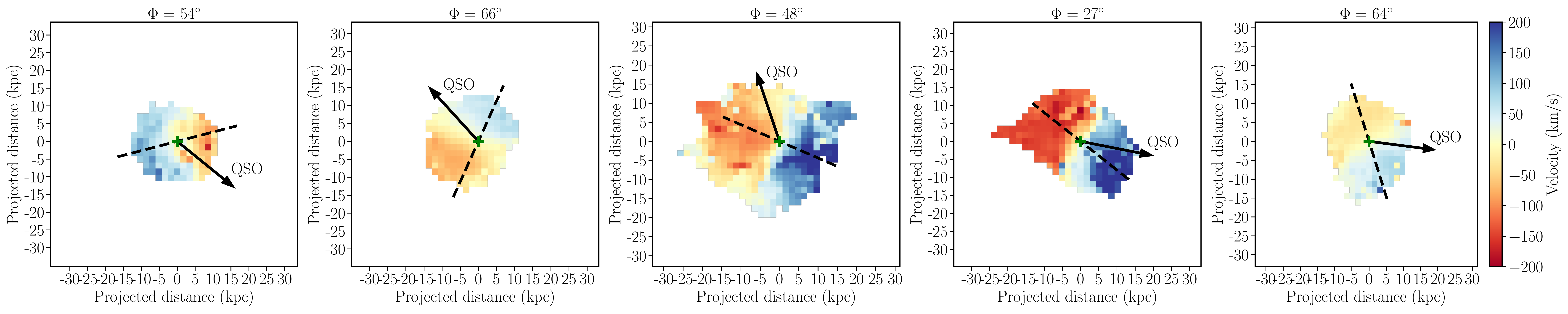}
    \caption{Examples of velocity maps of the galaxies in our sample. The direction of maximum velocity gradient is marked by dashed lines. The arrow indicates the direction towards the quasar. The angle between these two is defined as the azimuthal angle, which is indicated at the top of each map.}
    \label{fig:velocity_maps}
\end{figure*}

\subsection{Azimuthal distribution of \mgii\ absorbers}
\label{sec_results_azimuthal}

Following the analysis of the radial decline of \mgii\ absorbers around galaxies, we continue our exploration of the cool gas distribution in the CGM of $z\approx 1$ galaxies by examining  whether the distribution of \mgii\ depends on the relative projected angle with respect to the galaxy orientation. 
Several studies of \mgii-galaxy pairs have in fact reported that the distribution of \mgii\ absorption is not isotropic around galaxies \citep[e.g.][]{bordoloi2011,bouche2012,kacprzak2012,lan2018,martin2019,schroetter2019,zabl2019}, with \mgii\ absorption clustering either in the direction of the major or minor axis. These observations are often interpreted with models in which the \mgii\ absorbing gas traces either bi-conical outflows perpendicular to the galaxy disc along the minor axis or co-rotating/accreting gas along the galaxy major axis. However, some recent studies have examined in detail the metallicity of the CGM gas at $z<0.7$, finding a weak dependence upon the galaxy orientation, indicating that outflowing and inflowing gas cannot be completely separated by orientation angle, as they become well-mixed at low redshifts \citep{kacprzak2019,pointon2019}. Furthermore in a recent study, based on a sample of 211 isolated galaxies at $z<0.5$, \citet{huang2020} find no significant azimuthal dependence of the \mgii\ covering fraction and \wmg.

By leveraging our complete sample of galaxies, we explore whether there is any azimuthal dependence of \mgii\ absorption once we extend the analysis to larger impact parameters ($R>50$\,kpc) and do not limit to systems showing high-equivalent width (\wmg\ $>1$\,\AA) absorption.
Note that in this section, we revert to an absorber-centric study, restricting to a sub-sample of galaxies for which there is a positive detection of \mgii\ absorption. 

\begin{figure}
    \centering
    \includegraphics[width=0.5\textwidth]{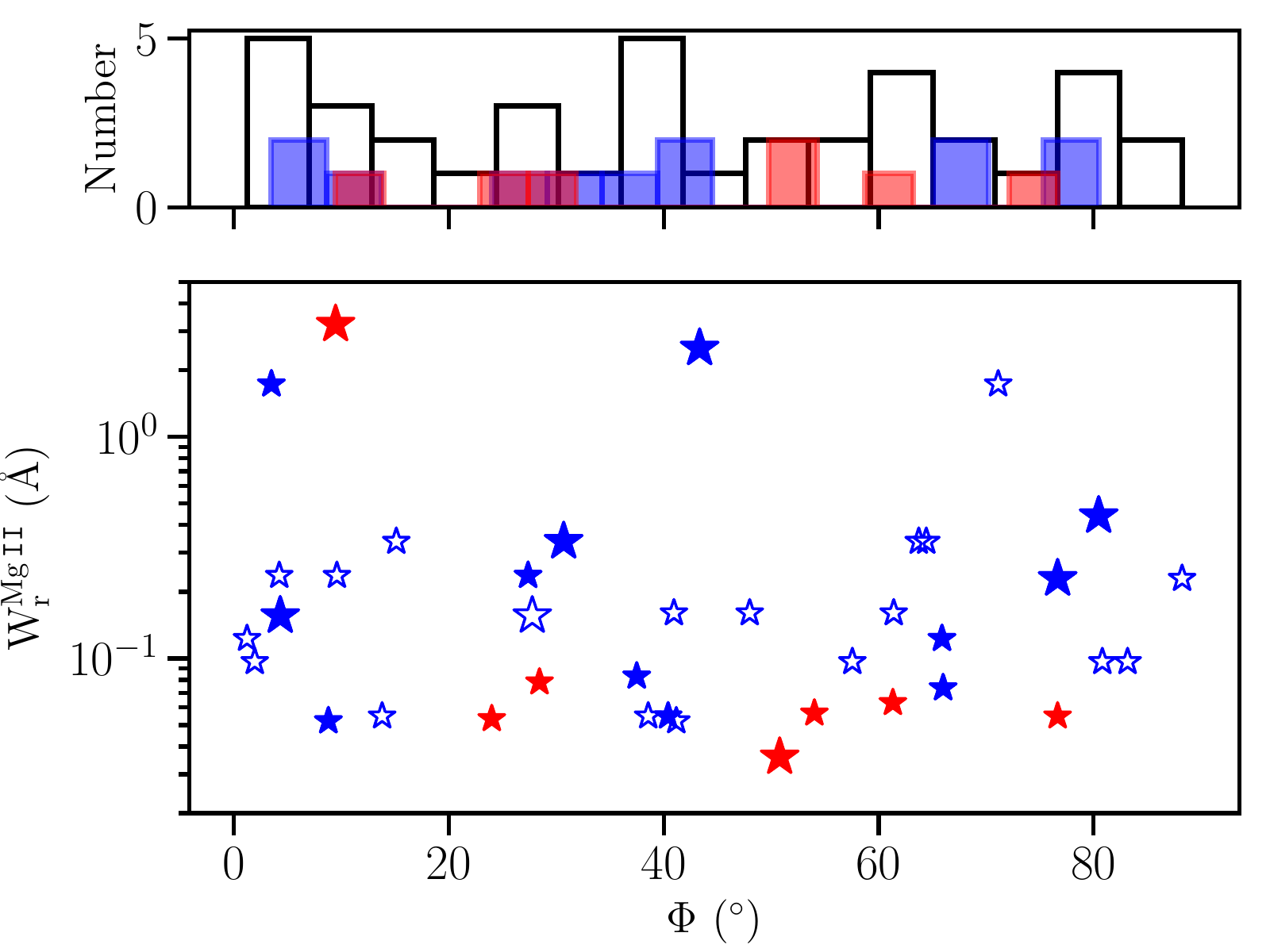}
    \caption{Rest-frame equivalent width of \mgii\ as a function of galaxy azimuthal angle ($\Phi$) with respect to the quasar. Single galaxies are plotted in red. Multiple galaxies are plotted in blue. The galaxy with smallest impact parameter in case of multiple galaxies is plotted as a filled symbol, while the remaining ones are plotted as open symbols. The larger symbols denote galaxies at $R\le100$\,kpc, while the smaller symbols denote galaxies at $R>100$\,kpc. The top panel shows the histograms of the azimuthal angles $-$ open black for all the galaxies, filled red for the single galaxies, and filled blue for the closest galaxies among multiple. We do not find any statistically significant trend of \wmg\ with $\Phi$.}
    \label{fig:ew_vs_pa}
\end{figure}

To establish the galaxy orientation, we first derive the velocity maps for the galaxies in our sample. We extract a sub-cube of $6\times6~$arcsec$^2$ (or $8\times8~$arcsec$^2$ if the \oii\ emission is more extended) and 100\,\AA\ centred around the \oii\ emission of each galaxy from the continuum-subtracted MUSE cubes. We then fit the \oii\ emission line with a double Gaussian profile in each pixel by averaging over the neighbouring $3\times3$ pixels (which corresponds to $0.6\times0.6~$arcsec$^2$, roughly equivalent to the seeing). While fitting the \oii\ emission lines, we fix the wavelength difference between the two Gaussian components to be equal to that of the \oii\ doublet, and require the line intensity ratio to be $0.2<$ \oii\,\l3729/\oii\,\l3727 $<2$. We set a S/N limit of 5 for the fit to be considered successful.
The S/N is determined by comparing the \chis\ of the Gaussian fit to that of a straight line fit, i.e. we accept the Gaussian fit if the improvement over a continuum-only fit is $>5\sigma$ \citep[see][]{swinbank2017}.
If the S/N of the fit is below this threshold, we increase the averaging region to $5\times5$ pixels and repeat the fit. Examples of the resulting velocity maps are shown in Fig.~\ref{fig:velocity_maps}. 

Next, we extract velocity profiles across the galaxy and find the direction through the continuum centre (i.e. the barycentre as estimated by {\sc SExtractor}) of the galaxy that maximizes the velocity gradient. We define this as the projected kinematic major axis of the galaxy. Then, the azimuthal angle ($\Phi$) is defined as the angle between this projected major axis and the projected vector from the galaxy's centre to the quasar, where $\Phi=0^\circ$ ($90^\circ$) represents the direction along the projected major (minor) axis, respectively. We are able to estimate the galaxy major axis and $\Phi$ for 38 of the 53 galaxies with \mgii\ absorption in our sample. These are shown in Fig.~\ref{fig:ew_vs_pa} along with \wmg\ of the absorption. The uncertainty in $\Phi$ ranges between 10$^\circ$ and 20$^\circ$.  

We do not find any clear trend between \wmg\ and $\Phi$, as the \mgii\ absorption is detected over the full range of $\Phi$.
Mindful of the fact that the majority of our sample is composed of galaxies that are in pairs or multiple associations, we also consider cases in which we restrict only to galaxy with the smallest impact parameter or with the highest stellar mass among multiple galaxies, finding that the \mgii\ absorption is located at $\Phi<45^\circ$ in 67\% cases and at $\Phi>60^\circ$ in 33\% cases, with the \wmg\ distributions being similar. 
Considering only the isolated galaxies in our sample, we again find an even distribution in equivalent width, perhaps with a hint of a gap between $30^\circ  \lesssim \Phi \lesssim 50^\circ$, but the numbers are too small for any reliable conclusion.

\begin{figure*}
    \centering
    \includegraphics[width=0.48\textwidth]{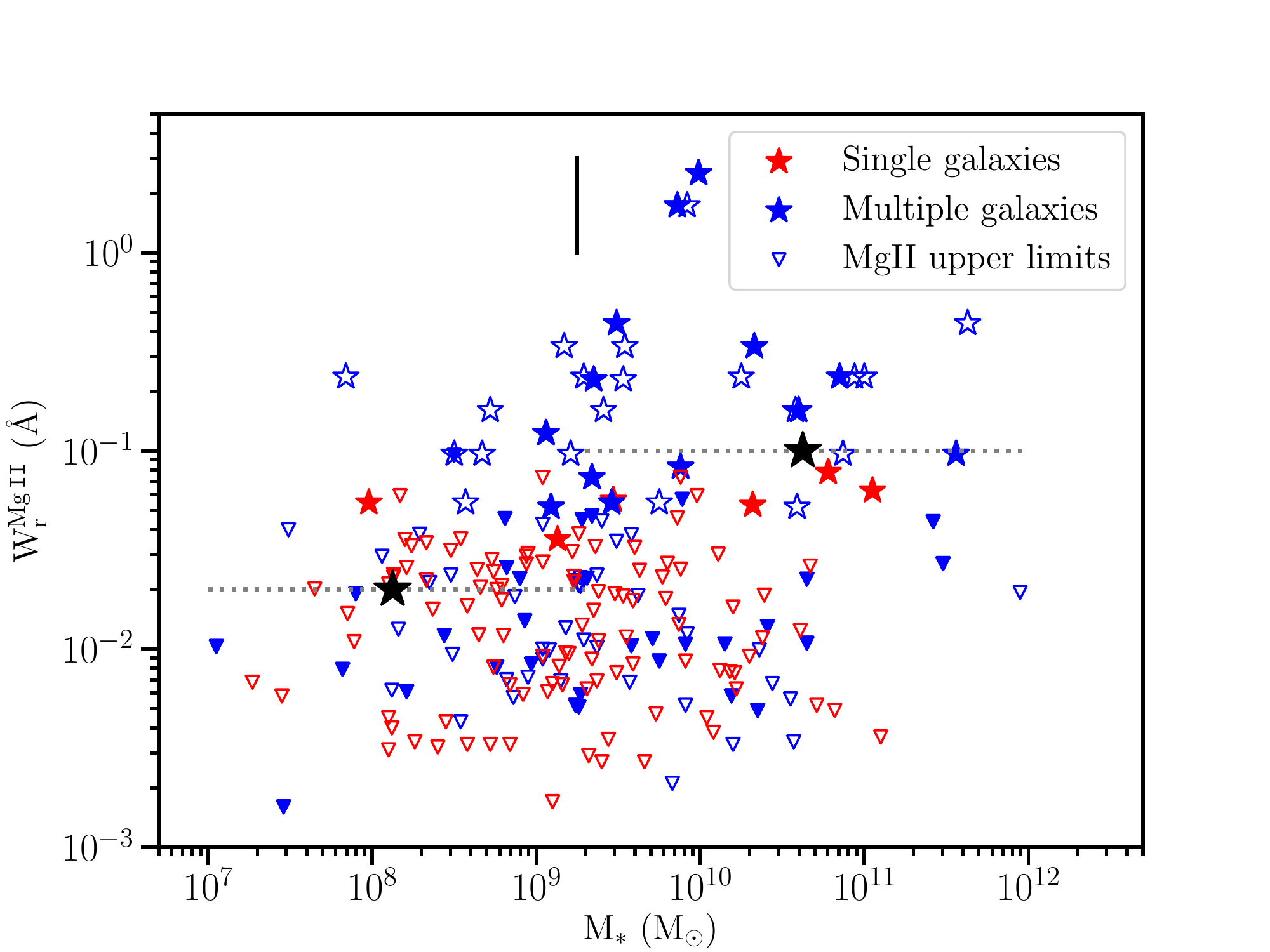}
    \includegraphics[width=0.48\textwidth]{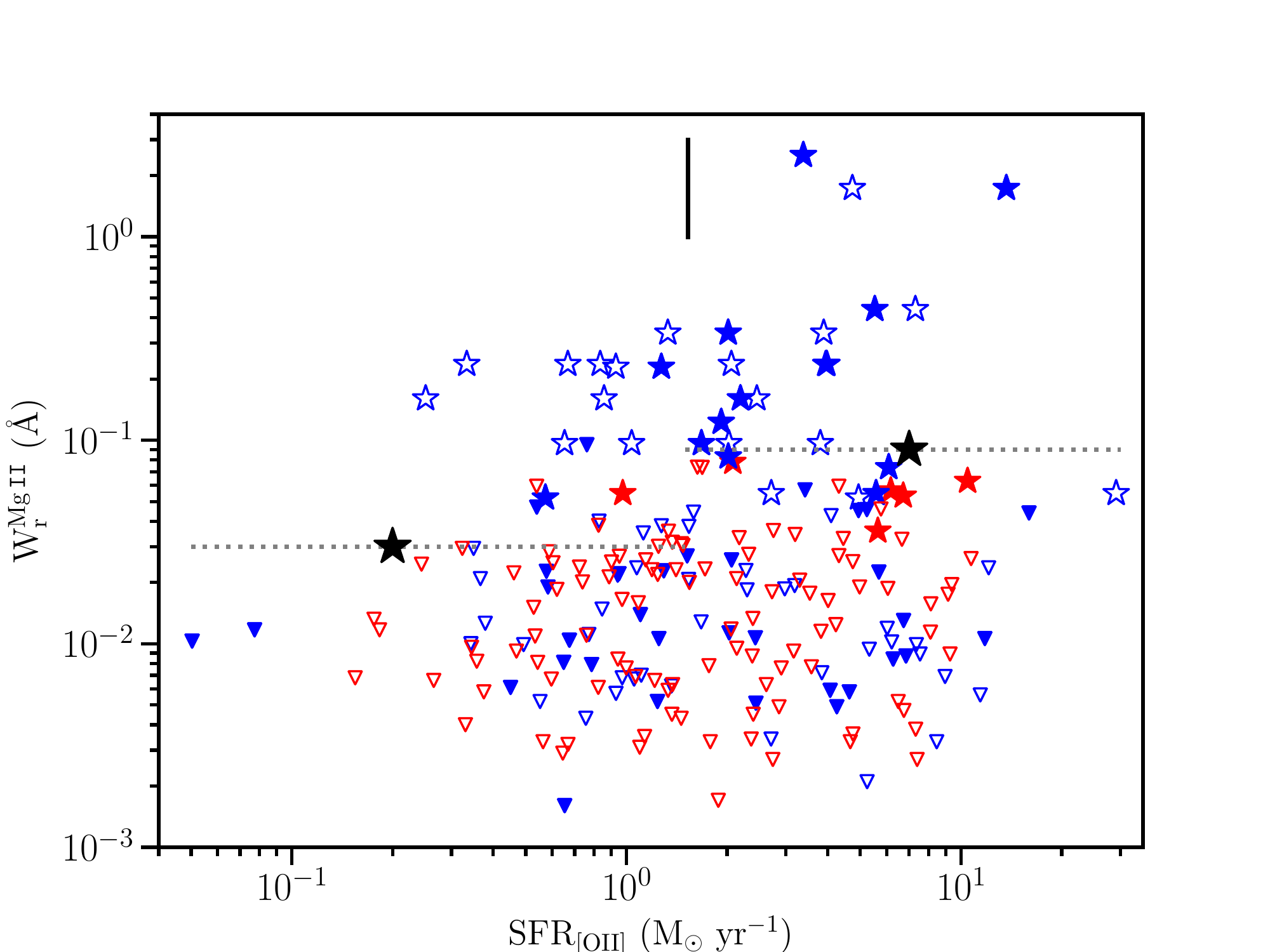}
    \caption{{\it Left:} Rest-frame equivalent width of \mgii\ as a function of stellar mass. \mgii\ detections are plotted as stars, while upper limits on \wmg\ in case of non-detections are plotted as downward triangles. Single galaxies are in red and multiple galaxies are in blue. The galaxy closest in impact parameter among multiple galaxies is shown as a filled symbol, while the remaining ones are shown as open symbols. The median stellar mass is indicated by a vertical tick. The distributions of \wmg\ below and above the median stellar mass are statistically different (see Table~\ref{tab:galaxy_prop}). The average \wmg\ (of detections and upper limits) in the two bins demarcated by the median value are plotted as black stars.
    {\it Right:} Rest equivalent width of \mgii\ as a function of SFR derived from \oii. Symbols are the same as in the left panel. As with stellar mass, \wmg\ shows an increasing trend with SFR, though at less statistical significance (Table~\ref{tab:galaxy_prop}).}
    \label{fig:ew_vs_galaxyprop}
\end{figure*}

The lack of a clear bimodal distribution of $\Phi$ in our sample can be primarily attributed to the presence of multiple galaxies associated with a single \mgii\ absorber. The above studies that have reported a bimodal distribution have been generally based on pairs of \mgii\ absorbers and galaxies defined as isolated on the basis of available data. 
As seen already in the \wmg\ vs $R$ relation above, multiple galaxies exhibit much more scatter compared to isolated galaxies due to environmental processes that perturb the CGM of individual galaxies (see the discussion in Section~\ref{sec_environment}).

A second likely reason for the lack of a strong non-isotropic distribution in our sample is the fact that we do not explicitly target strong absorbers, but rather we sample a typical population of \mgii\ absorbers for which the majority of the galaxies (91\%) are detected at $R>50$\,kpc. The azimuthal dependence of \mgii\ absorption has been found to be stronger at smaller impact parameters ($R\lesssim50$\,kpc), where the \mgii\ absorption is more likely to trace strong biconical outflows from the centre of the galaxy disk \citep{bordoloi2011,bordoloi2014,lan2018}. Hence, the distribution of gas traced by \mgii\ absorption further out in our sample appears to be different and more symmetric than that closer to the galaxies, implying that outflows do not retain a clear spatial coherence (or do not reach) beyond these projected radii. 

In addition, we look for trends between kinematics of the \mgii\ absorption and galaxy orientation. If some of the \mgii\ systems are tracing outflows along the galaxy minor axis, then that could manifest in more extended absorption profiles at $\Phi\gtrsim45^\circ$. We do not find any dependence of the velocity width (\v90) and number of absorption components on the galaxy azimuthal angle. Among the isolated galaxies and closest galaxies in case of multiple associations, the most widespread absorption systems (\v90\ $>300$\,\kms, number of components $=13-18$) occur close to the galaxy major axis ($\Phi\lesssim30^\circ$). In about half of the systems, the velocity offset between the absorption and galaxy centre is along the same direction as the velocity gradient of the galaxy along the direction to the absorber, while it is in the opposite direction in the other half. 

In short, we do not find any significant trend between the absorber kinematics and galaxy orientation in our sample, suggesting that a simple dichotomy between inflows/outflows and/or the presence of extended co-rotating disks appear insufficient to model the more complex distribution observed in the multiple associations uncovered by IFU surveys, especially at larger impact parameters.

\subsection{Dependence of \mgii\ absorption on galaxy properties}
\label{sec_results_galprop}
In this section, we explore the dependence of \mgii\ absorption on intrinsic galaxy properties like stellar mass, star formation rate and velocity dispersion. Several studies in the literature have found correlations between \wmg\ and host galaxy properties, although the reported trends have not been always consistent among different studies. 

\citet{chen2010} found that the extent of \mgii\ absorbing gas around galaxies increases with increasing galaxy $B$-band luminosity. However, they did not find any correlation between absorption strength and galaxy colours. On the other hand, \citet{bordoloi2011} found that \mgii\ absorption depends strongly on galaxy colours, with blue galaxies showing higher \wmg\ than red galaxies, especially at small impact parameters ($R<50$\,kpc). They also found that \wmg\ is correlated with \mstar\ for the blue galaxies. 

Using the MAG{\sc ii}CAT sample, \citet{nielsen2013b} reported that \wmg\ increases with the galaxy $B$- and $K$-band luminosity, but does not significantly depend on the galaxy colour. \citet{lan2014} presented evidence for difference in \mgii\ absorption properties between star forming and passive galaxies defined by their colours. They reported a lack of correlation of \wmg\ with impact parameter and stellar mass of passive galaxies, but an increase in \wmg\ with \mstar, SFR and specific SFR and decrease in \wmg\ with impact parameter of star forming galaxies. Based on a study utilizing background galaxies to probe foreground galaxies, \citet{rubin2018} reported that \wmg\ is higher around galaxies with higher stellar mass and SFR, especially at $R<50$\,kpc. Further, using stacked spectra of background quasars, \wmg\ has been shown to be correlated with the \oii\ luminosity surface density and hence the SFR \citep{noterdaeme2010,menard2011,joshi2018}.

With our complete and homogeneous sample, we analyse the connection between \wmg\ and stellar mass and SFR of the continuum-detected galaxies. Note that since the sample of continuum galaxies were identified independent of the \mgii\ systems, the analysis here is analogous to a galaxy-centric approach. The left panel of Fig.~\ref{fig:ew_vs_galaxyprop} shows \wmg\ measurements and upper limits as a function of galaxy stellar mass. More massive galaxies exhibit stronger \mgii\ absorption on average. Galaxies above the median \mstar\ of our sample have on average five times higher \wmg\ than those associated with lower mass galaxies. There is a positive correlation between \wmg\ (including upper limits) and \mstar\ ($\tau_{\rm k} = 0.11$, $p_{\rm k} = 0.02$). The probability that the distribution of \wmg\ at log \mstar\ $\le9.3$\,\msun\ is drawn from the same parent sample as that at higher \mstar\ is $P=0.003$. This probability decreases to 0.002 when selecting only the galaxy with the smallest impact parameter among multiple galaxies. Note that the impact parameter distributions of the galaxies below and above the median stellar mass are not statistically different (based on a two-sided Kolmogorov-Smirnov (K-S) test, $D_{\rm KS} = 0.12$ and $P_{\rm KS} = 0.40$). Our results are thus consistent with those in the literature discussed above in that more massive galaxies tend to show stronger \mgii\ absorption around them. For instance, \citet{rubin2018} find that the galaxies with log\,\mstar\ $>9.9$\,\msun\ exhibit five times higher \wmg\ at $30<R<50$\,kpc than the less massive ones.

By leveraging the wide range of halo masses probed in our sample ($3\times 10^{10}-4\times 10^{14}~\rm M_\odot$), we can further extend this comparison to the massive haloes that host LRGs. 
Even in this case, results are not always fully consistent in the literature: based on LRG-absorber cross-correlation techniques, \wmg\ have been reported to be anti-correlated with the galaxy halo mass over \wmg\ $\sim0.3-5$\,\AA\ \citep{bouche2006,lundgren2009,gauthier2009}, whereas, using halo abundance matching, \citet{churchill2013} do not find any anti-correlation between \wmg\ and M$_{\rm h}$. We note that there is insufficient number of strong absorbers (\wmg\ $\ge1$\,\AA) in our sample to directly compare to the results of these studies. But in general we find a trend of increasing \wmg\ with M$_{\rm h}$ ($\tau_{\rm k} = 0.11$, $p_{\rm k} = 0.02$), with the more massive haloes (M$_{\rm h}>3\times10^{11}$\,\msun) exhibiting five times higher \wmg\ on average.

\begin{figure}
    \centering
    \includegraphics[width=0.5\textwidth]{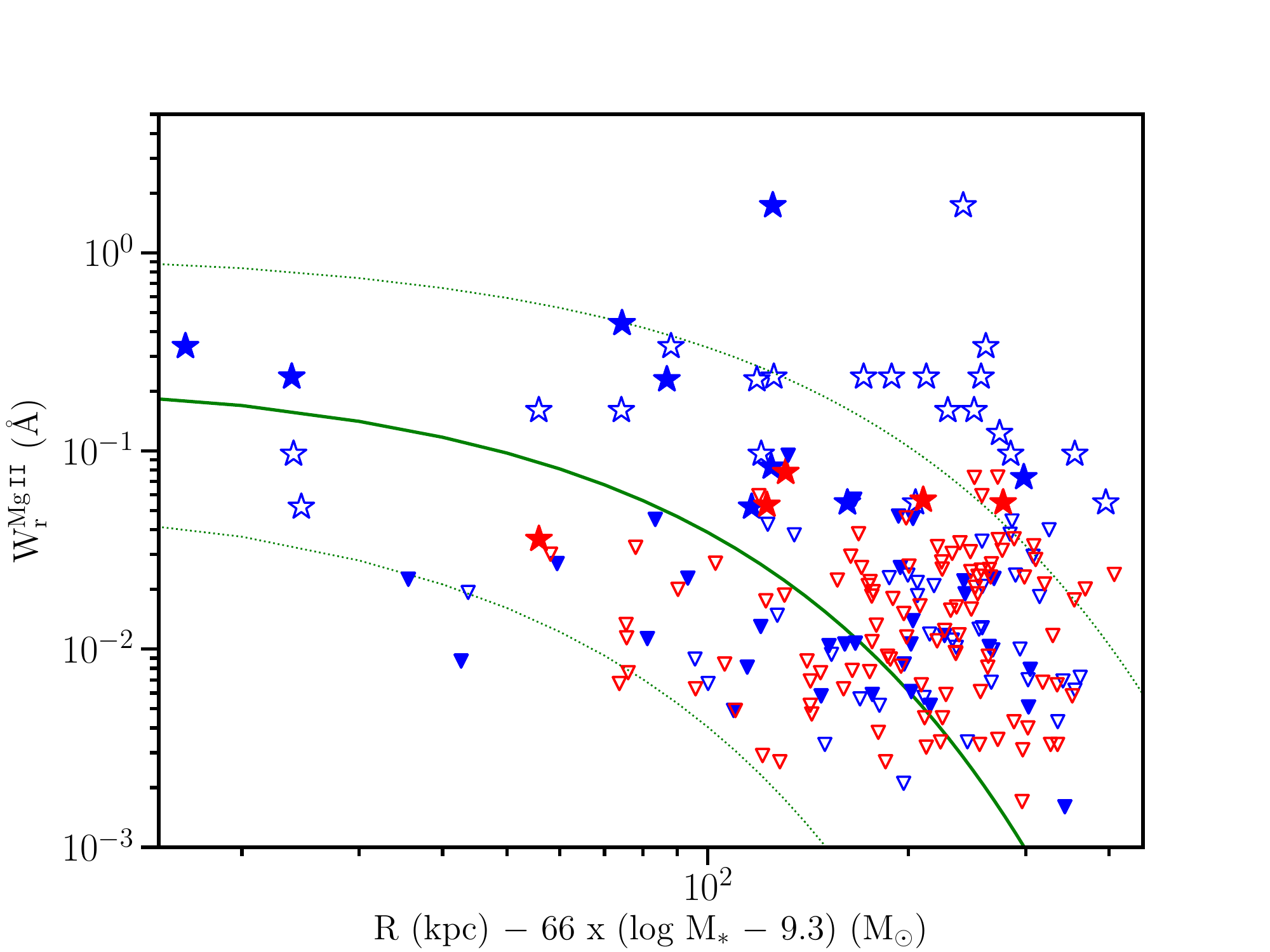}
    \caption{Rest-frame equivalent width of \mgii\ as a function of a combination of impact parameter and stellar mass. The best linear fit between log\,\wmg\ and this combination is shown as a solid line. Single galaxies (red points) and the galaxies closest in impact parameter (solid blue points) among multiple galaxies are considered for this fit. The dotted lines represent the $1\sigma$ uncertainty in this fit. Symbols are the same as in Fig.~\ref{fig:ew_vs_galaxyprop}.}
    \label{fig:ew_imp_mstar_fit}
\end{figure}

\begin{table*} 
\caption{Dependence of \mgii\ absorption on galaxy properties.}
\centering
\begin{tabular}{ccccccc}
\hline
\hline
\multicolumn{7}{c}{All galaxies} \\
\hline
~ & $R\le169$\,kpc & $R>169$\,kpc & log\mstar\ $\le9.3$\msun\ & log\mstar\ $>9.3$\msun\ & SFR $\le1.5$\,\msunyr\ & SFR $>1.5$\,\msunyr\ \\
\hline
$<$\wmg$>^a$ (\AA) & 0.17 & 0.05 & 0.02 & 0.10 & 0.03 & 0.09 \\
$P^b$ & \multicolumn{2}{c}{$4\times10^{-4}$} & \multicolumn{2}{c}{$3\times10^{-3}$} & \multicolumn{2}{c}{$4\times10^{-2}$} \\
\hline
\multicolumn{7}{c}{Single galaxies $+$ galaxies with smallest $R$ among multiple} \\
\hline
$<$\wmg$>$ (\AA) & 0.17 & 0.03 & 0.01 & 0.08 & 0.005 & 0.07 \\
$P$ & \multicolumn{2}{c}{$9\times10^{-5}$} & \multicolumn{2}{c}{$2\times10^{-3}$} & \multicolumn{2}{c}{$9\times10^{-3}$} \\
\hline
\hline
\end{tabular}
\begin{flushleft}
$^a$ Average of the \wmg\ distribution including upper limits for the subsets using the regression on order statistics method in {\sc r}. \\
$^b$ Probability of the \wmg\ distributions of the two subsets arising from the same parent population based on survival analysis using the Peto \& Peto modification of the Gehan-Wilcoxon test in {\sc r}.
\end{flushleft}
\label{tab:galaxy_prop}
\end{table*}

Next, we study the link between \wmg\ and SFR, finding that galaxies showing higher SFR also show stronger \mgii\ absorption (Fig.~\ref{fig:ew_vs_galaxyprop}; right), though the correlation is weaker ($\tau_{\rm k} = 0.05$, $p_{\rm k} = 0.24$) than that found between \mstar\ and \wmg. The average \wmg\ of galaxies above the median SFR of 1.5\,\msunyr\ is three times higher than that of the galaxies with lower SFR. The probability of the \wmg\ distributions with higher and lower SFR being from the same population is $P=0.04$ (0.008 when considering the galaxy closest in $R$ in case of multiple associations). However, we do not find any significant trend ($p_{\rm k} = 0.38$) of \wmg\ with the specific SFR (sSFR = SFR/\mstar) for the closest galaxies. 

For this reason, we attribute the existence of a correlation of \mgii\ absorption with SFR as not likely to be related to the underlying star formation physics, but more to a reflection of the correlation with stellar mass (or generally ``size" of a system) through the galaxy main sequence. Once this mass dependence is normalized out in the sSFR, there is no strong residual correlation.
This is not in agreement with the results of \citet{bordoloi2011} and \citet{lan2014} who find that blue galaxies show significantly stronger \mgii\ absorption, but is consistent with the lack of a strong dependence of \wmg\ on galaxy colours as found by \citet{chen2010} and \citet{nielsen2013b}. 
We note however that our sample predominantly comprises of star forming galaxies ($\sim$90\% have SFR $\gtrsim0.5$\,\msunyr, sSFR $\gtrsim10^{-10}$\,yr$^{-1}$). A lack of passive galaxies could explain the weaker correlations of \wmg\ with SFR and sSFR that we find compared to the studies of \citet{bordoloi2011} and \citet{lan2014}, which included a significant number of passive galaxies.

In Section~\ref{sec_results_radial}, we have seen that \wmg\ is dependent on $R$. To capture the combined dependence of \wmg\ on $R$ and \mstar\ as found above into a single relationship, we define a linear dependence of log\,\wmg\ on both $R$ and log\,\mstar\ as
\setcounter{equation}{2}
\begin{equation}
 \rm \log~W_{r}^{Mg\,\textsc{ii}} (\mbox{\normalfont\AA}) = a + b_1\times R~(\rm kpc) + \rm b_2\times[log~M_* (M_\odot) - 9.3].
\label{eqn:wmg_imp_mstar}
\end{equation}
Here, the dependence on log\,\mstar\ is offset by the median stellar mass of the sample. We define a likelihood function for the above model following Eqn.~{\color{blue}2} and estimate the best-fit parameters as described in Section~\ref{sec_results_radial}. As an attempt to isolate the effects of individual galaxies from the one of the environment, we fit to the single galaxies and closest galaxies among multiple galaxies. The best-fit relation is shown in Fig.~\ref{fig:ew_imp_mstar_fit} and is characterized by $a = -0.61^{+0.63}_{-0.58}$, $b_1 = -0.008^{+0.003}_{-0.004}$, $b_2 = 0.53^{+0.34}_{-0.28}$ and $\sigma_{sc} = 1.00^{+0.30}_{-0.20}$.

\begin{figure}
    \centering
    \includegraphics[width=0.5\textwidth]{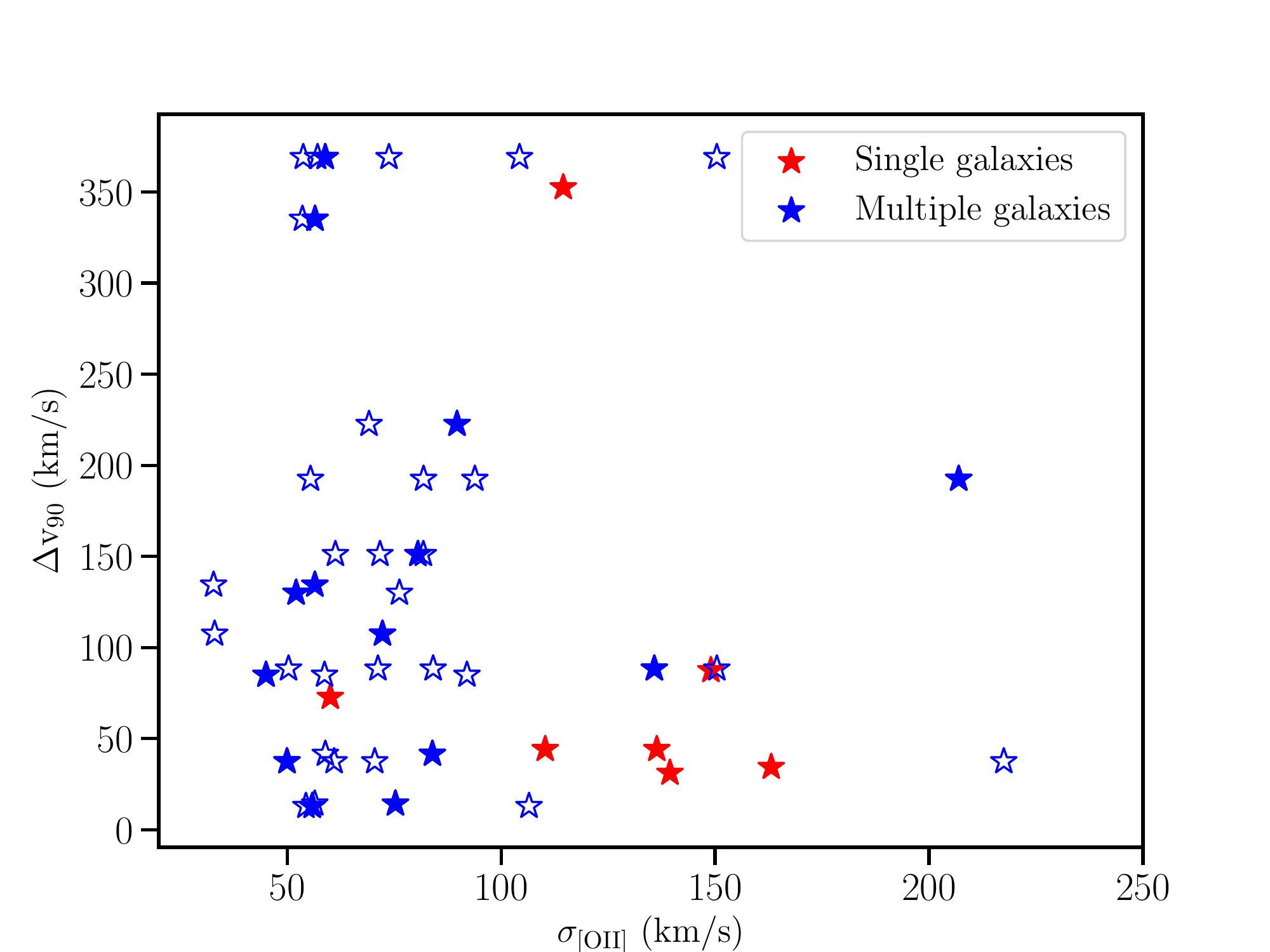}
    \caption{Velocity width of \mgii\ absorption which contains 90\% of the total optical depth (\v90) as a function of the velocity dispersion of the \oii\ emission ($\sigma_{\rm[O~\textsc{ii}]}$) from galaxies. Symbols are same as in Fig.~\ref{fig:ew_vs_galaxyprop}.}
    \label{fig:v90_vs_sigoii}
\end{figure}

Above, we have found a dependence of \wmg\ on \mstar\ and a weaker one on SFR, and we have argued that the lack of a dependence of \wmg\ on sSFR indicates that the trend with SFR is mainly driven by the dependence on \mstar. To investigate more this trend, we first estimate the \wmg\ values for our sample as predicted by the above relation of \wmg\ with $R$ and \mstar\ (Eqn.~\ref{eqn:wmg_imp_mstar}). Then, we check for a dependence of the difference in the predicted and observed \wmg\ values on the SFR of the galaxies. There is no significant trend ($p_{\rm k} = 0.22$) of the \wmg\ residuals with SFR, suggesting instead that the stellar mass (along with distance from galaxy centre) is the dominant factor influencing the \mgii\ absorption around the galaxies in our sample.

\begin{figure*}
    \centering
    \includegraphics[width=0.48\textwidth]{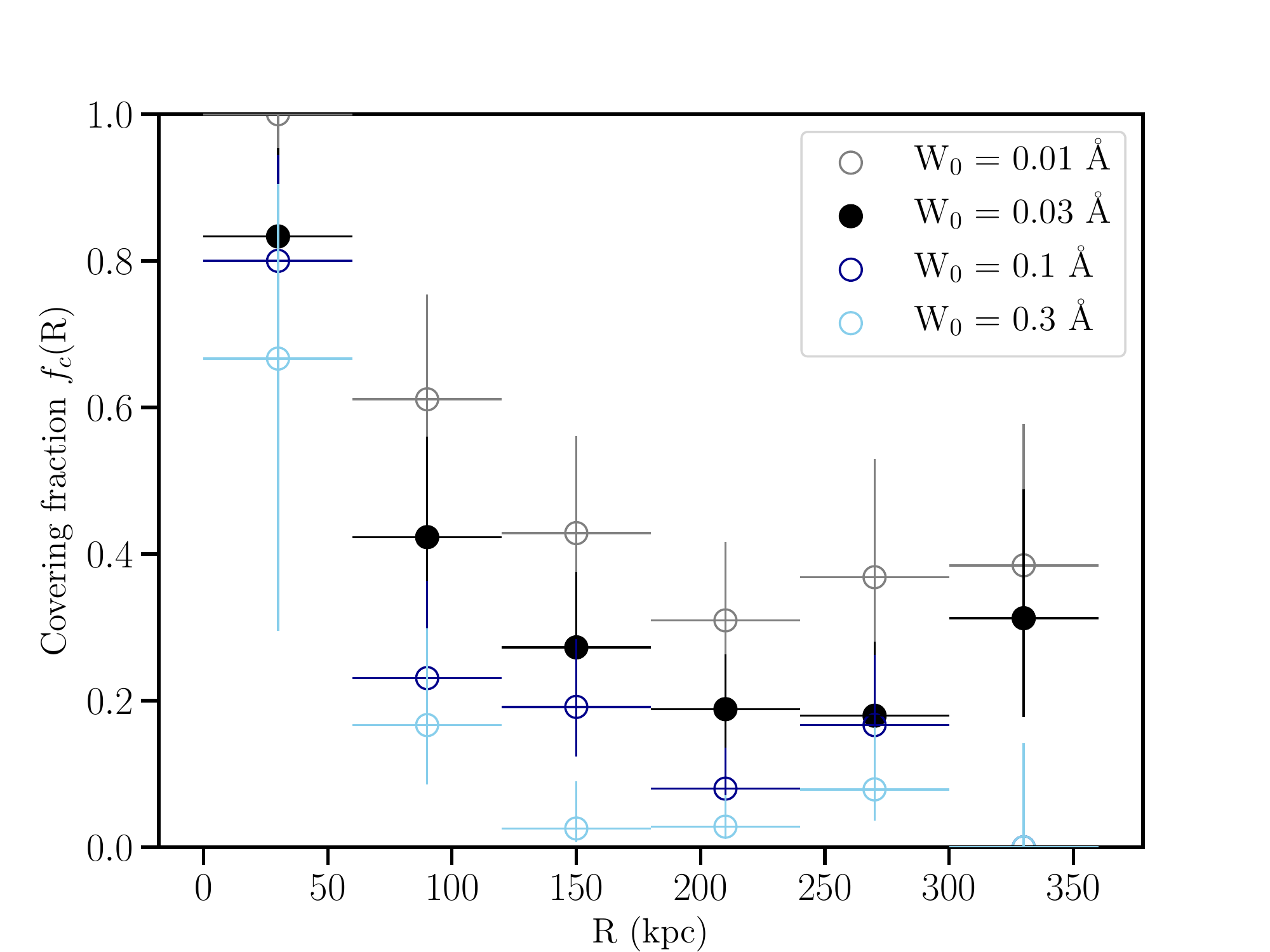}
    \includegraphics[width=0.48\textwidth]{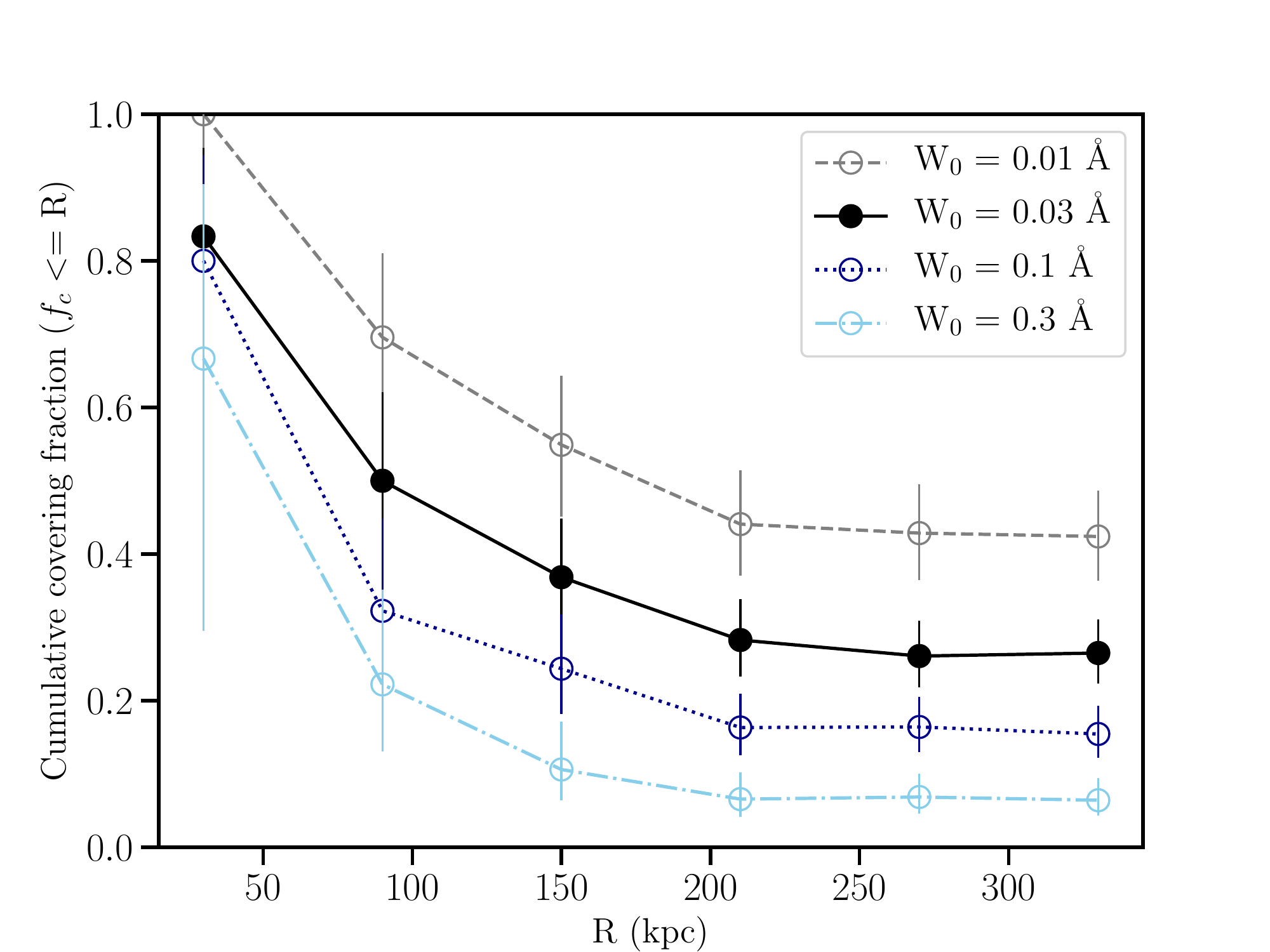}
    \caption{Differential {\it (left)} and cumulative {\it (right)} covering fraction of \mgii\ absorbing gas as a function of galaxy impact parameter for different \wmg\ sensitivity limits (W$_0$) considering all the galaxies in our sample. Error bars represent the 68\% confidence interval. Note that the cumulative covering fraction measurements are correlated.}
    \label{fig:covfrac_ewlimit}
\end{figure*}

We further check whether the absorption properties are in any way linked with other galaxy properties. For example, the kinematics of the \mgii\ absorption could be affected by the intrinsic galaxy properties. Strong \mgii\ absorption has been observed to trace powerful outflows from the host galaxies \citep{weiner2009,nestor2011,rubin2014}, as well as the intra-group medium \citep{kacprzak2010,gauthier2013,bielby2017}. Hence, the velocity extent of the absorption could be driven by the host galaxy's star formation, or it could reflect the disturbed gas dynamics in the galaxies due to interaction within a group (see below).

We check for correlations between the velocity width of the \mgii\ absorption, as defined by \v90, with the SFR and velocity dispersion of the \oii\ emission ($\sigma_{\rm[O~\textsc{ii}]}$). We do not find any correlations between these parameters. As it can be seen from Fig.~\ref{fig:v90_vs_sigoii}, \v90\ of the \mgii\ absorption does not show any specific trend with $\sigma_{\rm[O~\textsc{ii}]}$ of the galaxies. Further, the number of absorption components in a \mgii\ system reflects the kinematic complexity of the gas, which could be influenced by the galaxy properties. We do not find any correlation of the number of absorption components (from Voigt profile fits; Section~\ref{sec_analysis_mgii}) with impact parameter and velocity separation from the galaxy, although the most kinematically complex \mgii\ absorption system (number of components = 18) occurs closest to the host galaxy. Besides the correlation of \wmg\ with \mstar\ discussed above, the other strong trend we find is that of \wmg\ with the galaxy environment, as we discuss in Section~\ref{sec_environment}. The dependence of \wmg\ on galaxy properties discussed here is summarized in Table~\ref{tab:galaxy_prop}.

\subsection{Covering fraction of cool gas}
\label{sec_results_covfrac}

Having explored correlations between galaxies and \mgii\ absorption in the above sections, we now turn to a more statistical description of the cool gas distribution around galaxies by means of the covering fraction of \mgii\ absorption and its dependence on galaxy properties, taking again a more galaxy-centric approach.
The covering fraction profile of absorbing gas with distance from galaxies can in fact be used to place constraints on the extent and patchiness of the CGM. In particular, the covering fraction of \mgii\ absorbing gas can be used to infer the extent of cool ($T\sim10^4$\,K) metal-enriched gas around galaxies. 

We define covering fraction as
\begin{equation}
 f_c = \frac{ N_{\rm det}({\rm W}^{\rm Mg~\textsc{ii}}_{\rm det} \ge {\rm W_0}) } { N_{\rm tot}({\rm W}^{\rm Mg~\textsc{ii}}_{\rm sens} \le {\rm W_0}) },
\label{eqn:covfrac} 
\end{equation}
which is the fraction of galaxies showing \mgii\ absorption with \wmg\ above the equivalent width limit W$_0$ out of all the galaxies with sufficient sensitivity to detect \mgii\ absorption above this limit. We show the differential (estimated in different $R$ annuli) and cumulative (estimated inside a radius $R$) covering fraction as a function of the impact parameter for different W$_0$ in the left and right panel of Fig.~\ref{fig:covfrac_ewlimit}, respectively. The error bars represent $1\sigma$ Wilson score confidence intervals. 

The covering fraction profiles show a decrease with impact parameter as expected. Also, the covering fraction increases as we go to lower sensitivity limits, declining more steeply beyond $R\sim100$\,kpc for W$_0$ $\ge0.1$\,\AA\ than for W$_0$ $\le0.03$\,\AA. In line with the trend of \wmg\ with $R$, this implies that strong \mgii\ absorbers originate from the inner CGM of individual galaxies, while the weaker absorbers also trace the outer CGM and, in some instances, the extended intra-group medium. The covering fraction of \mgii\ absorbers with \wmg\ $\ge0.1$\,\AA\ is $\approx 0.8$ at $R\le 50$\,kpc and decreases to $\lesssim0.2$ at $R>100$\,kpc. This is consistent within the uncertainties with the reported covering fractions of \mgii\ absorbers above the same detection threshold in literature samples \citep{chen2010,nielsen2013b}, which are primarily at lower redshifts. For the rest of the paper, we adopt W$_0$ = 0.03\,\AA\ for the covering fraction analysis in this work, since $\sim$90\% of our galaxies have sensitivity to this equivalent width.

For W$_0$ = 0.03\,\AA, the covering fraction estimated within $R=50~$kpc is $\sim0.8$, but has large uncertainties due to the small number statistics. \mgii\ absorption signal is present out to the highest impact parameter probed in our sample, with $f_c$ on an average being $\sim0.2$ at $R\gtrsim200$\,kpc. The total covering fraction decreases from $\sim0.5$ within $R\le 100$\,kpc to $\sim0.3$ within $R<300$\,kpc. The differential $f_c$ shows an increase at the outermost impact parameter ($R>300$\,kpc) that, although not statistically significant, could arise out of the superposition of individual \mgii\ haloes of the galaxies in multiple associations. Note that for this analysis, we have considered all galaxies with associated \mgii\ absorption. The results do not change within the $1\sigma$ uncertainties when we consider only the closest galaxies in presence of multiple association.

As explained above, in our full sample of 228 galaxies, 14 are identified on the basis of line emission alone. Unlike the continuum-detected galaxies these were not identified blindly but based on the presence of associated \mgii\ absorption. Furthermore, we are most probably missing a small fraction of continuum-faint line emitting galaxies, especially near the bright quasar PSF, from the full galaxy sample. To test for the effect of this, we repeat the covering fraction estimates excluding the line emitting-galaxies. The estimates change by $\le20$\% for W$_0$ = 0.03\,\AA, but are consistent within the $1\sigma$ errors with the ones shown in Fig.~\ref{fig:covfrac_ewlimit}. 

\begin{figure*}
    \centering
    \includegraphics[width=0.48\textwidth]{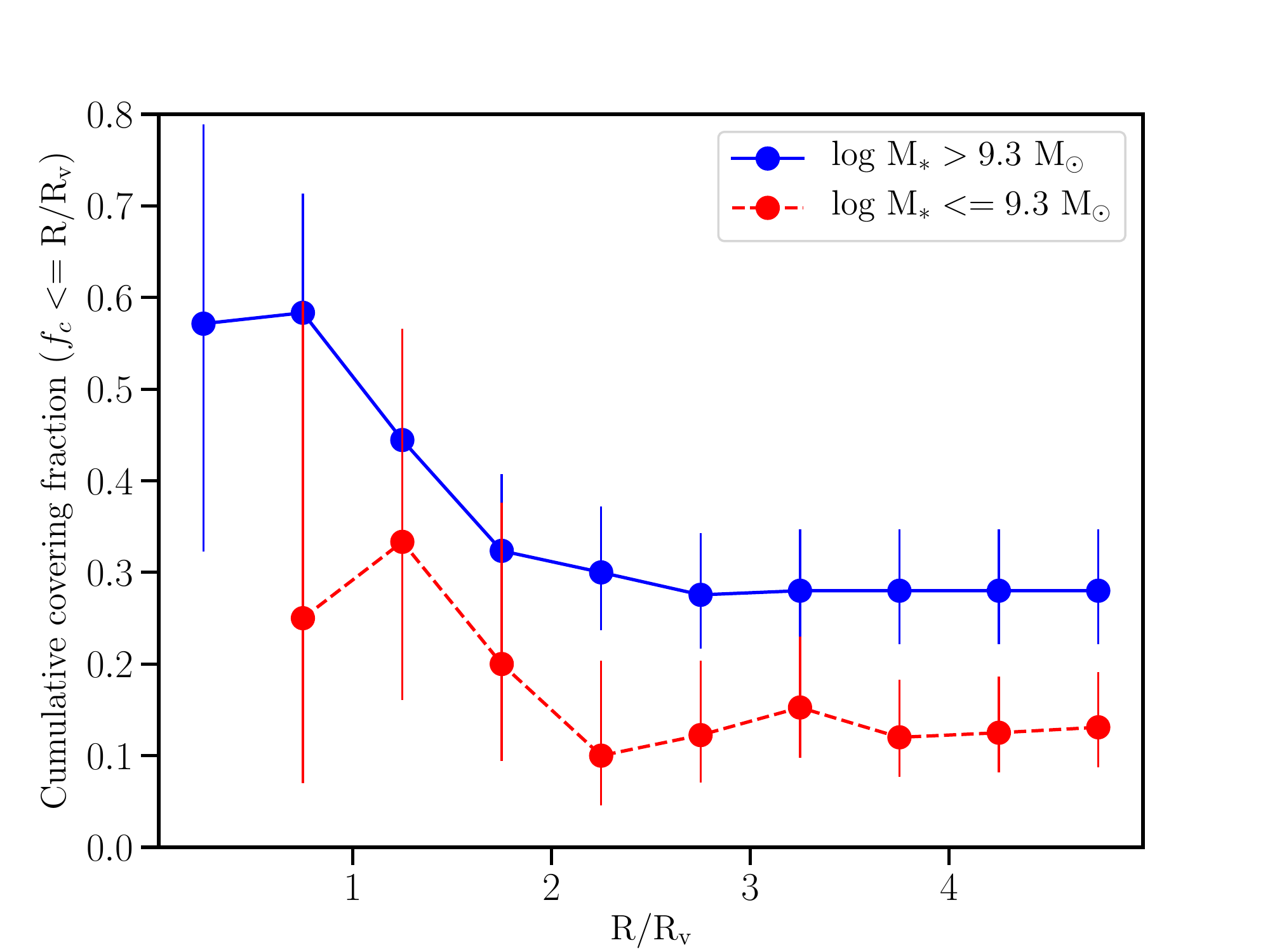}
    \includegraphics[width=0.48\textwidth]{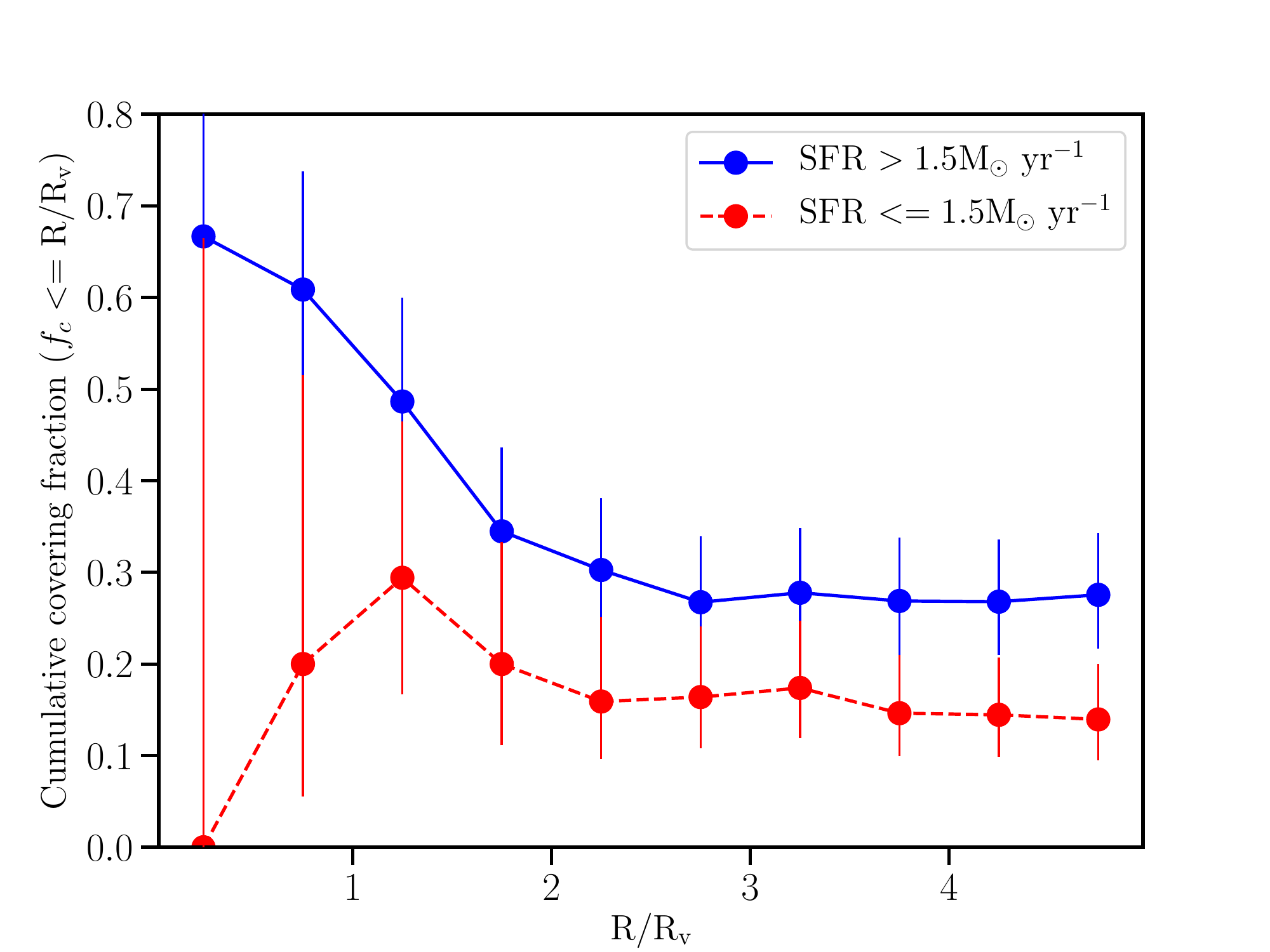}
    \caption{Dependence of the covering fraction of \mgii\ absorbing gas on galaxy properties $-$ stellar mass {\it (left)} and SFR {\it (right)}, as a function of the impact parameter normalized by the virial radius, for a \wmg\ limit of W$_0$ = 0.03\,\AA. Blue (red) symbols connected with solid (dashed) lines are estimates for galaxies with higher (lower) \mstar\ and SFR, respectively. Error bars represent the 68\% confidence interval.}
    \label{fig:covfrac_galprop}
\end{figure*}

Since the \mgii\ absorption strength was found to be correlated with \mstar\ and marginally with SFR in Section~\ref{sec_results_galprop}, next we investigate the dependence of the covering fraction profile on these properties in the continuum-detected galaxy sample, which is unbiased with respect to absorption. Since the CGM of a more massive galaxy is expected to have a larger radial extent, we normalise the impact parameter with the galaxy virial radius. In Fig.~\ref{fig:covfrac_galprop} we plot the cumulative covering fraction as a function of the scaled impact parameter, and this time divide the sample into two subsets based on the median \mstar\ and SFR values. The decline in the covering fraction profile is more pronounced for the higher mass and more star forming galaxies. More massive galaxies show two times higher $f_c$ compared to the less massive galaxies within the virial radius, while higher star forming galaxies show three times higher $f_c$ within this radius. We note, however, that these trends are not strongly statistically significant due to the appreciable uncertainty arising from small number statistics. The above trends hold when we consider only the closest galaxy in the case of multiple associations. These results are compatible with studies in the literature that report higher covering fractions around more luminous star forming galaxies \citep{nielsen2013b,lan2014}.

It is interesting to note that the more massive/star-forming galaxies still exhibit on an average $f_c$ $\sim0.3$ beyond the virial radius, indicating the presence of extended metal-enriched gas haloes. This could arise due to the overlapping of haloes in groups or the extended intra-group medium (see e.g. Section~\ref{sec_environment}). Indeed, \citet{bordoloi2011} have reported more extended \mgii\ haloes around group galaxies. Hence, the more massive/star-forming galaxies in our sample could tend to be intrinsically more clustered and their higher covering fraction could reflect effects of the denser environment in which they reside. We look at the effect of the environment on the \mgii\ absorption in detail within the next part of this work. We conclude this part by looking next at redshift evolution in \mgii-galaxy trends.

\subsection{Redshift evolution}
\label{sec_results_redshift}

The majority of the \mgii-galaxy correlations we have discussed at this point based on MAGG are generally at higher redshift compared to most literature studies. For example, 90\% of the galaxies in the MAG{\sc ii}CAT sample \citep{nielsen2013a} lie at $z<0.8$, and the median redshift of the sample is 0.36. The study of \citet{lan2014} focuses on the redshift range $0.4<z<0.6$, while that of \citet{rubin2018} is at $0.35<z<0.8$. In general, due to observational constraints, CGM studies have focused either on lower redshifts ($z\le0.5$) using UV absorption lines and \mgii\ absorption lines, or higher redshifts ($z\ge2$) using mainly \lya\ absorption lines. Hence, the intermediate redshift range ($z\sim0.8-1.5$) probed here is all the more interesting since it allows us to study the evolution of CGM properties over a redshift period where the cosmic star formation rate density declines from its peak \citep{madau2014}. Given the heterogeneous nature of \mgii-galaxy samples in the literature $-$ different depth and sensitivity of galaxy surveys and absorption spectra, and methodology (single-/multi-slit/IFU, absorption probed against background quasar/galaxy, individual/stacked, galaxy/\mgii-centric) $-$ it is not trivial to undertake direct quantitative comparison of our results with those in the literature. 

However, as discussed in a few instances above, most of the trends seen in our sample, (e.g. the radial decline of \wmg\ and $f_c$, and the increase in \wmg\ and $f_c$ with stellar mass) are at least qualitatively similar to what have been reported in the literature for the lower redshift samples.
Therefore, we do not find obvious evidence for evolution in the properties of the cool metal enriched CGM around galaxies from $z\sim0.3$ to $z\sim1.5$. Within our sample we also do not find any significant dependence of \wmg\ and $f_c$ on redshift, though the sample size and redshift range probed is still small to appreciate a mild evolution. Further, the distributions of impact parameter and stellar mass, and the fraction of isolated galaxies is not different below and above the median redshift of $z=1.25$ in our sample.
Overall, there seem to be only weak (if at all present) evolution of the CGM as probed by \mgii\ at $z\approx 0.5-1.5$.
This is consistent with the results of \citet{chen2012}, who find that the spatial extent and absorption strength of the CGM around galaxies of comparable mass have not changed significantly over $z\sim0-2$. \citet{lan2020} also find no redshift evolution over $0.4<z<1.3$ in the covering fraction of \mgii\ absorbers with $0.4<$ \wmg\ $<1$\,\AA, which is similar to the \wmg\ distribution of our sample, although they do find that strong absorbers (\wmg\ $>1$\,\AA) show significant redshift evolution in their covering fraction around both star-forming and passive galaxies.
Such trends in redshift evolution of the cool CGM will be soon better explored by ongoing surveys such as QSAGE \citep{bielby2019} and CUBS \citep{Chen2020}.

\begin{figure*}
    \centering
    \includegraphics[width=0.48\textwidth]{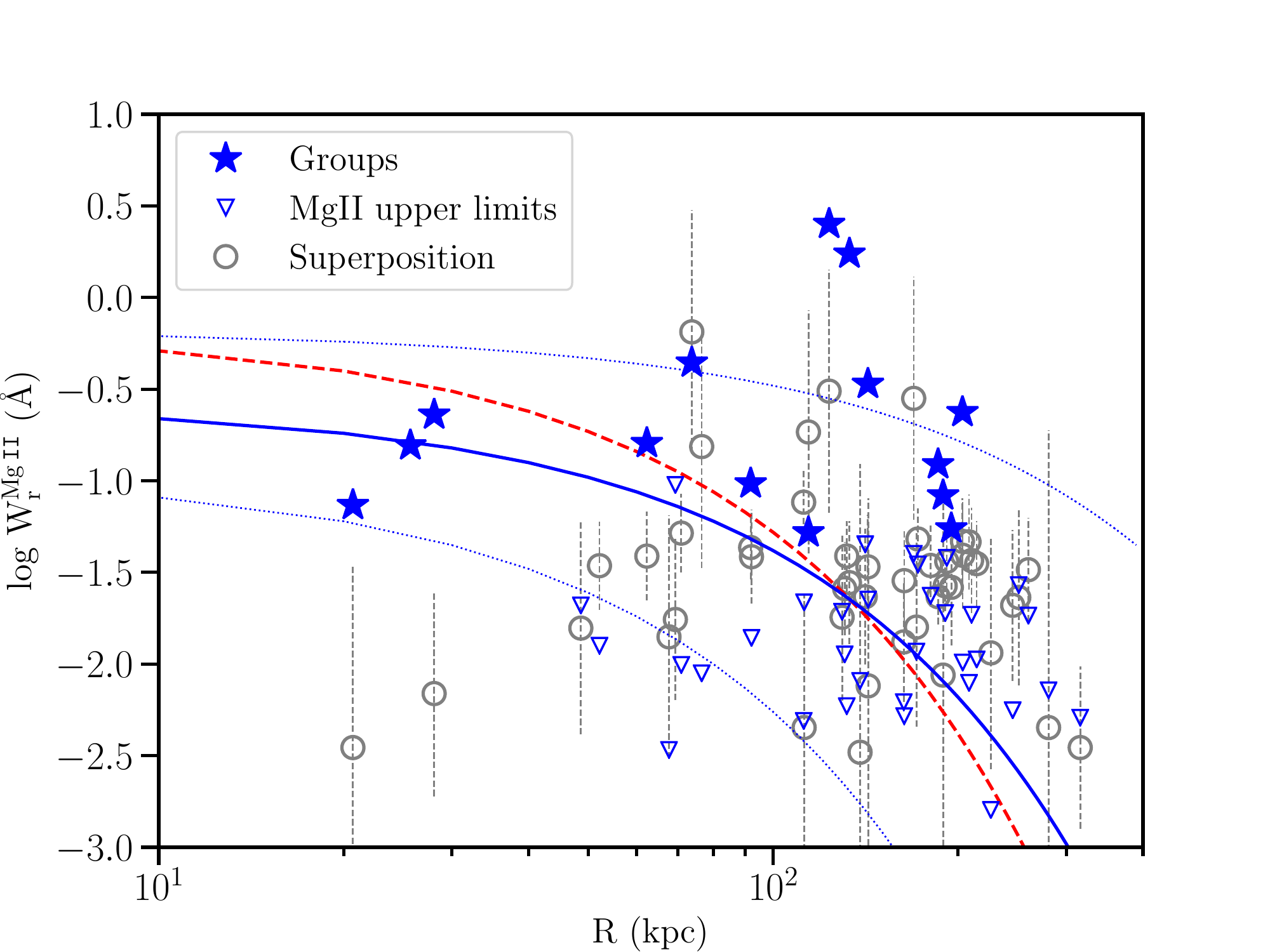}
    \includegraphics[width=0.48\textwidth]{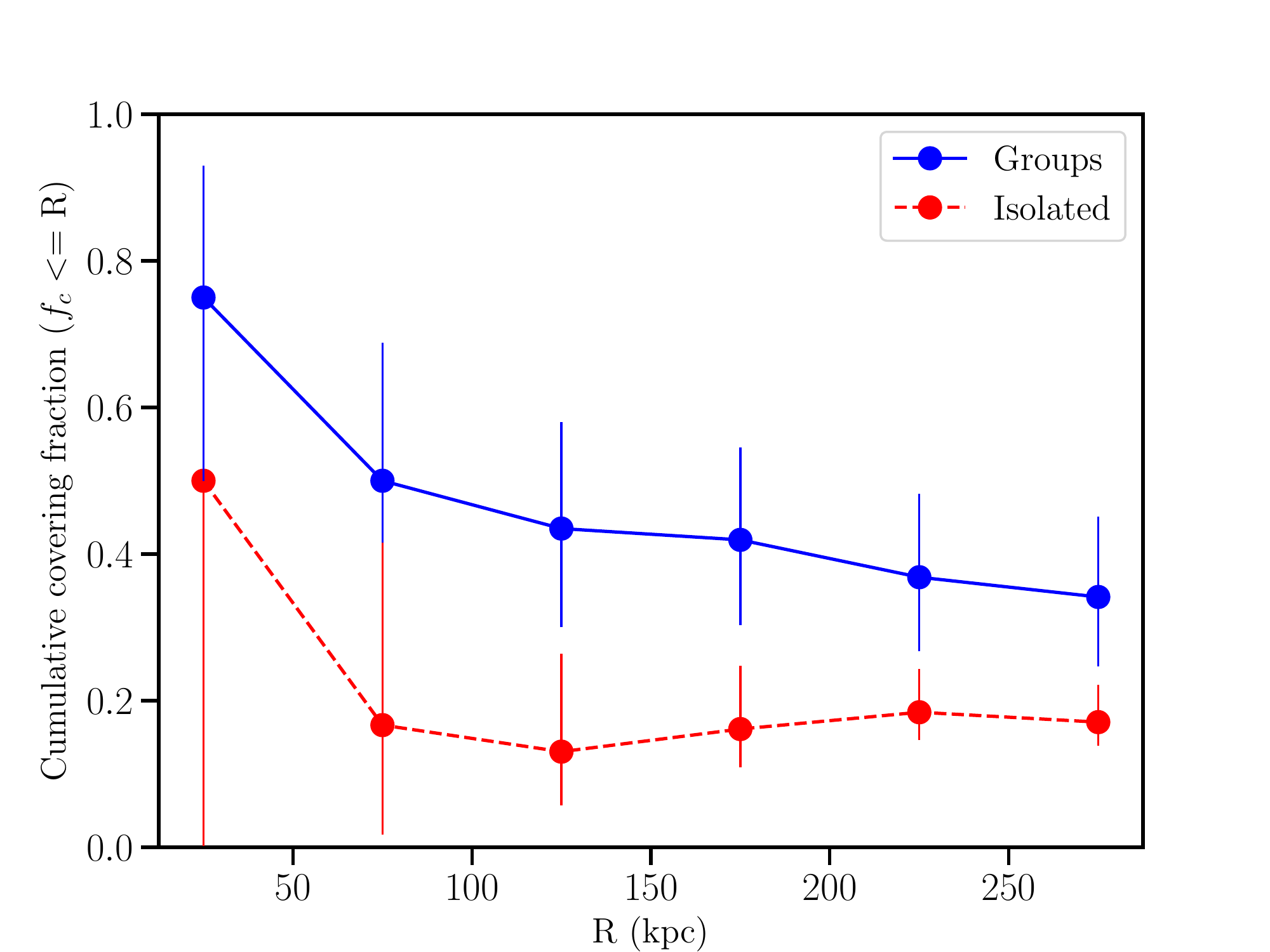}
    \caption{{\it Left:} Rest-frame equivalent width of \mgii\ as a function of impact parameter for the groups in our sample. Measurements of \wmg\ are shown as stars while upper limits are shown as downward triangles. Estimates of \wmg\ for the groups from the superposition model are shown in grey symbols. The estimate for each group is shown at the impact parameter from the group centre. The best linear fit to log\,\wmg\ and $R$ for groups is plotted in blue solid line. The dotted blue lines indicate the $1\sigma$ uncertainty in the fit. We show the best fit to isolated galaxies in red dashed line for reference.
    {\it Right:} Cumulative covering fraction of \mgii\ absorbing gas as a function of impact parameter for the isolated galaxies (red) and groups (blue). Error bars represent 68\% confidence interval.}
    \label{fig:ew_vs_imp_groups}
\end{figure*}

\section{The role of environment}
\label{sec_environment}

In the previous sections, we have explored the properties of \mgii\ absorption in correlation to galaxies, highlighting general trends characteristic of individual galaxies. As noted in Section~\ref{sec_results_detection}, the majority (67\%) of the \mgii\ absorption systems in our sample is associated with more than one galaxy.
Moreover, in several instances, it has already become clear that the properties of the cool gas in the CGM of galaxies in richer environments differ from those of isolated galaxies. In this section, we explore more deeply the role of environment in shaping the \mgii\ gas in the CGM.

\subsection{Results from the literature}
\label{sec_environment_lit}

There is already significant literature on the study of \mgii\ absorption in denser environment, e.g. starting from the discoveries of galaxy groups associated with single \mgii\ systems in the past \citep[e.g.][]{whiting2006,kacprzak2010,gauthier2013,fossati2019b}. \citet{gauthier2013} find that a $z=0.5$ LRG associated with ultra-strong \mgii\ absorption (\wmg\ $>4$\,\AA) resides in a group environment, and they argue that gas stripped from the gravitational potential of the group members could lead to such large \wmg, a result echoed in \citet{fossati2019b} on the basis of a larger sample. 
Indeed, the extended presence of \mgii\ absorption ($f_c\approx15$\% at 100\,kpc) around passive LRGs \citep{huang2016} suggests that some of the \mgii\ absorption comes from the CGM or stripped material of satellite galaxies around LRGs.

\begin{table} 
\caption{Dependence of \mgii\ absorption on the environment.
Reported are the mean of the \mgii\ absorption property (\wmg, \v90, and number of components) for the isolated and group sub-samples, and the probability of these distributions arising from the same parent sample. In case of \wmg, upper limits are included in the analysis.
}
\centering
\begin{tabular}{ccc}
\hline
\hline
 & Isolated & Groups \\
\hline
\wmg\ (\AA) & 0.03 & 0.14 \\
$P$ & \multicolumn{2}{c}{$2\times10^{-5}$} \\
\hline
\v90\ (\kms) & 95 & 137 \\
$P$ & \multicolumn{2}{c}{0.13} \\
\hline
$N_{\rm comp}$ & 4 & 6 \\
$P$ & \multicolumn{2}{c}{0.07} \\
\hline
\hline
\end{tabular}
\label{tab:group_prop}
\end{table}

There have also been a few statistical studies that have investigated environmental effects in shaping the CGM of group galaxies. \citet{chen2010} did not find any correlation between \wmg\ and $R$ for galaxies in groups, defined as a galaxy having at least one neighbour within its $B$-band luminosity-scaled gas radius and within line-of-sight velocity separation, $\rm \Delta v = 300$\,\kms. 
Recently, \citet{huang2020} expanded the sample studied by \citet{chen2010}, and they similarly find no clear trend between \wmg\ and $R$ for the non-isolated galaxies, defined as having at least one neighbour at $R \le 500$\,kpc and $\rm \Delta v \le 1000$\,\kms.

Using stacked spectra of background galaxies, \citet{bordoloi2011} found a flatter and more extended radial profile around group galaxies, identified on the basis of the zCOSMOS group catalogue of \citet{knobel2012}. They further found that the \mgii\ radial profiles around the geometric centre of the groups or around the most massive galaxy in the groups are more extended than that around isolated galaxies. However, they were able to reproduce the extended radial profile around groups with a superposition model of the absorption profiles of isolated galaxies, indicating that group environment may not have a strong effect on \mgii\ absorption profile of individual group galaxies.
\citet{nielsen2018}, on the other hand, based on a similar superposition model combined with kinematic analysis of 29 groups (defined as two or more galaxies within $R=200$\,kpc and $\rm \Delta v = 500$\,\kms), with a \wmg\ completeness cut of 0.04\,\AA, suggested that the \mgii\ absorption is associated with the intra-group medium rather than individual galaxies in a group. While they found marginally extended radial profile of \wmg\ around galaxies in their group sample, this was consistent with that found around their isolated sample within the uncertainties. They additionally reported enhanced \wmg\ and covering fraction around groups compared to isolated galaxies. A recent study by \citet{pointon2020} finds instead no difference in the CGM metallicity of isolated galaxies and groups at $z=0.25$.

We caution that one should allow for uncertainty in the classification of isolated galaxies and groups in the literature studies that lack complete galaxy spectroscopy of the absorber fields. Furthermore, there is no unambiguous classification of what constitute a group in the literature, e.g. based on halo mass or number of members. 
Recently, however,  with the advent of MUSE, more sensitive and complete analysis of the galaxy environment around absorbers is becoming possible. \citet{bielby2017} studied the \mgii\ absorption associated with a single group (defined as M$_{\rm h}\approx6\times10^{12}$\,\msun) using MUSE and hypothesized that the absorption is most likely to originate from a superposition of cool gas clouds that form the intra-group medium. The MUSE-ALMA haloes study \citep{peroux2019,hamanowicz2020} also suggests that the absorption is associated with more complex galaxy structures, rather than a single galaxy, with most \mgii\ systems in their sample being associated with multiple galaxies. However, the radial profile of \mgii\ absorption around their galaxies in groups is consistent with that around isolated galaxies in the literature. 

\citet{chen2019} revisited the group environment around a $z=0.3$ \mgii\ system, that was observed initially by \citet{kacprzak2010} and more recently with MUSE by \citet{peroux2019}. They detected a giant nebula of ionized gas pervading the group in the MUSE data and suggested that kinematics of the absorption is consistent with it originating from tidally stripped gas. \citet{fossati2019b} studied the correlation between galaxy groups detected in the MUDF and \mgii\ absorption in the spectra of the two background quasars in this field. The strength of the absorption in these groups is higher than what has been found for isolated galaxies in the literature at comparable impact parameters and stellar mass. Based on an analysis of correlated absorption towards both the quasars and in the stacked spectra of background galaxies probing a foreground group, they found no evidence for widespread cool gas tracing the intra-group medium. Rather they suggested that either the strong \mgii\ absorption is arising out of the CGM of individual galaxies in a group that lie at small impact parameters, or it could have been stripped from the CGM of these galaxies due to gravitational interactions within the group.

\begin{figure*}
    \centering
    \includegraphics[width=0.48\textwidth]{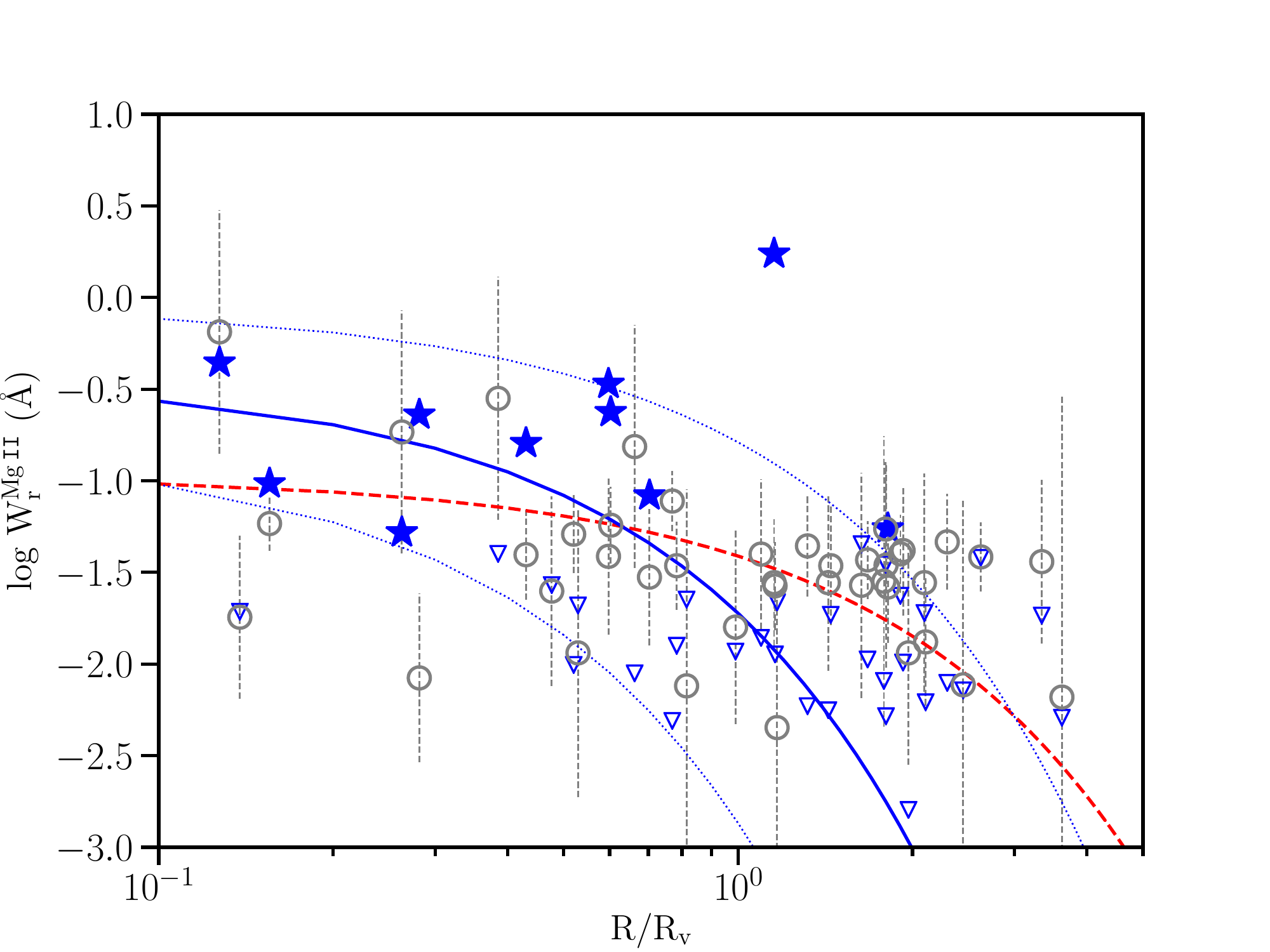}
    \includegraphics[width=0.48\textwidth]{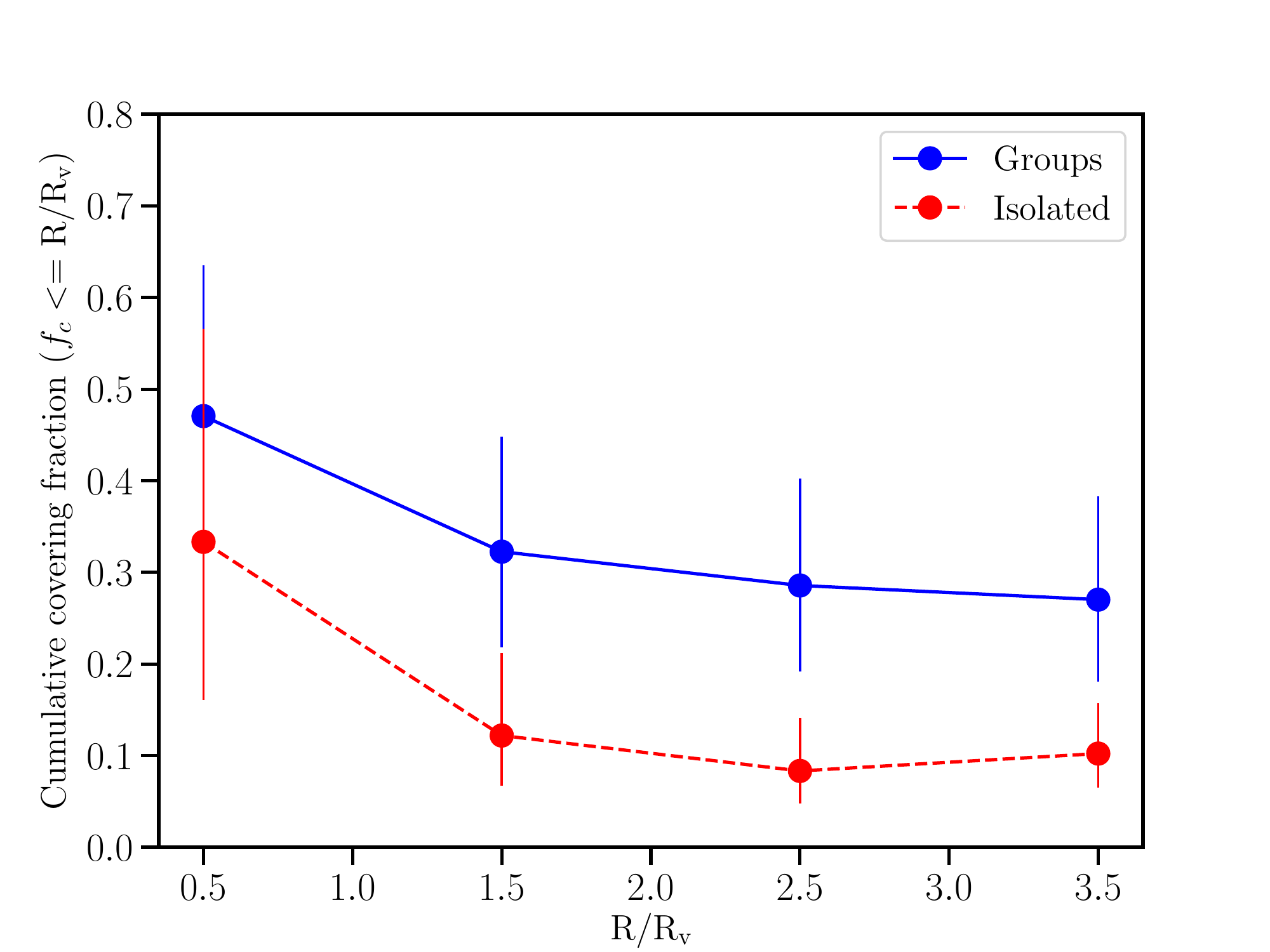}
    \caption{Same as in Fig.~\ref{fig:ew_vs_imp_groups}, but with the impact parameter normalized by the virial radius for the continuum-detected galaxies.}
    \label{fig:ew_vs_imp_rvir}
\end{figure*}

\subsection{\mgii\ absorption in groups}
\label{sec_environment_mgii_groups}

In Table~\ref{tab:group_prop}, we list the properties of \mgii\ absorption associated with single or isolated galaxies and with multiple associations and groups of galaxies as a whole. In our definition, galaxies are isolated when there is no other galaxy detected down to the flux limit of our sample in the MUSE FoV and within $\rm \Delta v = 500$\,\kms, otherwise they are taken as part of a multiple galaxy association. We use the term ``group" for these galaxy associations in this work, with the caveat that no halo mass constraint has been included. In other words, we adopt a more loose definition of ``group" as a non-isolated environment.

The \wmg\ distributions (including upper limits) of isolated galaxies and groups are significantly different (probability of arising from same population, $P=2\times10^{-5}$), with groups showing five times higher \wmg\ than isolated galaxies on average. The distributions of velocity width and number of absorption components of \mgii\ detections in the isolated and group subsets are marginally different. Groups tend to be associated with more extended \mgii\ absorption, having on average slightly higher \v90\ and number of absorption components than isolated galaxies.

Next, we analyse the radial profile of \mgii\ absorption around the groups in our sample. We define the impact parameter for each group as the projected separation from the quasar to the geometric centre of the group. \wmg\ as a function of this group-centric impact parameter is shown in the left panel of Fig.~\ref{fig:ew_vs_imp_groups}. 
The best-fit linear relation between log\,\wmg\ and $R$ for groups (slope $b=-0.008\pm0.005$) is slightly flatter but consistent with that obtained for the isolated galaxies alone ($b=-0.011\pm0.005$). These relations are also consistent with that obtained for the combined sample of isolated galaxies and closest galaxies in groups ($b=-0.010\pm0.003$) in Section~\ref{sec_results_radial}.
As with the isolated and closest galaxies, there is considerable scatter in the \wmg\ versus $R$ plane for the group-centric measurements as well, especially at higher impact parameters, indicating that a simple relation is not likely to fully capture the complexities of the radial distribution of metals in groups.

Further, we investigate whether the \mgii\ covering fraction is different for the isolated galaxies and groups. The cumulative covering fraction profiles around isolated galaxies and group centres as a function of impact parameter are shown in the right panel of Fig.~\ref{fig:ew_vs_imp_groups}. The covering fraction declines with increasing distance from both galaxies and group centres. However, groups generally show higher covering fraction at any given impact parameter. The average covering fraction within $R<200$\,kpc of groups is three times higher than that around isolated galaxies, indicating that \mgii\ absorbing gas is more prevalent in the group environment.

\subsection{Superposition models}
\label{sec_environment_superposition}

As mentioned above, both \citet{bordoloi2011} and \citet{nielsen2018} have compared superposition models resulting from summing \wmg\ of isolated galaxies with the group \wmg. The assumption behind such models is that the enhanced absorption seen around groups is due to the quasar sightline tracing haloes of multiple galaxies, and that the gas radial profile around individual galaxies in a group is similar to that around isolated galaxies. We investigate here whether a simple superposition model can explain the \mgii\ absorption we associate to groups in our sample. 

To construct such a model, for each galaxy in a group we identify isolated galaxies that are at similar impact parameter from the quasar (i.e. $R$ differs by $\le20$\,kpc), and that differ in stellar mass by $\le0.5$\,dex. Then we randomly select the \wmg\ measurement of one of them if there are at least three such isolated galaxies. We sum the associated \wmg\ of the member galaxies under the assumption that the upper limits correspond to measurements at the $3\sigma$ upper limit value. We repeat this process 1000 times and consider the median and $1\sigma$ uncertainty values from the bootstrap analysis as the final estimates for each group. In case there are no or less than three control isolated galaxies that can be associated with a group galaxy, we rely on the best-fit relation obtained between log\,\wmg, $R$ and \mstar (Eqn.~\ref{eqn:wmg_imp_mstar}). In this case, for each galaxy in a group, we estimate the \wmg\ expected from the model fit for its impact parameter and stellar mass. Then, similar to the above, we sum up the \wmg\ obtained for all the galaxies in the group. The estimates based on the superposition model are shown as grey symbols in the left panel of Fig.~\ref{fig:ew_vs_imp_groups}.

About half of the superposition model estimates are consistent within the uncertainties with the observed values. A two-sided K-S test between the model and observed values gives $D_{\rm KS} = 0.24$ and $P_{\rm KS} = 0.13$, indicating that they are not very different.
Hence, on an average the superposition model is consistent with the group measurements. However, such a simplistic model, which does not take into account interactions in the group, cannot explain all the observed \wmg\ measurements associated with groups in our sample, especially the highest and lowest values of the distribution. This is in contrast with the results of \citet{bordoloi2011}, and more in line with those of \citet{nielsen2018}, who found that the superposition model could explain about half of their group measurements but not the ones with low \wmg.

\begin{figure*}
    \centering
    \includegraphics[width=0.48\textwidth]{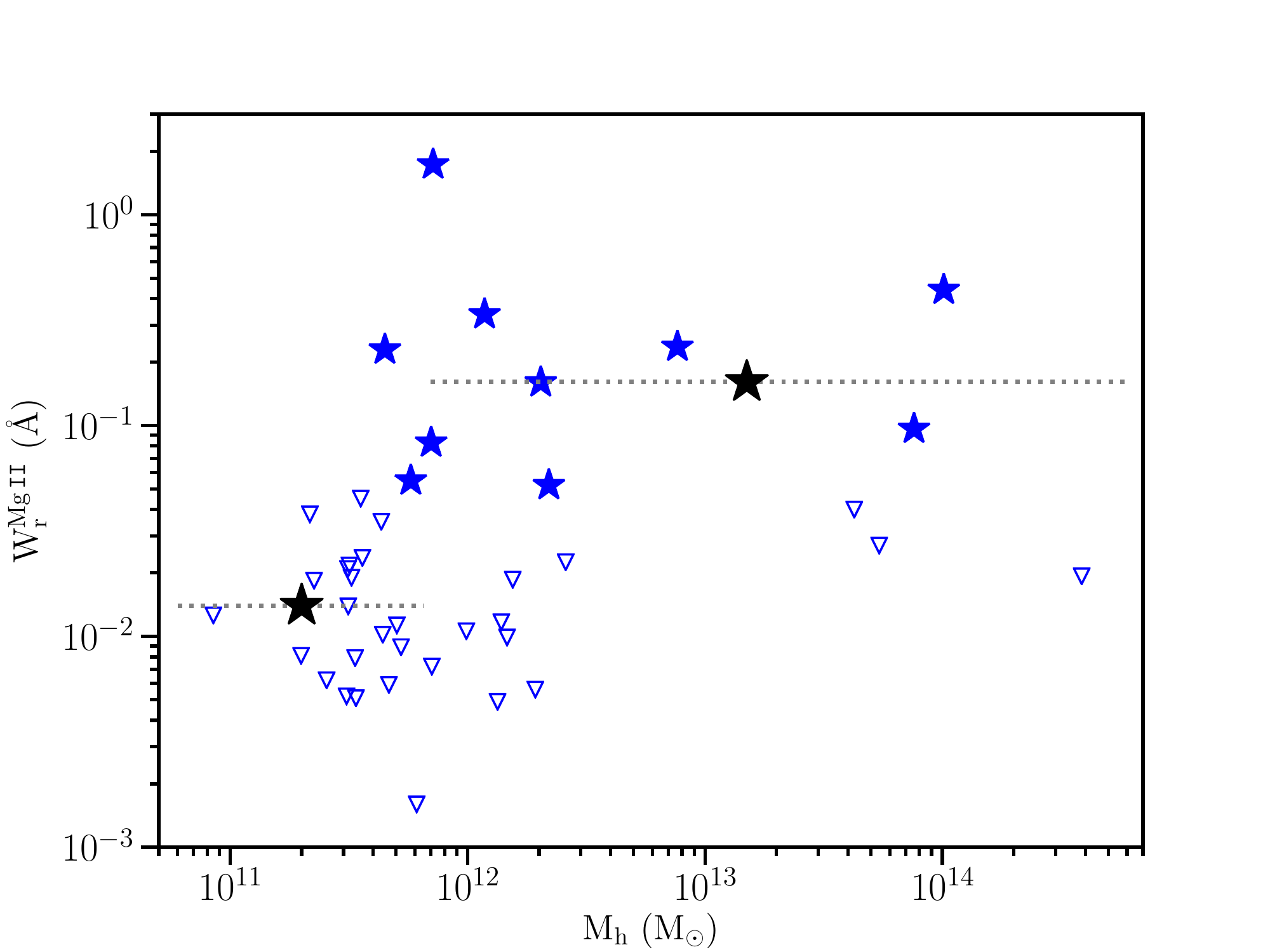}
    \includegraphics[width=0.48\textwidth]{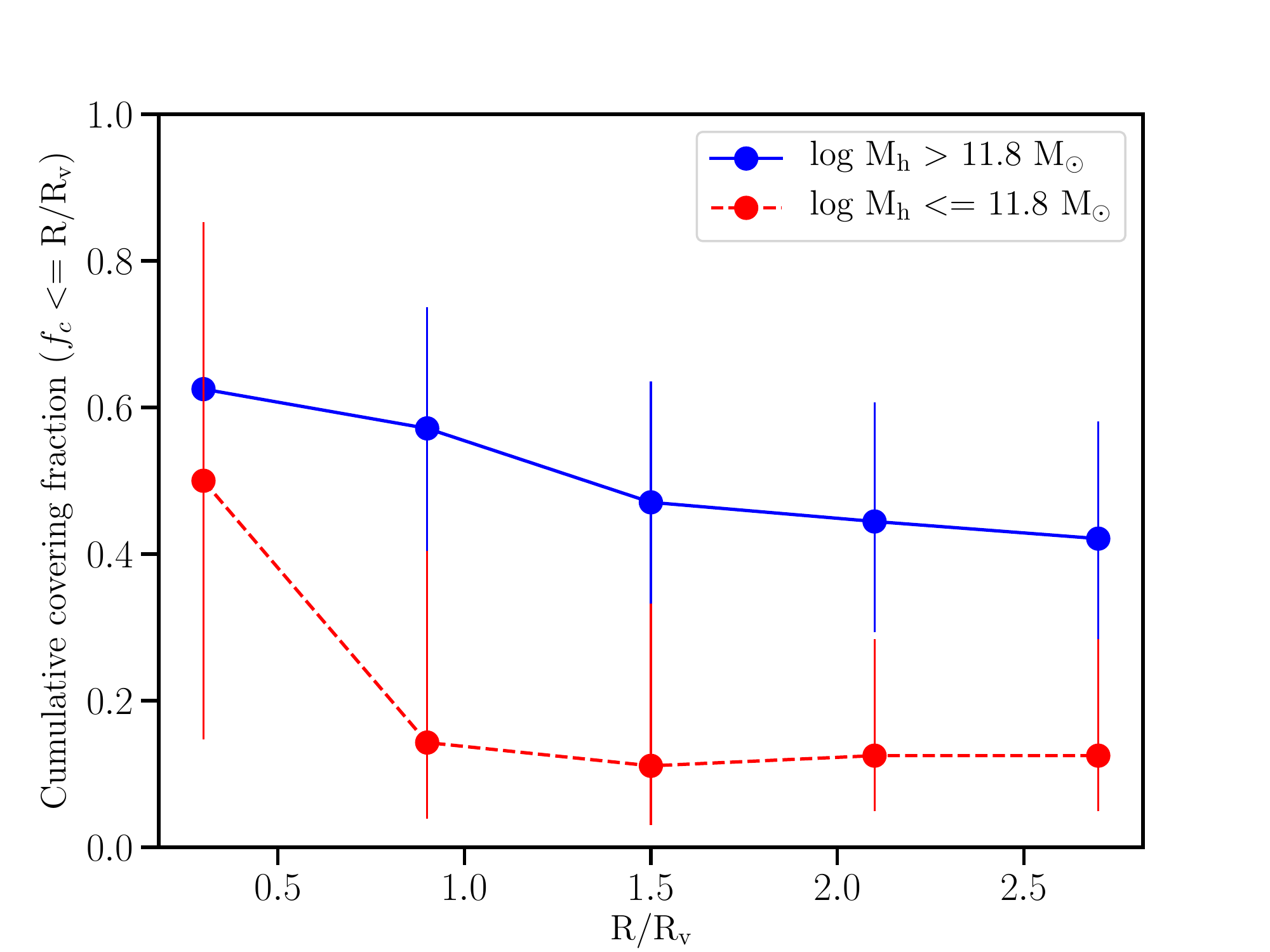}
    \caption{{\it Left:} Rest equivalent width of \mgii\ systems associated with groups as a function of the halo mass. Measurements of \wmg\ are shown as stars while upper limits are shown as downward triangles. The average \wmg\ values above and below the median halo mass are plotted as black stars.
    {\it Right:} Cumulative covering fraction of \mgii\ absorbing gas in groups as a function of impact parameter scaled by the group virial radius. Estimates for groups above the median halo mass are shown in blue while those for less massive groups are shown in red. Error bars represent 68\% confidence interval.}
    \label{fig:ew_vs_mhalo_groups}
\end{figure*}

\subsection{Trends with group size and mass}
\label{sec_environment_trends_groups}

To get a perspective on the physical properties of the groups in our sample, we next run a Friends-of-Friends algorithm on the MAGG continuum-detected galaxy catalogue. These algorithms use a separation in physical distance ($\Delta$r) and line-of-sight velocity ($\Delta$v) to link together galaxies in structures \citep[e.g.][]{knobel2009,knobel2012,diener2013}. We use $\Delta$r = 500\,kpc and $\Delta$v = 500\,\kms\ to link two or more galaxies into groups. 
For the purpose of this analysis, we consider the most massive galaxy in the group as the central galaxy \citep[see e.g.][]{yang2008} and use its stellar mass to obtain the group halo mass using the stellar-halo mass relation of \citet{moster2010}. 
The halo masses of the groups lie in the range $9\times10^{10}-4\times10^{14}$\,\msun\ and the median halo mass is $\sim6\times10^{11}$\,\msun.
Number of members in a group range between two and seven. We associate the groups to \mgii\ absorption if the group redshift (average of that of the members) lies within $\pm500$\,\kms\ of the \mgii\ redshift, otherwise we estimate an upper limit on \wmg.

We plot in the left panel of Fig.~\ref{fig:ew_vs_imp_rvir} the \wmg\ associated with each group versus the impact parameter scaled by the virial radius ($R_{\rm v}$) of the group as estimated from the halo mass. Note that for this analysis, the galaxies are defined into groups based on only the continuum-detected sample. \mgii\ absorption is detected out to twice the virial radius of groups. We show in the same figure the best linear fit to log\,\wmg\ and $R/R_{\rm v}$ for groups and isolated galaxies. The radial profile of \wmg\ around groups becomes steeper (slope $b=-1.3^{+0.5}_{-0.8}$) compared to that around isolated galaxies ($b=-0.44^{+0.3}_{-0.4}$) once the impact parameter is normalized with the virial radius. The \mgii\ absorption associated with groups tend to be more concentrated within the virial radius compared to isolated galaxies, reducing the number of detections at large projected separation which appear as outliers in Fig.~\ref{fig:ew_vs_imp_groups}.

We further compare \wmg\ for the groups defined on the basis of continuum-detected galaxies with the estimates from the superposition model described above. These model estimates are plotted as grey symbols in the left panel of Fig.~\ref{fig:ew_vs_imp_rvir}. 
About 60\% of the model values are consistent with the observed ones within the uncertainties. However, a two-sided K-S test between them gives $D_{\rm KS} = 0.37$ and $P_{\rm KS} = 0.01$, indicating that they are likely to be drawn from different distributions.
This is most likely because the model is unable to explain all the observed \wmg\ values associated with these groups. 

We have seen above that the covering fraction around groups is higher compared to isolated galaxies at a given impact parameter. To additionally correct for the effect of mass, we normalize the impact parameter with the virial radius. We plot the covering fraction as a function of impact parameter scaled with the virial radius of groups and isolated galaxies in the right panel of Fig.~\ref{fig:ew_vs_imp_rvir}. Even after scaling with virial radius, groups tend to show higher covering fraction of \mgii\ gas. The covering fraction within twice the virial radius is three times higher around groups as compared to around isolated galaxies. 

Till now we have seen that groups show on an average stronger and more prevalent \mgii\ absorption than isolated galaxies. We now look at whether there are any trends among groups themselves. Fig.~\ref{fig:ew_vs_imp_rvir} shows that stronger \mgii\ absorption arises closer to the group centres. In the left panel of Fig.~\ref{fig:ew_vs_mhalo_groups}, we plot the \wmg\ associated with groups as a function of the group halo mass. 
More massive groups tend to be associated with stronger absorption, and \wmg\ shows a positive correlation with the halo mass ($\tau_{\rm k} = 0.21$, $p_{\rm k} = 0.04$).
\mgii\ absorption is also more widespread around the more massive groups. 
The cumulative covering fractions of \mgii\ for groups above and below the median halo mass of $6\times10^{11}$\,\msun, as a function of impact parameter scaled by the group virial radius, are shown in the right panel of Fig.~\ref{fig:ew_vs_mhalo_groups}. The more massive groups show four times the \mgii\ covering fraction within the virial radius than the less massive groups.
However, we do not find any significant trend of the absorption width (\v90) with the halo mass and velocity dispersion of the groups.

\subsection{Stacking of \oii\ and \mgii\ emission}
\label{sec_environment_stack}

\begin{figure*}
    \centering
    \includegraphics[width=1.0\textwidth]{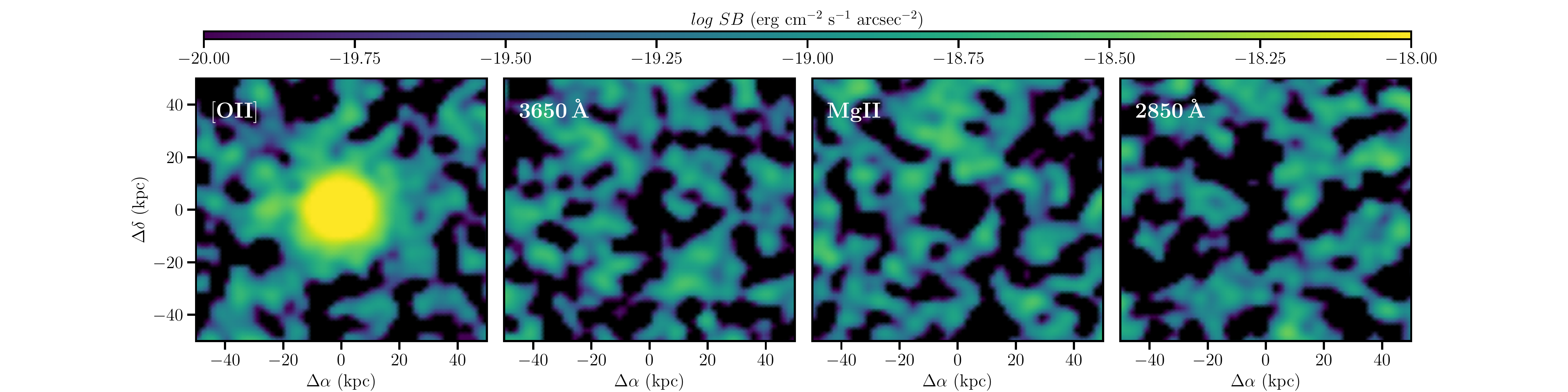}
    \includegraphics[width=1.0\textwidth]{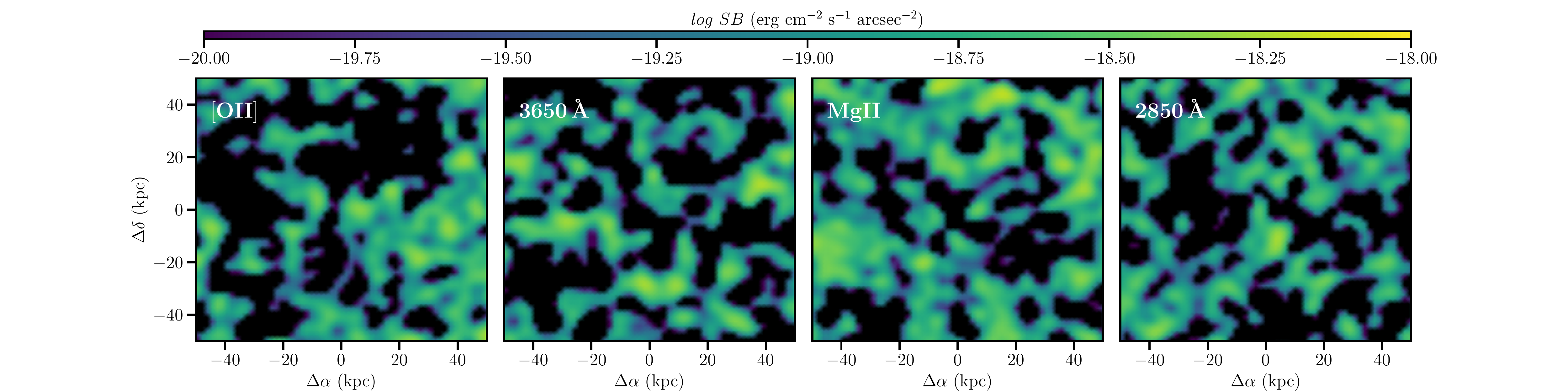}
    \caption{{\it Top:} Median stacked pseudo-narrow band maps (width of 15\,\AA) centred at the position of the galaxies (53) with associated \mgii\ absorption, around the \oii\ and \mgii\ emission lines and two nearby control wavelengths (3650\,\AA\ and 2850\,\AA).
    {\it Bottom:} Same as above, but the median stacks in this case are centred at the geometric centres of the groups (14) with associated \mgii\ absorption.}
    \label{fig:stack_images}
\end{figure*}

\begin{figure}
    \centering
    \includegraphics[width=0.5\textwidth]{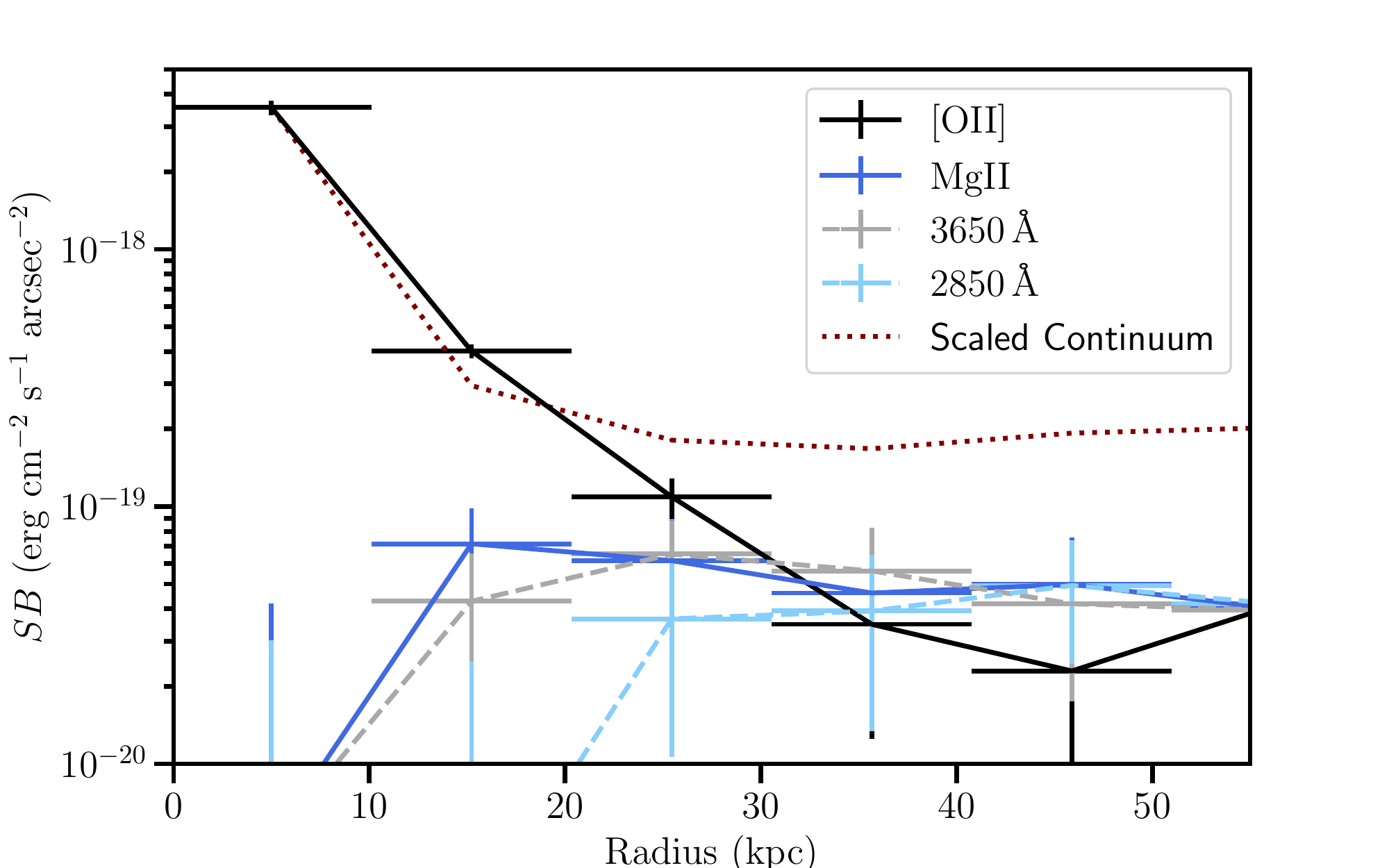}
    \caption{The average surface brightness profiles of \oii\ and \mgii\ emission lines and two nearby control wavelengths in circular annuli of 10 kpc centred on the galaxies with associated \mgii\ absorption (see top panel of Fig.~\ref{fig:stack_images}). The radial profile of the average continuum emission around the galaxies is shown in dotted lines, scaled to the peak \oii\ emission for comparison.}
    \label{fig:stack_profiles}
\end{figure}

We explore the possibility that the \mgii\ absorption could be tracing an underlying extended medium by checking for spatially extended emission. In Section~\ref{sec_analysis_emitters}, we have seen that we do not find extended \oii\ emission around the individual galaxies associated with \mgii\ absorption in the narrow band images. Here we probe the presence of an average extended medium in emission by stacking the continuum-subtracted MUSE cubes in the rest frame of the galaxies. We first obtain a stack using median statistics at the position of all the galaxies (53) with associated \mgii\ absorption detection. Then we generate pseudo-narrow band surface brightness maps of width 15\,\AA\ centred around the \oii\ and \mgii\ lines from the stacked cube. These are shown in the top panel of Fig.~\ref{fig:stack_images}, along with the maps around two control wavelengths (2850\,\AA\ and 3650\,\AA) near the emission lines to check for effects of continuum subtraction residuals. The average radial surface brightness profiles are shown in Fig.~\ref{fig:stack_profiles}. We do not detect any \mgii\ emission around the galaxies in the stacked image, down to an average surface brightness level of $\approx6\times10^{-20}$\,\ergscmarc\ between 10 and 50\,kpc. The stacked \oii\ emission is detected around the galaxies out to $\sim20$\,kpc, beyond which the emission is on average $\approx5\times10^{-20}$\,\ergscmarc. In Fig.~\ref{fig:stack_profiles}, we also plot for reference the radial profile of the average continuum emission around the galaxies, obtained by stacking the MUSE white-light images centred at the galaxy positions. The average \oii\ emission does not appear to extend beyond the continuum emission from the galaxies.   

We repeat the above stacking exercise for the closest galaxies and the most massive galaxies in case of groups, with similar results. Next, we stack the MUSE cubes at the location of the geometric centres of the groups (14) which have detection of \mgii\ absorption. The pseudo-narrow band images generated from this stacked cube are shown in the bottom panel of Fig.~\ref{fig:stack_images}. In this case we do not detect any significant \oii\ and \mgii\ emission on an average compared to the control images in the vicinity of the group centres (average surface brightness $\approx(5-8)\times10^{-20}$\,\ergscmarc\ within 50\,kpc). The results from our stacking analysis indicates that \mgii\ emission is not common around the galaxies in our sample, and that the average \oii\ emission does not extend beyond few tens of kpc from galaxies. Further, there is no evidence for the presence of an extended emission in the group environment down to the surface brightness limit probed here. 


\subsection{Origin of \mgii\ absorption in groups}
\label{sec_environment_origin}

The above results point towards a picture where the \mgii\ absorption associated with groups is stronger, more kinematically complex and spatially widespread than that associated with isolated galaxies, and within groups themselves, the absorption is stronger and more prevalent in the more massive groups. While it is difficult with pencil-beam measurements to definitively identify the origin of the \mgii\ absorption associated with groups in our sample, we discuss different scenarios that can lead to the observed enhancements of \mgii\ in groups: (i) superposition of haloes of individual galaxies in groups; (ii) outflows in the halo of a single massive galaxy within the group that is close to the quasar sightline; (iii) halo of a galaxy that is affected by gravitational or hydrodynamic interactions in the group; and (iv) a widespread diffuse intra-group medium. 
We note that the term ``intra-group medium" is usually broadly defined in the literature and there is no clear consensus on the observational signature of such a medium. For the purpose of this discussion, we use it to denote a diffuse gaseous medium beyond individual galaxy haloes, that pervades the group environment, and we further differentiate it from stripped gaseous structures around individual galaxies.
Of course, it is not expected that a single effect operates in isolation, and it is instead more likely that multiple concurrent processes give rise to the observed distribution of \mgii\ absorption in groups.

Above, we have compared the \wmg\ of groups with predictions from a superposition model which assumes that the total \mgii\ absorption in a group results from superposition of absorption by individual galaxy haloes in the groups. Since we have found a dependence of \wmg\ on the impact parameter and stellar mass of individual galaxies in our sample, we have tried to account for these in the model. Our analysis shows that a simple superposition model could explain about half ($\sim50-60$\%) of the \mgii\ absorption measurements and upper limits associated with groups. In the remaining systems, the model under-predicts the observed values in about half of the cases, while it over-predicts in the other half. In cases where it under-predicts, there could be additional factors like interactions and gas flows among group members that boost the \mgii\ absorption \citep{fossati2019b}. In cases where the model over-predicts the measured signal, the absorption could be associated predominantly with a single galaxy in the group or a more diffuse intra-group medium. 

When we compare the distributions of impact parameters and stellar masses of isolated galaxies and all the individual galaxies in groups, we find that they are consistent with being drawn from the same parent population ($P_{\rm KS}=0.5$). However, the impact parameter distribution of the galaxies closest to the quasar in groups is different from that of the isolated galaxies ($P_{\rm KS}=4\times10^{-4}$), with the closest galaxies in groups having smaller impact parameters on an average than the isolated galaxies. Similarly, the stellar masses of the most massive galaxies in groups are higher than that of the isolated galaxies ($P_{\rm KS}=0.002$). Further, among groups, those that comprise of galaxies closer to the quasar sightline or more massive galaxies tend to show stronger absorption. \wmg\ is negatively correlated with impact parameter of the closest galaxies in groups ($\tau_{\rm k} = -0.25$, $p_{\rm k} = 0.01$), and it is positively correlated to a lesser extent with the stellar mass of the most massive galaxies in groups ($\tau_{\rm k} = 0.16$, $p_{\rm k} = 0.11$). The above suggests that the stronger absorption associated with groups could be driven by a single massive member galaxy close to the quasar. These galaxies could be launching galactic-scale winds and depositing more metals into their CGM. 
However, group galaxies do not fit individually the scaling relations for isolated galaxies, in line with a scenario in which, e.g., gravitational interaction among the group members or hydrodynamic interaction of the galaxy's gas with the intra-group medium, are at play in enhancing the \mgii\ absorption strength and cross section.

There is now growing evidence, from both observations and simulations, that galaxies undergo pre-processing in groups, i.e. much of the morphological modification and gas stripping occur in the group environment before galaxies fall into the cluster centres \citep{hess2013,odekon2016,marasco2016,brown2017,jung2018}. The prominent physical mechanisms that can directly remove gas from group galaxies are ram-pressure stripping as the galaxies move through the intra-group medium and tidal stripping due to gravitational interaction among group members. Tidal interactions are more efficient in perturbing the stellar and gaseous structures of galaxies within the inner haloes of mass \mstar$<10^{14}$\,\msun, i.e. in a group-like environment, while ram-pressure stripping is believed to be more effective in denser environment \citep{boselli2006,marasco2016}. Indeed, in the local Universe, gas stripping phenomenon has been observed around galaxies in groups in the form of tidal tails, plumes and bridges \citep[e.g.][]{yun1994,cortese2006,mihos2012,rasmussen2012,serra2013,taylor2014,deblok2018,osullivan2018,for2019}.

The environmental dependence of galaxy properties has been observed out to higher redshifts ($z\sim1.5$) as well \citep{peng2010,wetzel2012,balogh2016,lemaux2019,old2020}, though the relative importance of different physical mechanisms in removing gas from galaxies in groups is still uncertain at $z\sim1$. However, it is likely that the one or more processes in the group environment causes the CGM to be perturbed and metals to be dislodged further out, leading to enhancement in the \mgii\ absorption cross section. Hence, some fraction of the \mgii\ absorption detected in our groups could arise from stripped gas, analogous to what has been observed in groups at lower redshifts \citep[$z\sim0.3$;][]{chen2019}. Indeed, based on a hydrodynamic simulation study of cold ($10^4$\,K) gas clouds in the CGM of massive haloes ($\sim10^{13}$\,\msun) at $z\sim0.5$, \citet{nelson2020} find that the cold clouds are seeded by strong local density perturbations, which could initially arise out of tidally and ram-pressure stripped gas structures in group environment.

Further, the stripped gas could be in the process of mixing and eventually forming an intra-group medium. Studies of individual \mgii\ systems at $z<0.5$ have suggested that they are tracing the cool intra-group medium \citep{gauthier2013,bielby2017}. On the other hand, \citet{fossati2019b} have not found significant evidence for a widespread cool intra-group medium via tomography with multiple sightlines. Rather, their results point towards \mgii\ absorption arising out of tidally stripped gas in groups. 
In line with these results and with direct imaging of the warm enriched gas in groups \citep{fossati2019a,chen2019}, we do not find sufficiently compelling evidence for a widespread (over $\sim$100\,kpc) intra-group medium giving rise to the observed \mgii\ absorption ($\gtrsim0.03$\,\AA).
Absorption arising out of such a medium would be distributed everywhere in the group and not be strongly correlated with individual galaxies in the groups. The correlations discussed above between \mgii\ absorption and galaxy properties in our sample seem to disfavour such a scenario. 
While a diffuse medium could still pervade the groups, it is not likely to statistically dominate the \mgii\ cross section at the equivalent width limit ($\gtrsim0.03$\,\AA) we probe here.
Emission line mapping and tomographic studies are required to actually map the distribution of such a medium. However, based on stacking analysis of the MUSE cubes, we do not detect on average \oii\ emission beyond $\sim$20\,kpc from the galaxies and detect no significant \mgii\ emission around the galaxies, down to a surface brightness limit of $\approx6\times10^{-20}$\,\ergscmarc. Further detailed kinematic analysis of galaxies and the associated absorption in individual systems would help to shed light on the true origin of the \mgii\ absorbing gas associated with multiple galaxies in our sample.
%
%=========================== SUMMARY ====================================================================================
%
\section{Summary and conclusions} 
\label{sec_summary} 
In this work, we have presented results from the MUSE Analysis of Gas around Galaxies survey \citep[MAGG;][]{lofthouse2020} on the cool metal-enriched CGM around $z\sim1$ galaxies. MAGG is based on a MUSE large programme with the primary goal of studying the connection between strong \hi\ absorbers and \lya\ emitting galaxies at $z\sim3-4$ in 28 quasar fields. The wide spatial and spectral coverage of MUSE and the availability of high S/N, high-resolution quasar spectra in all the fields, render the MAGG dataset valuable for a variety of CGM studies. Here we use the MAGG survey to build a sample of 228 continuum- and \oii\ emission line-detected galaxies at $0.8<z<1.5$ for which we could measure the corresponding \mgii\ absorption properties using the background quasar spectra. The stellar masses of the galaxies range between $10^7$ and $10^{12}$\,\msun, with a median value of $2\times10^{9}$\,\msun. The 27 \mgii\ absorbers, identified in a blind search, show rest-frame equivalent widths in the range, \wmg\ $=0.02-3.2$\,\AA, with a median value of 0.1\,\AA, and column densities between $10^{12}$\,\cms\ to $\gtrsim10^{15}$\,\cms. We probe the CGM up to impact parameter, $R\sim250-350$\,kpc and typically twice the virial radius. The main results of this study are summarized below.

\begin{itemize}

 \item[--] We identify 53 galaxies associated with 21 of the 27 \mgii\ absorption line systems in our sample (galaxy detection rate $78^{+9}_{-13}$\%). Among these galaxies, seven are in an isolated environment, the rest make up 14 associations and groups of two to six members. The majority ($67^{+12}_{-15}$\%) of the \mgii\ absorption thus arise in non-isolated environment.
 
 \item[--] The \mgii\ absorption strength exhibits a declining radial profile around galaxies. The relationship of \wmg\ (including upper limits for non-detections) with impact parameter of the isolated galaxies and the closest/most massive galaxies in groups and associations is similar to that shown by reportedly more isolated galaxy-\mgii\ absorber pairs in the literature, albeit with a larger scatter, indicating that a simple relation may not be sufficient to characterize the complex radial distribution of metals in denser environment. 
 
 \item[--] We do not find any significant azimuthal dependence of \wmg\ around the galaxies in our sample. This is contrary to studies of \mgii\ absorption that have reported an increase in \wmg\ close to the galaxy's minor axis along which biconical outflows are believed to take place. Since the majority of the absorbers are probing the CGM at $R>50$\,kpc, we propose that the distribution of metals is likely to be more symmetric further out from the galaxies, with collimated outflows not extending beyond tens of kiloparsecs.

 \item[--] The covering fraction of cool gas decreases with increasing impact parameter and increases with increasing \wmg\ sensitivity at a given impact parameter. The average covering fraction for \mgii\ absorbers with \wmg\ $\ge0.03$\,\AA\ is $\sim0.5$ ($\sim0.3$) within $R\sim100$\,kpc ($300$\,kpc).
 
 \item[--] We find a trend of stronger \mgii\ absorption being associated with more massive, and to a lesser extent, higher star forming galaxies. \wmg\ is five times higher for the sub-sample of galaxies with \mstar\ $>2\times10^9$\,\msun\ and three times higher for the sub-sample with SFR $>1.5$\,\msunyr\ compared to that for the lower mass and less star forming galaxies, respectively. Furthermore, the covering fraction is also two to three times higher on an average within the virial radius for the more massive and more star forming galaxies. This is consistent with studies in the literature that have reported stronger \mgii\ absorption and covering fraction around  more luminous galaxies. We also observe a lack of a strong trend of \wmg\ with specific star formation, suggesting that mass (and size) of a galaxy is more relevant than the actual impacts of star formation in driving these correlations.

 \item[--] We do not find any qualitative evolution in the studied trends between \mgii\ absorption and galaxy properties from lower redshift ($z<0.8$) studies in the literature to the higher redshift range ($z=0.8-1.5$) probed by our sample. Hence, there does not appear to be an important evolution in the properties of the cool metal-enriched CGM up to $z\sim1.5$.
 
 \item[--] We find a difference in the \mgii\ absorption associated with groups and isolated galaxies in our sample. Groups (here loosely defined as systems of multiple galaxies) show on an average five times the \wmg\ of isolated galaxies, and marginally higher velocity spread. The \mgii\ absorbing gas is also more prevalent in group environment. The covering fraction within $R\sim200$\,kpc (twice the virial radius) is three times higher in groups compared to isolated galaxies. Among groups, \wmg\ and covering fraction show an increasing trend with the group halo mass.
 
 \item[--] A simple superposition model where we add up the equivalent widths of isolated galaxies at similar impact parameter and stellar mass to group member galaxies is able to explain on average $\sim50-60$\% of the \mgii\ absorption associated with groups. However, it does not explain all the observed measurements, in particular at the higher and lower ends, suggesting that additional processes in the group environment play a role in enhancing the \wmg.
 
 \item[--] Galaxies in groups with smallest impact parameter are on average closer to the quasar sightline compared to the isolated galaxies. Hence, the observed enhancement of \wmg\ in groups could also be due to, e.g. starburst-driven outflows in the CGM of the galaxies closest to the sightlines. However, as these galaxies do not fit individually the correlations for more isolated galaxies, cool stripped gas that arises from environmental processes appears to contribute to the observed \mgii\ absorption in denser environment. Taking into account the dependence of \mgii\ absorption on galaxy properties and the lack of significant extended emission in stacking analysis, altogether there seems to be no strong evidence for a widespread (beyond $\gtrsim$20\,kpc from galaxies) intra-group medium that statistically dominates the \mgii\ signal ($\gtrsim0.03$\,\AA) we observe.
\end{itemize}

From this work it is clear that IFU galaxy surveys in quasar fields with complete mapping of the environment is crucial for understanding the complex interplay between galaxies and their gaseous haloes. Deeper surveys like that being carried out in the MUDF \citep{fossati2019b} will push such studies to lower mass galaxies. Further, NIR spectroscopic coverage of our fields or in fields part of recent Hubble Space Telescope surveys \citep{bielby2019,Chen2020} will help to extend the study of cool gas and of environmental effects on the metals in CGM to higher redshifts.
%
%========================================= Acknowledgment ==================================================================
%
\newline\newline
\noindent \textbf{ACKNOWLEDGEMENTS} \newline 
\noindent 
We thank the anonymous reviewer for their comments that helped improve the quality of this paper.
This project has received funding from the European Research Council (ERC) under the European Union's Horizon 2020 research and innovation programme (grant agreement No 757535). 
This work has been supported by Fondazione Cariplo, grant number 2018-2329.
RD thanks Mark Swinbank for helpful discussions about \oii\ kinematics.
SC gratefully acknowledges support from Swiss National Science Foundation grant PP00P2\_163824.
MTM thanks the Australian Research Council for Discovery Project grants DP130100568 and DP170103470, which supported this work.
This work is based on observations collected at the European Organisation for Astronomical Research in the Southern Hemisphere under ESO programmes ID 197.A-0384, 065.O-0299, 067.A-0022, 068.A-0461, 068.A-0492, 068.A-0600, 068.B-0115, 069.A-0613, 071.A-0067, 071.A-0114, 073.A-0071, 073.A-0653, 073.B-0787, 074.A-0306, 075.A-0464, 077.A-0166, 080.A-0482, 083.A-0042, 091.A-0833, 092.A-0011, 093.A-0575, 094.A-0280, 094.A-0131, 094.A-0585, 095.A-0200, 096.A-0937, 097.A-0089, 099.A-0159, 166.A-0106, 189.A-0424. 
This work used the DiRAC Data Centric system at Durham University, operated by the Institute for Computational Cosmology on behalf of the STFC DiRAC HPC Facility (www.dirac.ac.uk). This equipment was funded by BIS National E-infrastructure capital grant ST/K00042X/1, STFC capital grants ST/H008519/1 and ST/K00087X/1, STFC DiRAC Operations grant ST/K003267/1 and Durham University. DiRAC is part of the National E-Infrastructure. 
This research made use of Astropy, a community-developed core Python package for Astronomy \citep{astropy2013}. 
This research has made use of the NASA/IPAC Extragalactic Database (NED) which is operated by the Jet Propulsion Laboratory, California Institute of Technology, under contract with the National Aeronautics and Space Administration. 
This research has made use of data from the Sloan Digital Sky Survey IV (www.sdss.org), funded by the Alfred P. Sloan Foundation, the U.S. Department of Energy Office of Science, and the Participating Institutions. 
Some of the data presented herein were obtained at the W. M. Keck Observatory, which is operated as a scientific partnership among the California Institute of Technology, the University of California and the National Aeronautics and Space Administration. The Observatory was made possible by the generous financial support of the W. M. Keck Foundation. 
This research has made use of the Keck Observatory Archive (KOA), which is operated by the W. M. Keck Observatory and the NASA Exoplanet Science Institute (NExScI), under contract with the National Aeronautics and Space Administration. 
The authors wish to recognise and acknowledge the very significant cultural role and reverence that the summit of Maunakea has always had within the indigenous Hawaiian community. We are most fortunate to have the opportunity to conduct observations from this mountain. 
Some of the data presented in this work were obtained from the Keck Observatory Database of Ionized Absorbers toward QSOs (KODIAQ), which was funded through NASA ADAP grant NNX10AE84G. The Keck Observatory Archive (KOA) is a collaboration between the NASA Exoplanet Science Institute (NExScI) and the W. M. Keck Observatory (WMKO). NExScI is sponsored by NASA’s Exoplanet Exploration Program, and operated by the California Institute of Technology in coordination with the Jet Propulsion Laboratory (JPL). 
\newline\newline
\noindent \textbf{DATA AVAILABILITY} \newline
\noindent
The data used in this work are available from \url{https://archive.eso.org/} and the online supplementary material, while the codes used in this work are available at \url{http://www.michelefumagalli.com/codes.html}.
%======================================== Bibliography =====================================================================
% 
\def\aj{AJ}%
\def\actaa{Acta Astron.}%
\def\araa{ARA\&A}%
\def\apj{ApJ}%
\def\apjl{ApJ}%
\def\apjs{ApJS}%
\def\ao{Appl.~Opt.}%
\def\apss{Ap\&SS}%
\def\aap{A\&A}%
\def\aapr{A\&A~Rev.}%
\def\aaps{A\&AS}%
\def\azh{A$Z$h}%
\def\baas{BAAS}%
\def\bac{Bull. astr. Inst. Czechosl.}%
\def\caa{Chinese Astron. Astrophys.}%
\def\cjaa{Chinese J. Astron. Astrophys.}%
\def\icarus{Icarus}%
\def\jcap{J. Cosmology Astropart. Phys.}%
\def\jrasc{JRASC}%
\def\mnras{MNRAS}%
\def\memras{MmRAS}%
\def\na{New A}%
\def\nar{New A Rev.}%
\def\pasa{PASA}%
\def\pra{Phys.~Rev.~A}%
\def\prb{Phys.~Rev.~B}%
\def\prc{Phys.~Rev.~C}%
\def\prd{Phys.~Rev.~D}%
\def\pre{Phys.~Rev.~E}%
\def\prl{Phys.~Rev.~Lett.}%
\def\pasp{PASP}%
\def\pasj{PASJ}%
\def\qjras{QJRAS}%
\def\rmxaa{Rev. Mexicana Astron. Astrofis.}%
\def\skytel{S\&T}%
\def\solphys{Sol.~Phys.}%
\def\sovast{Soviet~Ast.}%
\def\ssr{Space~Sci.~Rev.}%
\def\zap{$Z$Ap}%
\def\nat{Nature}%
\def\iaucirc{IAU~Circ.}%
\def\aplett{Astrophys.~Lett.}%
\def\apspr{Astrophys.~Space~Phys.~Res.}%
\def\bain{Bull.~Astron.~Inst.~Netherlands}%
\def\fcp{Fund.~Cosmic~Phys.}%
\def\gca{Geochim.~Cosmochim.~Acta}%
\def\grl{Geophys.~Res.~Lett.}%
\def\jcp{J.~Chem.~Phys.}%
\def\jgr{J.~Geophys.~Res.}%
\def\jqsrt{J.~Quant.~Spec.~Radiat.~Transf.}%
\def\memsai{Mem.~Soc.~Astron.~Italiana}%
\def\nphysa{Nucl.~Phys.~A}%
\def\physrep{Phys.~Rep.}%
\def\physscr{Phys.~Scr}%
\def\planss{Planet.~Space~Sci.}%
\def\procspie{Proc.~SPIE}%
\let\astap=\aap
\let\apjlett=\apjl
\let\apjsupp=\apjs
\let\applopt=\ao
\bibliographystyle{mnras}
\bibliography{mybib}
\bsp
\label{lastpage}
\end{document}